\documentclass[11pt,a4paper]{scrartcl}
\usepackage{ILD}
\usepackage[symbol]{footmisc}

\newcommand{\PUREeLpR}{\ensuremath{\mathrm{e^-_L e^+_R}}}

\newcommand{\PUREeRpL}{\ensuremath{\mathrm{e^-_R e^+_L}}}

\newcommand{\eLpR}{\ensuremath{\mathrm{e^-_{L80} e^+_{R30}}}}
\newcommand{\eLpL}{\ensuremath{\mathrm{e^-_{L80} e^+_{L30}}}}
\newcommand{\eRpL}{\ensuremath{\mathrm{e^-_{R80} e^+_{L30}}}}
\newcommand{\eRpR}{\ensuremath{\mathrm{e^-_{R80} e^+_{R30}}}}
\newcommand{\eett}{\ensuremath{e^- e^+ \to \tau^- \tau^+}}
\newcommand{\tpn}{\ensuremath{\tau^\pm \to \pi^\pm \nu}}
\newcommand{\trn}{\ensuremath{\tau^\pm \to \pi^\pm \pi^0 \nu}}
\newcommand{\tAn}{\ensuremath{\tau^\pm \to \pi^\pm \pi^0 \pi^0 \nu}}

\title{ILD benchmark: a study of \eett\ at 500 GeV}

\def\MyNoteNumber{ILD-PHYS-PUB--2019-004}

\date{10 October 2019}
\addauthor{Daniel Jeans}{\institute{1}}
\addauthor{Keita Yumino}{\institute{2}}
\addinstitute{1}{IPNS, KEK}
\addinstitute{2}{Sokendai}

\abstract{
The process \eett\ is of particular interest because the tau lepton polarisation can
be reconstructed, allowing its chiral nature to be probed.
This note reports on a study of the reconstruction of the di-tau final state at ILC-500, its selection
and the reduction of backgrounds, the identification of the tau lepton's decay mode, and on the extraction of the tau leptons' polarisation.
The performance of this analysis is studied in two models of the ILD detector, one larger (IDR-L) the other smaller (IDR-S),
which differ in the outer radius of the TPC and of the 
subdetectors beyond, and in the magnetic field strength of the detector solenoid.

We find that the high-mass tau-pair events in which at least one tau decays haronically can be selected with an efficiency of
around 60\%, with a remaining background from non-di-tau processes at the few-\% level. Single-prong decay modes \tpn, \trn, \tAn\ can be correctly
identified in around 60-90\% of cases, with sample purities in the range 50-90\%, depending on decay mode.

The sensitivity to tau polarisation was estimated in the four beam polarisation datasets envisaged for the $4~\mathrm{ab}^{-1}$ of data forseen for ILC-500. 
Statistical precisions on the polarisation in the different datasets are predicted to be between 0.5 and 2\%.
While some small performance differences between the two detector models are seen, they have very similar final sensitivity to the polarisation measurement.
}

\titlecomment{This work was carried out in the framework of the ILD detector concept group}

\begin{document}

\titlepage







\section{Introduction}

In this note we study high invariant mass tau lepton pairs at ILC--500. 
We use simple methods to reconstruct and select a high purity sample of such events.
The tau lepton, with its rather short lifetime, allows reconstruction of its spin direction by the 
distribution of its decay products. Maximum sensitivity to the spin orientation
requires reconstruction of the tau decay mode and the kinematics of its decay.
We develop a cut-based procedure to distinguish decay modes, and 
reconstruct the tau decay kinematics, allowing reconstruction of the tau polarisation.
The performance of this analysis is studied in two models of the ILD detector, one larger (``IDR-L'') the other smaller (``IDR-S''),
which differ in the outer radius of the TPC (IDR-L: 1770~mm, IDR-S: 1427~mm) and of the subdetectors beyond, 
and in the strength of the detector's solenoidal magnetic field (IDR-L: 3.5~T, IDR-S: 4.0~T).



\section{Simulation setup}

\begin{figure}
\centering
\includegraphics[width=0.45\textwidth]{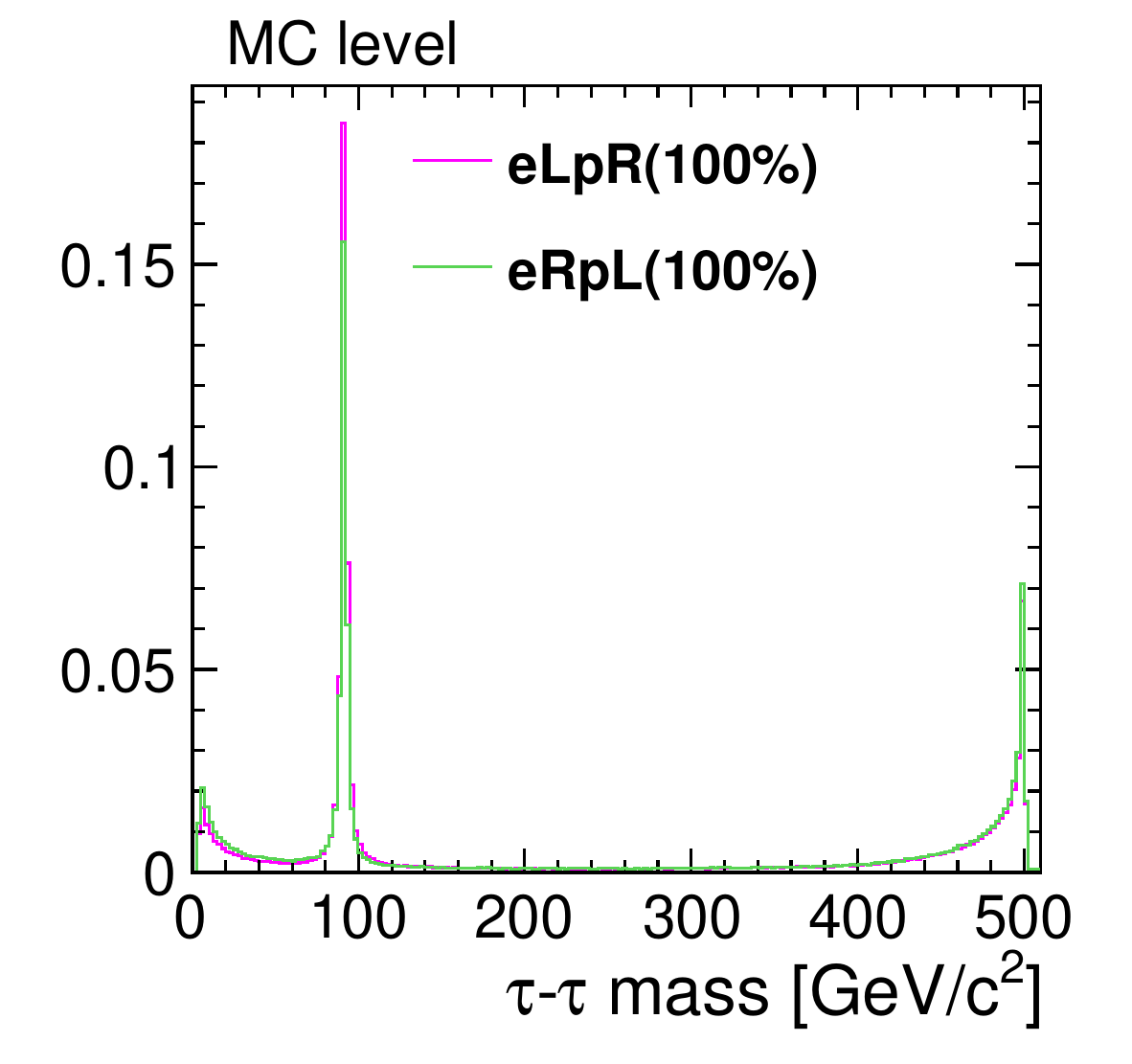}
\includegraphics[width=0.45\textwidth]{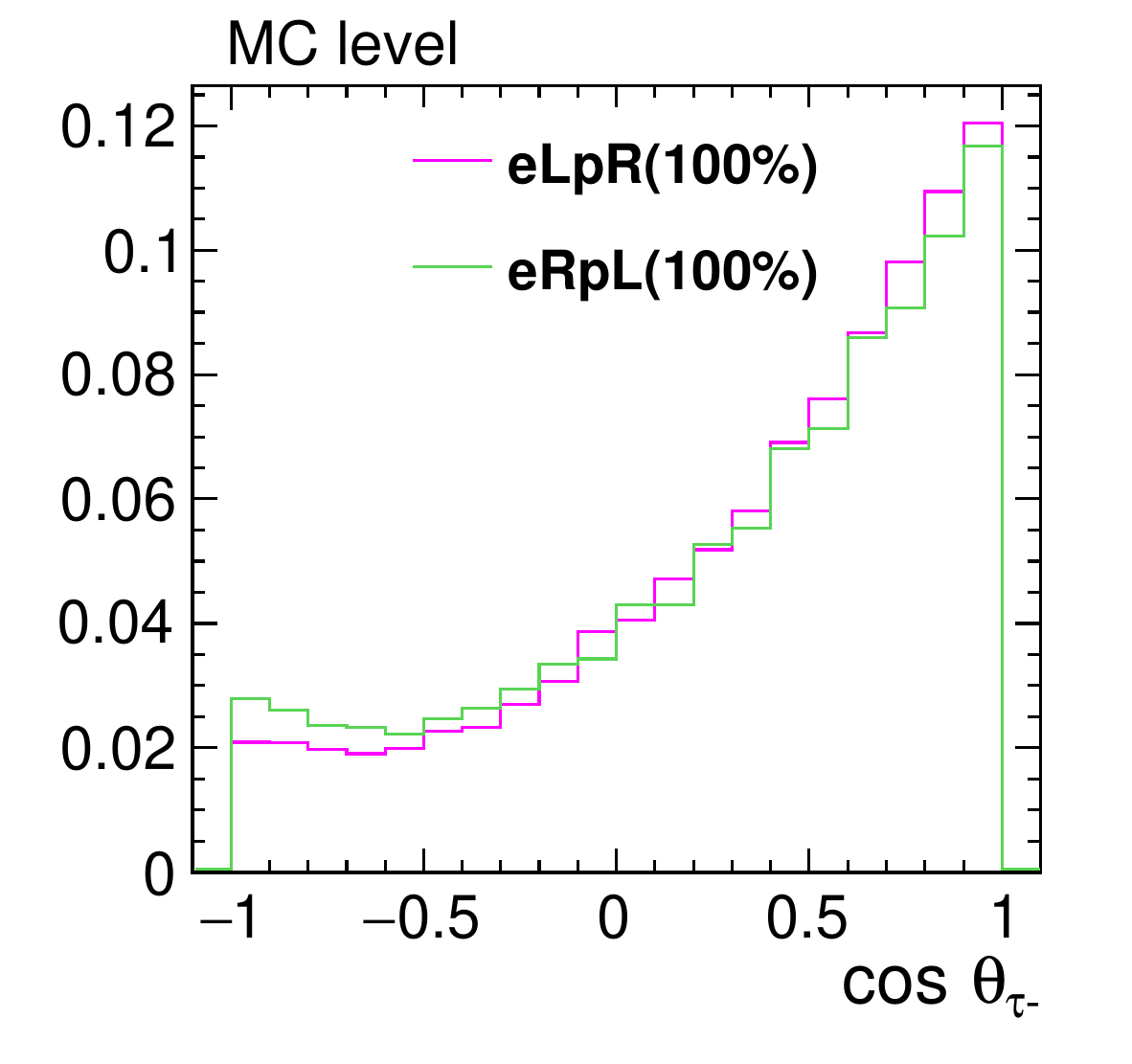}
\caption{MC distributions: 
  the tau-tau invariant mass $m_{\tau\tau}$, and
  $\cos \theta_{\tau -}$ for events with $m_{\tau\tau} >$ 480 GeV.
}
\label{fig:mcdists}
\end{figure}

\begin{figure}
\centering
\includegraphics[width=0.9\textwidth]{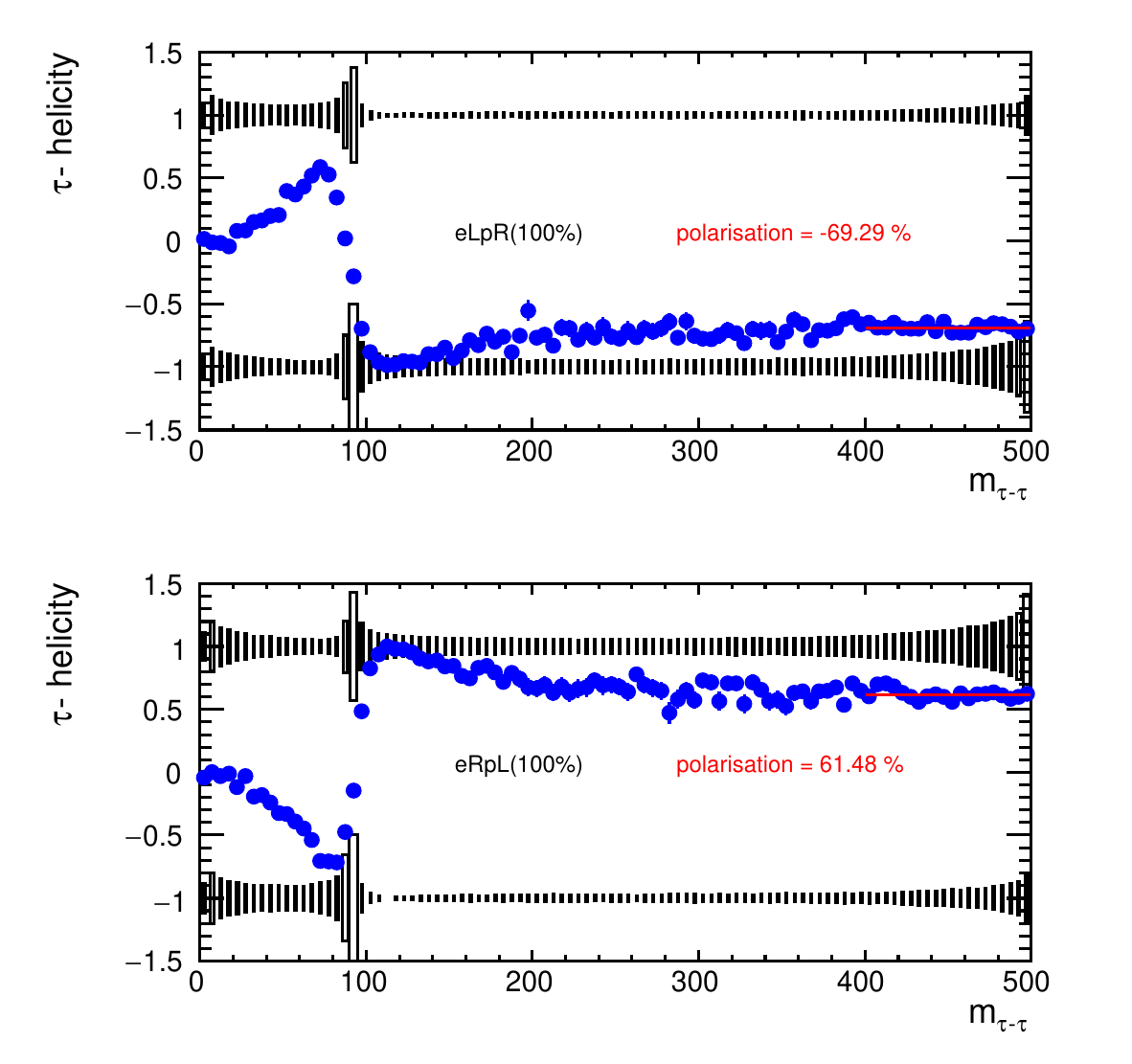}
\caption{The box histograms show the distribution of the helicity of the $\tau^-$ as a function of the invariant
mass of the $\tau-$pair $m_{\tau-\tau}$ in the input MC samples. The blue points show the average helicity (in other words the polarisation).
The red line is a fit to the points to a constant value in the high mass region. Upper [lower] plot: pure \PUREeLpR\ [\PUREeRpL] initial state.
}
\label{fig:mcdists1a}
\end{figure}

Signal event samples were generated using \texttt{WHIZARD} version 1.95~\cite{whizard, omega}, 
producing pairs of polarised tau leptons from polarised beams, taking into account the ILC beam energy spread, beamstrahlung and initial state radiation. 
The decay of the polarised tau leptons was done using \texttt{TAUOLA}~\cite{Tauola}. Two samples were used, with 
100\% polarised left-handed electron/right-handed positron (\PUREeLpR) and right-handed electron/left-handed positron (\PUREeRpL) beams.
In the \eett\ process, we plot the tau-tau invariant mass  and scattering angle in Fig.~\ref{fig:mcdists}. 
These are shown at MC level, separately for samples with (100\%) \PUREeLpR\ and \PUREeRpL\ beam polarisations. 
Figure~\ref{fig:mcdists1a} shows the tau polarisation as a function of the tau-pair invariant mass. 
The initial beam polarisation affects the relative contributions of $\gamma$ and $Z$ (or at these energies $B$ and $W^0$) to the process.
For high invariant mass tau pairs, the tau polarisations with 100\% polarised beams are consistent with 
the expected $ ( 1 - 4 \sin^4 \theta_W ) / (  1 + 4 \sin^4 \theta_W ) \sim 67\% $ for \PUREeLpR\ 
and $ ( (Y^\tau_R)^2 - (Y^\tau_L)^2 ) / ( (Y^\tau_R)^2 + (Y^\tau_L)^2 ) \sim 60\% $ for \PUREeRpL.

A full set of Standard Model background processes were produced using the same \texttt{WHIZARD} version.
Events were simulated and reconstructed using standard ILD tools based on \texttt{ddsim/DD4hep}~\cite{dd4hep} and \texttt{MarlinReco}~\cite{marlinreco}.
The principal output of the reconstruction is a collection of Particle Flow Objects (PFO), corresponding to reconstructed
final state particles.

The analysis is performed assuming the 500~GeV portion of data provided in the ``H-20'' ILC running scenario~\cite{ILCrunning}.
The electron (positron) beam is 80\% (30\%) polarised, either left or right--handed. The $4.0~\mathrm{ab^{-1}}$ of integrated luminosity
forseen at 500~GeV is split among different beam polarisation combinations (\eLpR\ : \eRpL\ : \eLpL\ : \eRpR) as (40\%\ : 40\%\ : 10\%\ : 10\%).

\section{Polarimeters}
\label{sec:pol}


Optimal tau polarimeter vectors can be rather simply defined in the case of \tpn\ (which we sometimes abbreviate as ``$\tau \rightarrow \pi$'') and 
\trn\  (``$\tau \rightarrow \rho$'') decays, see e.g.~\cite{Tauola}.
The polarimeter vectors are defined in the tau rest frames as follows: for \tpn, it is the direction of the neutrino momentum,
while for \trn\ it is the direction of the vector $\mathbf{P} = 2 (\mathbf{q} \cdot \mathbf{p_\nu}) \mathbf{q} - m_q^2 \mathbf{p_\nu}$, where 
$\mathbf{q} = \mathbf{p}_{\pi^\pm} - \mathbf{p}_{\pi^0}$, and $\mathbf{p_\nu}, \mathbf{p}_{\pi^\pm}, \mathbf{p}_{\pi^0}$ are respectively
the 3-momenta of the neutrino, charged and neutral pions. 
To distinguish taus of different helicity, we consider the cosine of the angle this polarimeter vector makes to the tau flight direction: we call this the ``polarimeter''.
We refer to this form of the polarimeters as ``optimal''.

This optimal form of the polarimeter requires knowledge of the tau neutrino momentum, which is obviously not directly measureable.
We leave for a future study the reconstruction of full tau momenta (including the neutrino component) in this di-tau final state.

\begin{figure}
\centering
\includegraphics[width=0.45\textwidth]{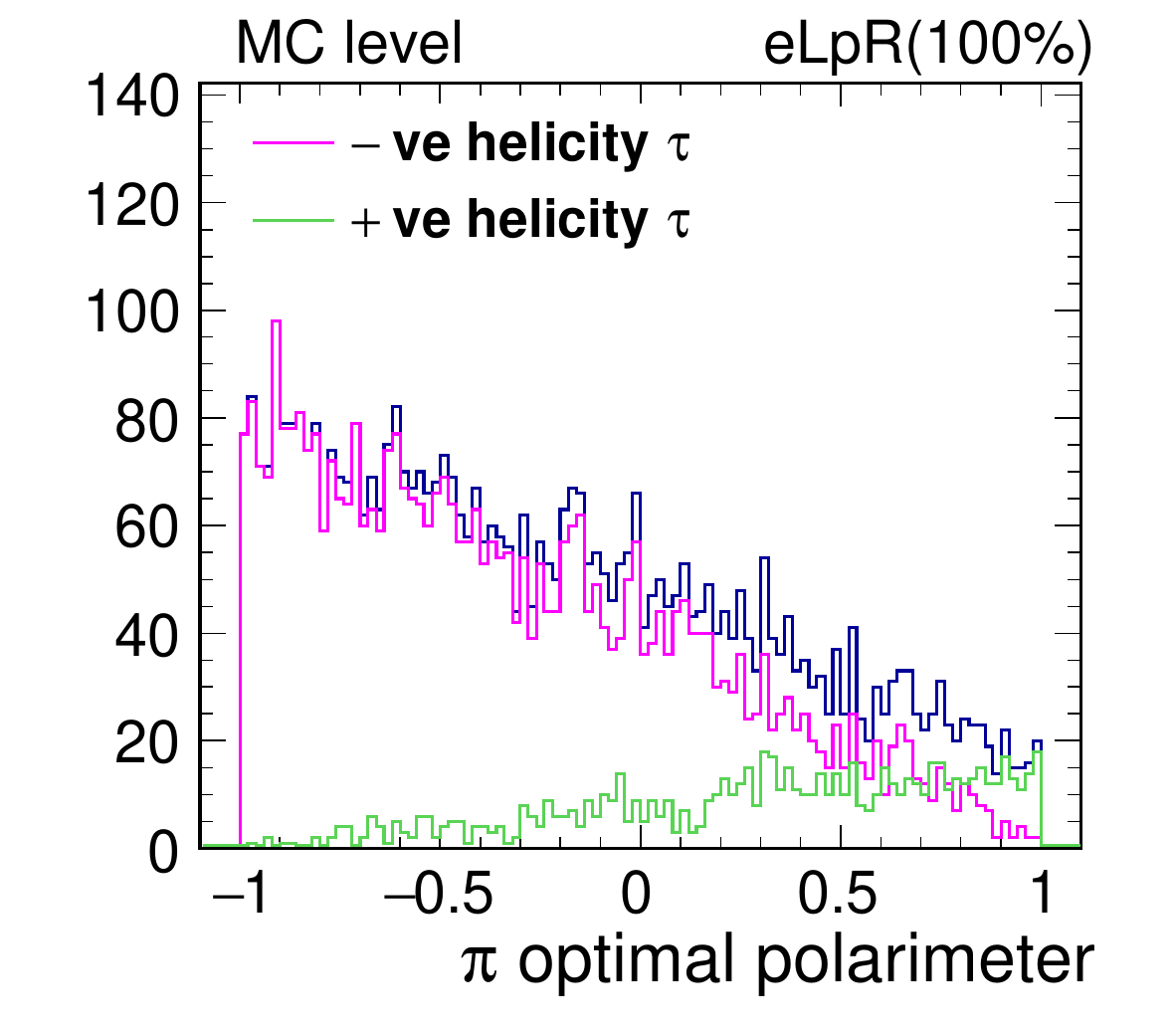} 
\includegraphics[width=0.45\textwidth]{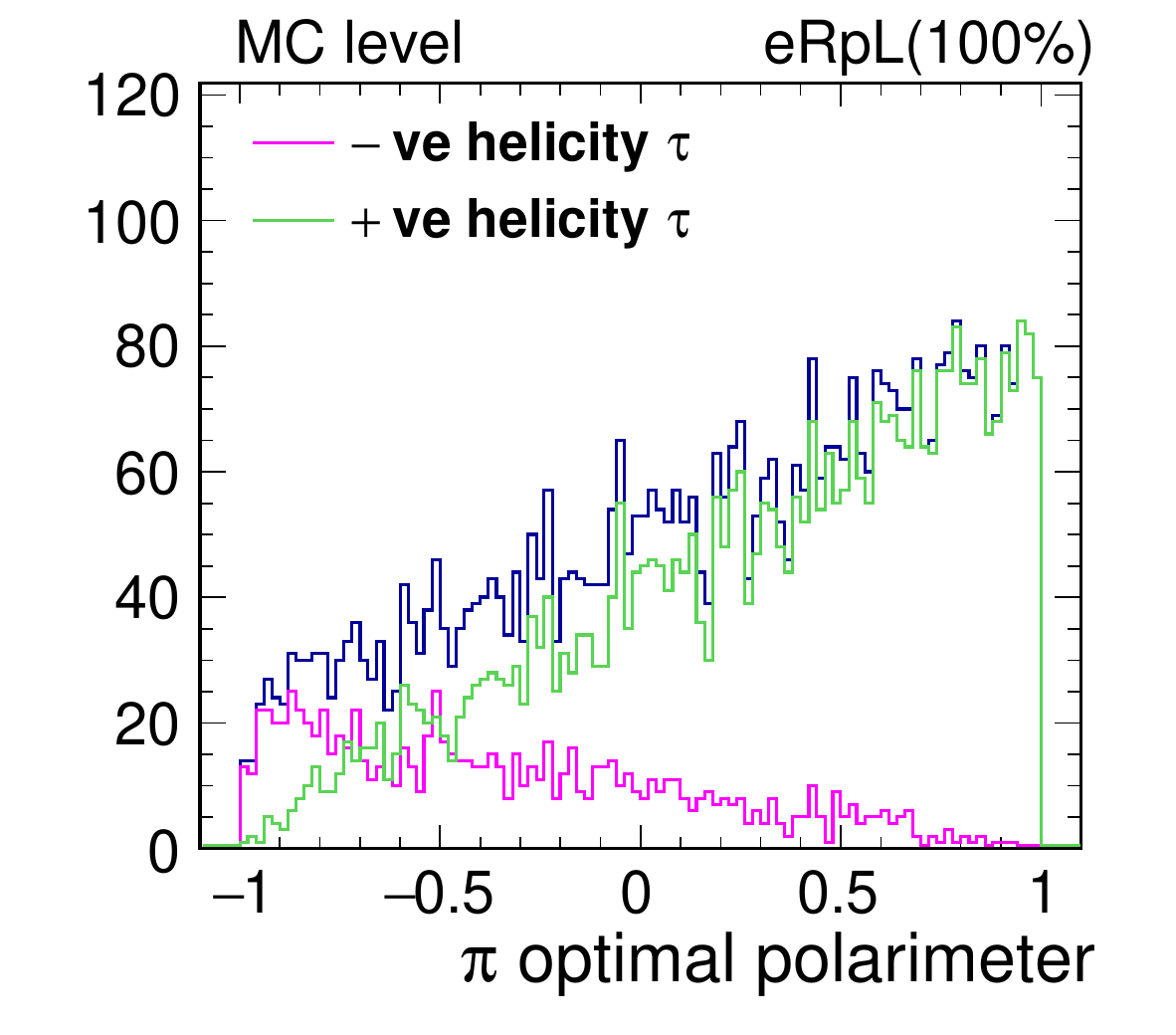} \\
\includegraphics[width=0.45\textwidth]{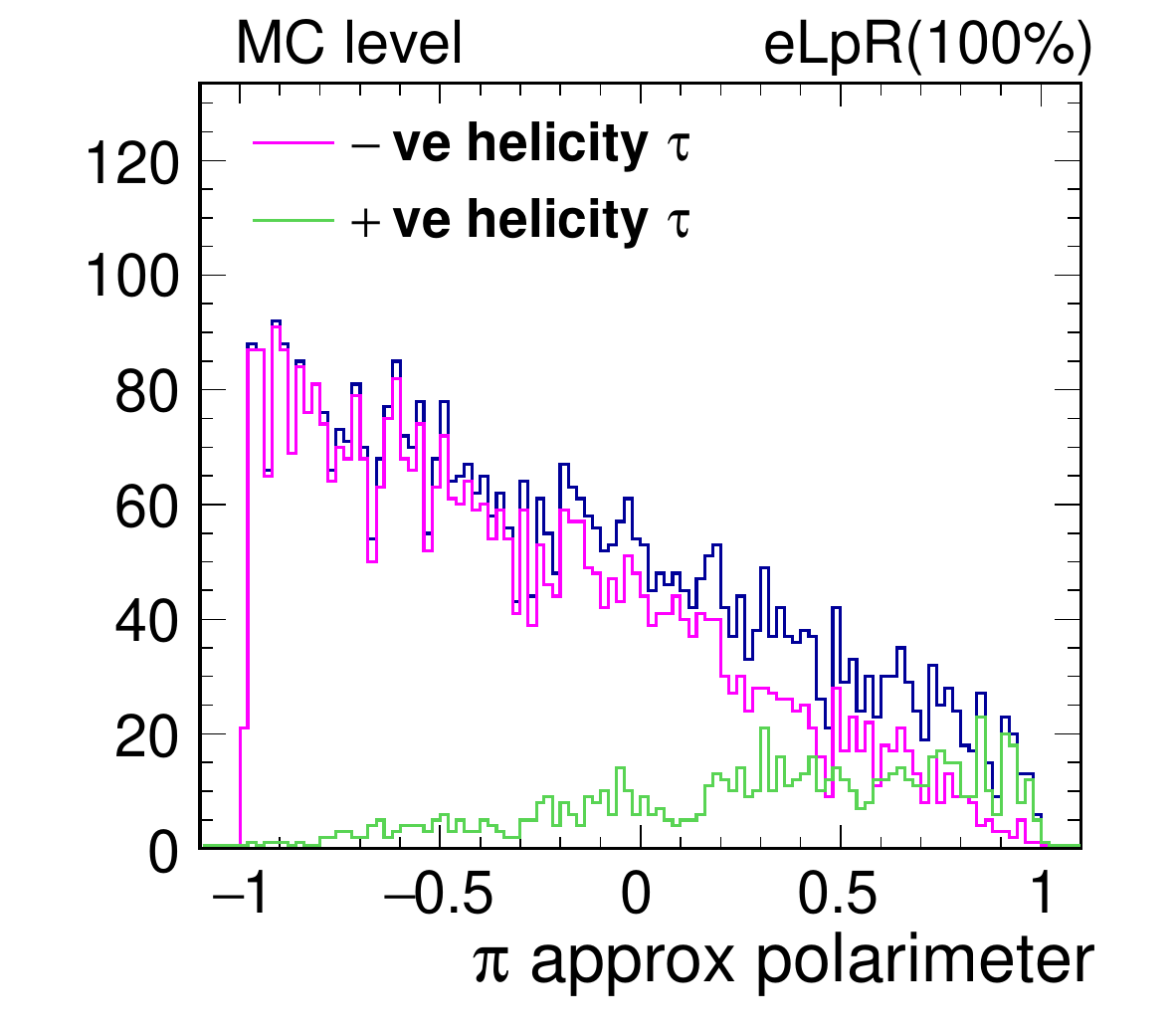}
\includegraphics[width=0.45\textwidth]{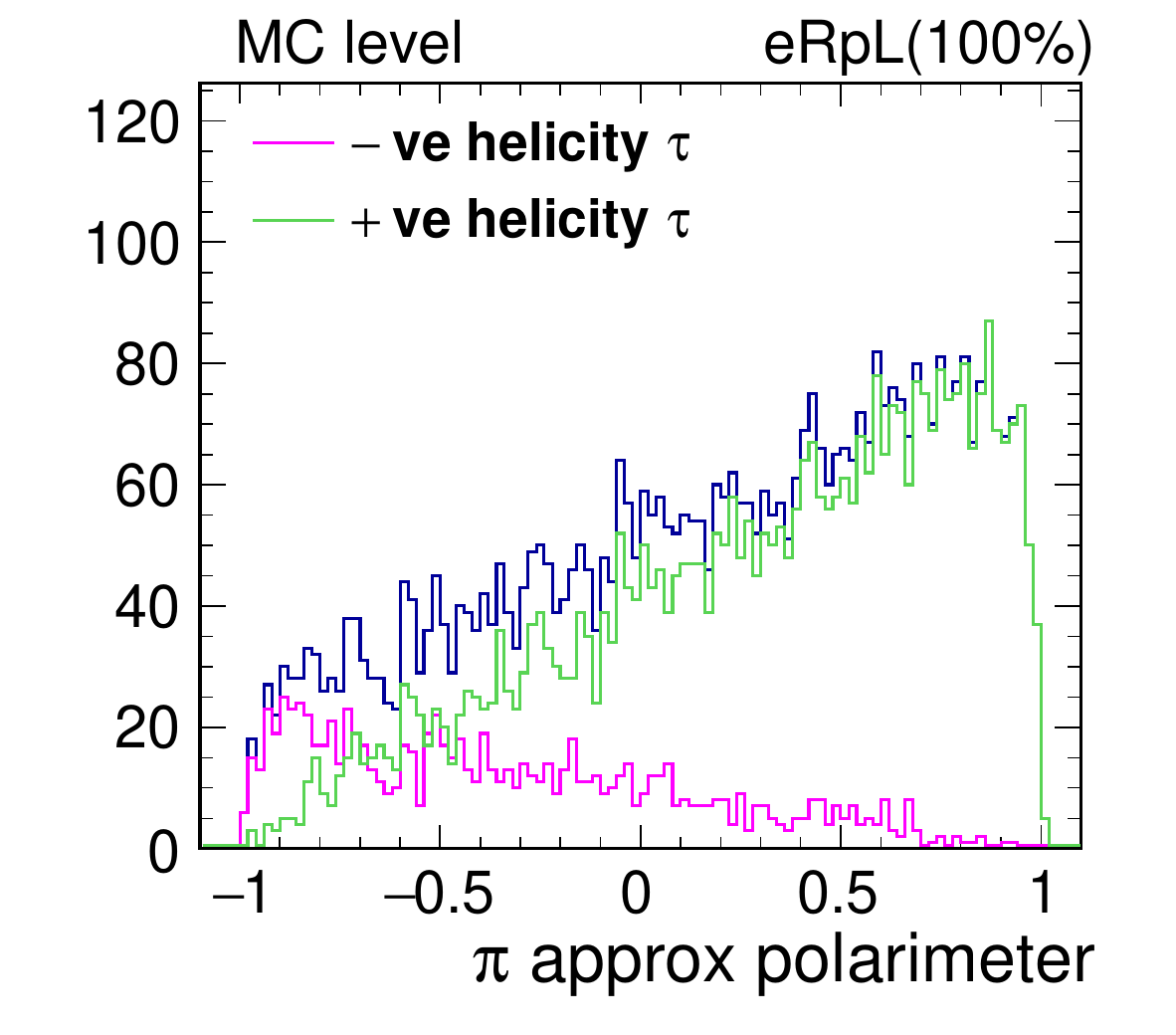}
\caption{
Polarimeter distributions for \tpn\ decays in events with $m_{\tau\tau} >$ 480 GeV, calculated using the true MC momenta of tau decay products. 
The upper (lower) plots show the ``optimal'' (``approximate'') forms of the polarimeters for taus of positive and negative helicity, 
while left and right plots are for different initial beam polarisations.
}
\label{fig:mcdists2}
\end{figure}

\begin{figure}
\centering
\includegraphics[width=0.45\textwidth]{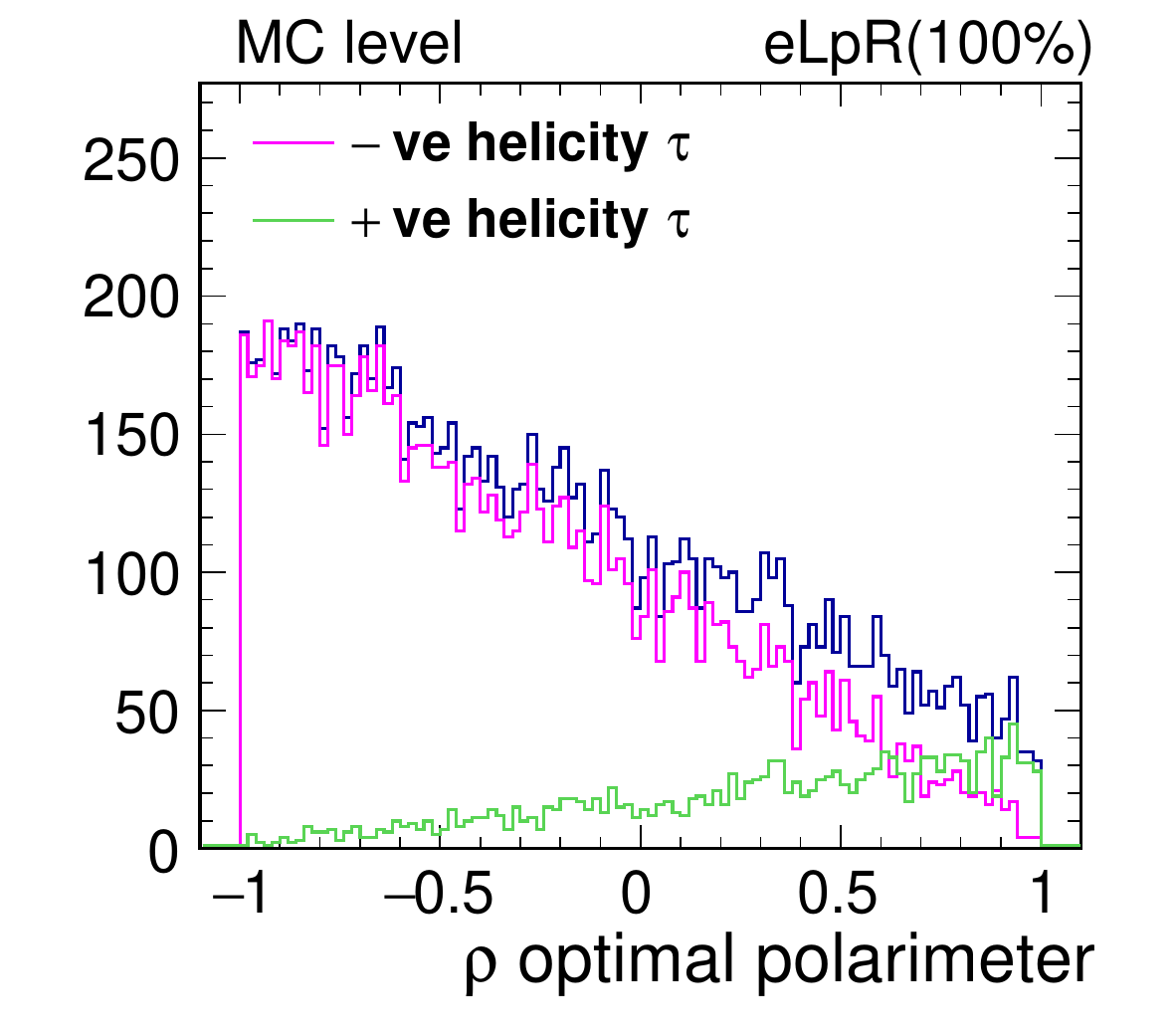}
\includegraphics[width=0.45\textwidth]{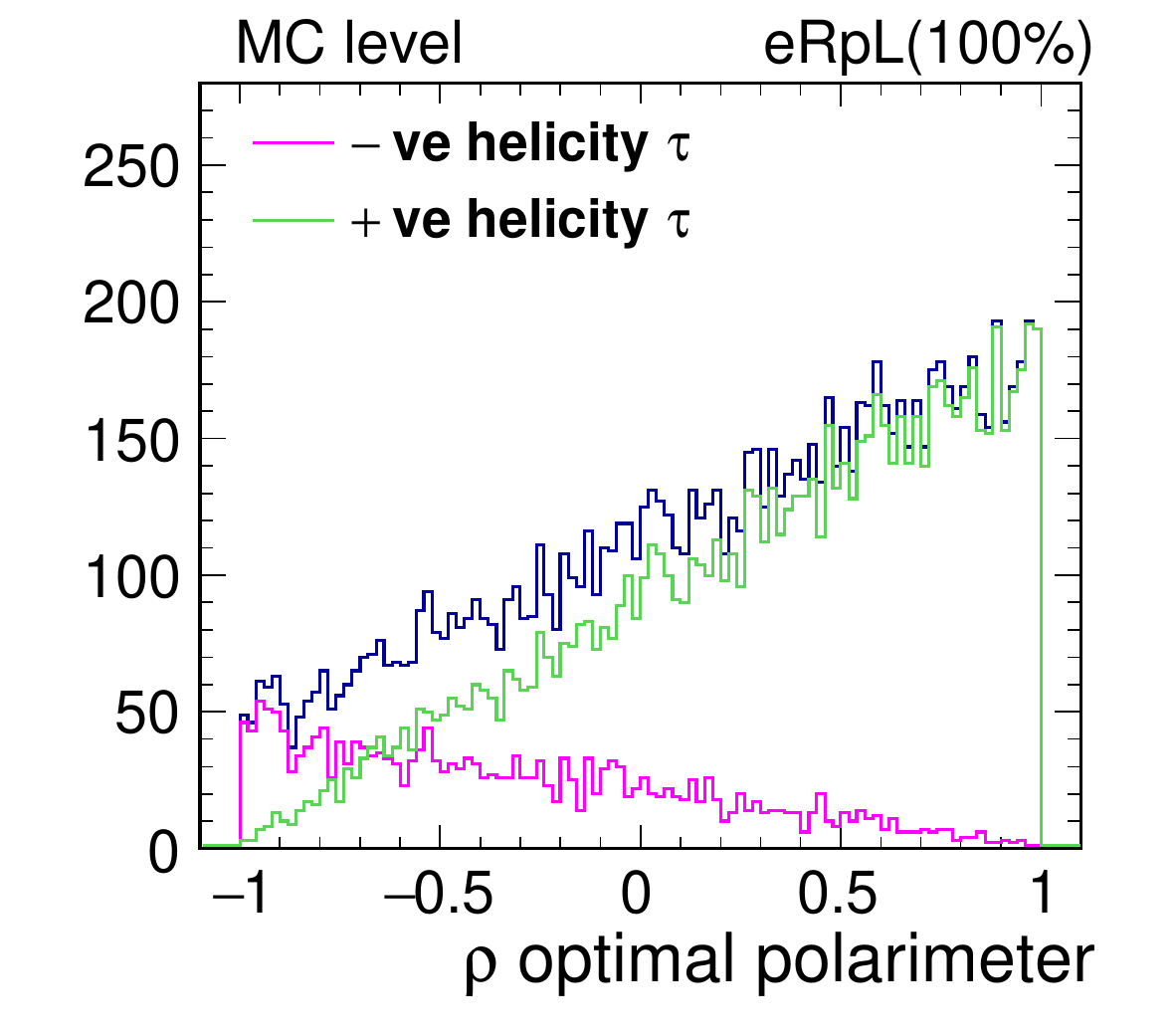} \\
\includegraphics[width=0.45\textwidth]{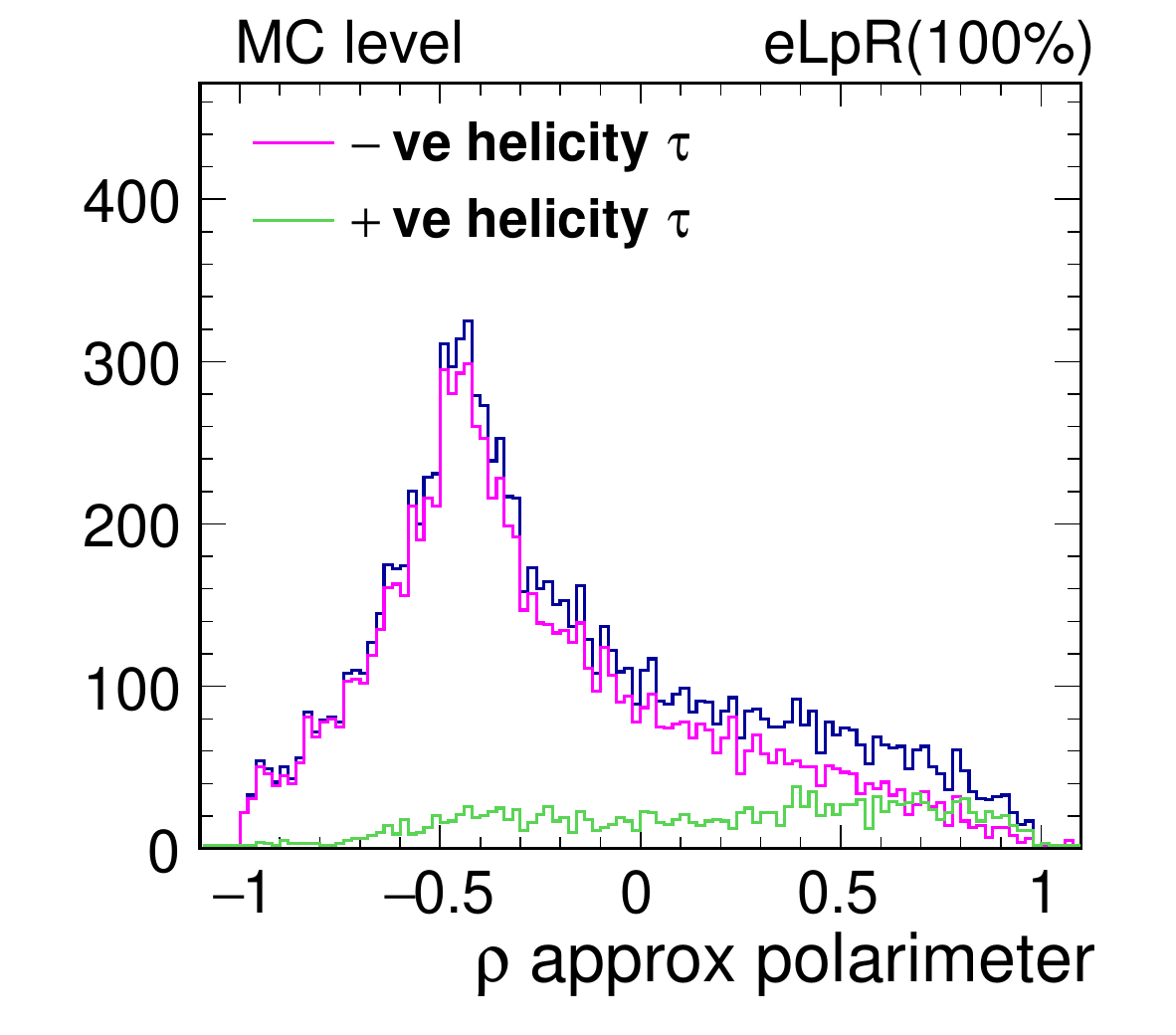}
\includegraphics[width=0.45\textwidth]{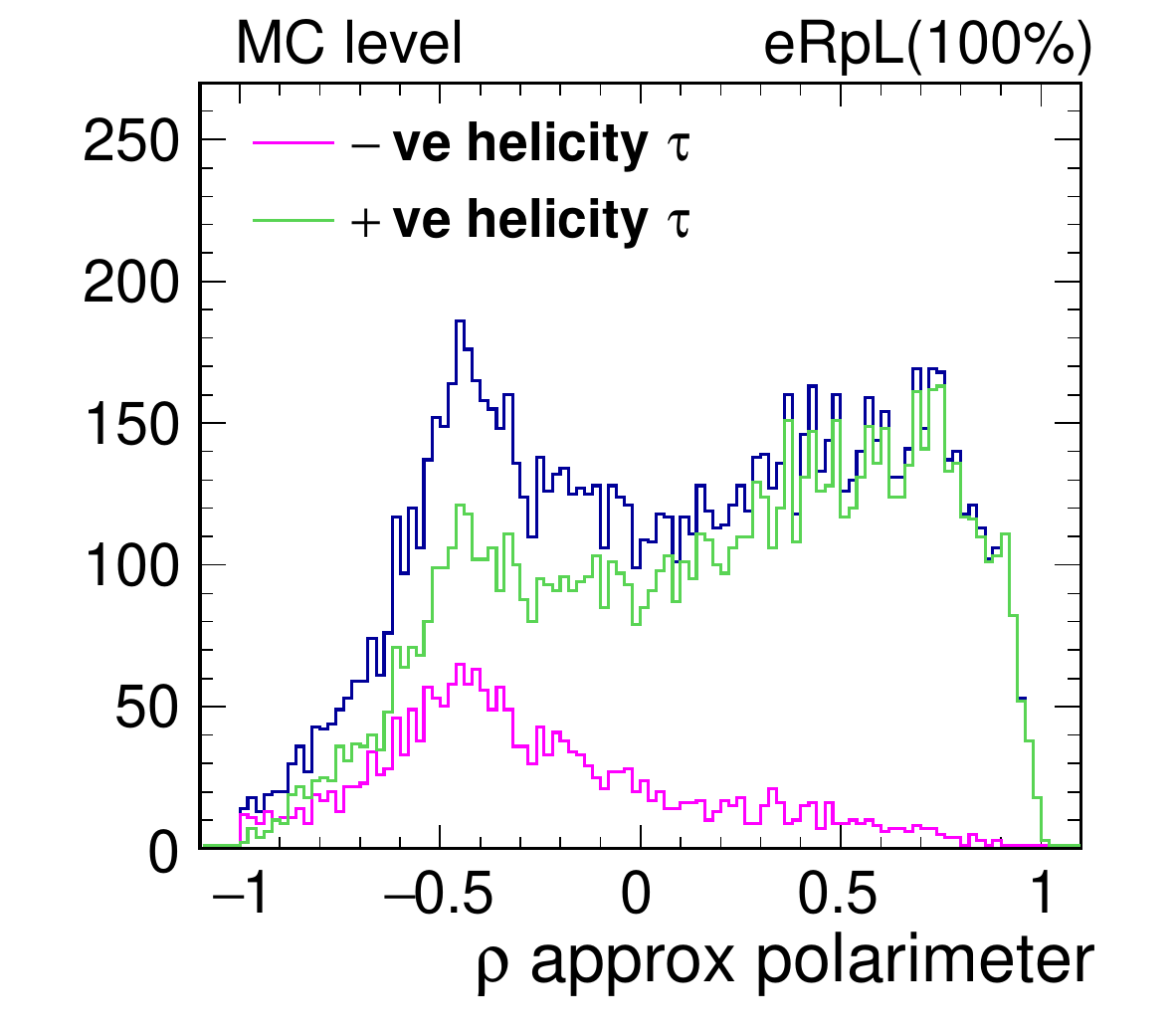}
\caption{
Polarimeter distributions for \trn\ decays in events with $m_{\tau\tau} >$ 480 GeV, calculated using the true MC momenta of tau decay products. 
The upper (lower) plots show the ``optimal'' (``approximate'') forms of the polarimeters for taus of positive and negative helicity, 
while left and right plots are for different initial beam polarisations.
}
\label{fig:mcdists3}
\end{figure}

``Approximate'' polarimeters can be defined, which are reconstructed based only on the momenta of visible tau decay products. In the 2-body decay 
\tpn, 
the energy of the pion is an optimal polarimeter for taus of known energy, however due to 
the spread in beam energies at a real collider and the resulting spread in tau energies, its sensitivity is slightly decreased.
In the case of \trn, one can arrive at an approximate polarimeter by integrating over possible
neutrino momenta, as described in \cite{duflot}. The resulting form of the polarimeters is reproduced in appendix~\ref{appendix:1}.
The resulting ``approximate'' polarimeters, calculated using MC truth information on the decay products' momenta, 
are compared to the ``optimal'' ones in figs.~\ref{fig:mcdists2} and \ref{fig:mcdists3} respectively for 
\tpn\ and \trn\ decays.
In the case of \tpn\ decays, the approximate method retains almost all the sensitivity of the ``optimal'' analysis, 
while for \trn\ decays the sensitivity of the ``approximate'' method is significantly smaller than the ``optimal'' one.

The aim of this analysis is to estimate how well we can reconstruct these polarimeter distributions using fully simulated, reconstructed, and
selected events, comparing two the models of ILD, ``IDR-L'' and ``IDR-S''.

\section{Event selection}

It is the semi-leptonic tau decays (in which the tau decays to a single neutrino plus hadrons) which are most sensitive to tau 
polarisation (fully leptonic modes suffer from the presence of two neutrinos per tau decay).
We therefore emphasise hadronic decays, in particular \tpn\ (the cleanest hadronic decay) and 
\trn\ (which has the largest decay branching ratio, accounting for around 26\% of tau decays).
We concentrate our efforts on events in which the tau pair invariant mass is close to the nominal collision energy of 500 GeV.

From the detector point of view, the identification and measurement of the charged hadron is rather easy.
The most sensitive aspect is probably the reconstruction and measurement of the $\pi^0$ decay products in the highly boosted tau decays.

\begin{figure}
\centering
\includegraphics[width=0.45\textwidth]{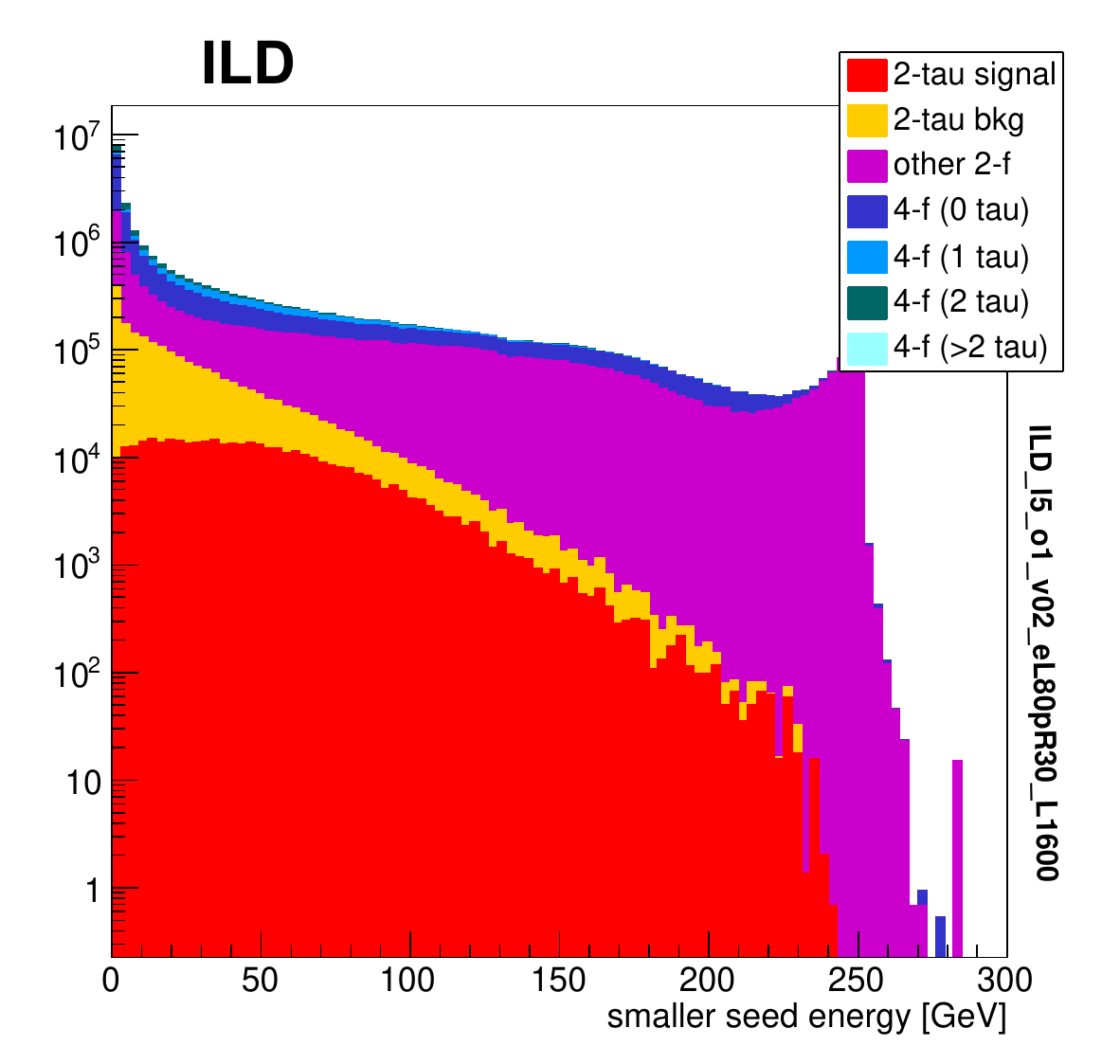}
\includegraphics[width=0.45\textwidth]{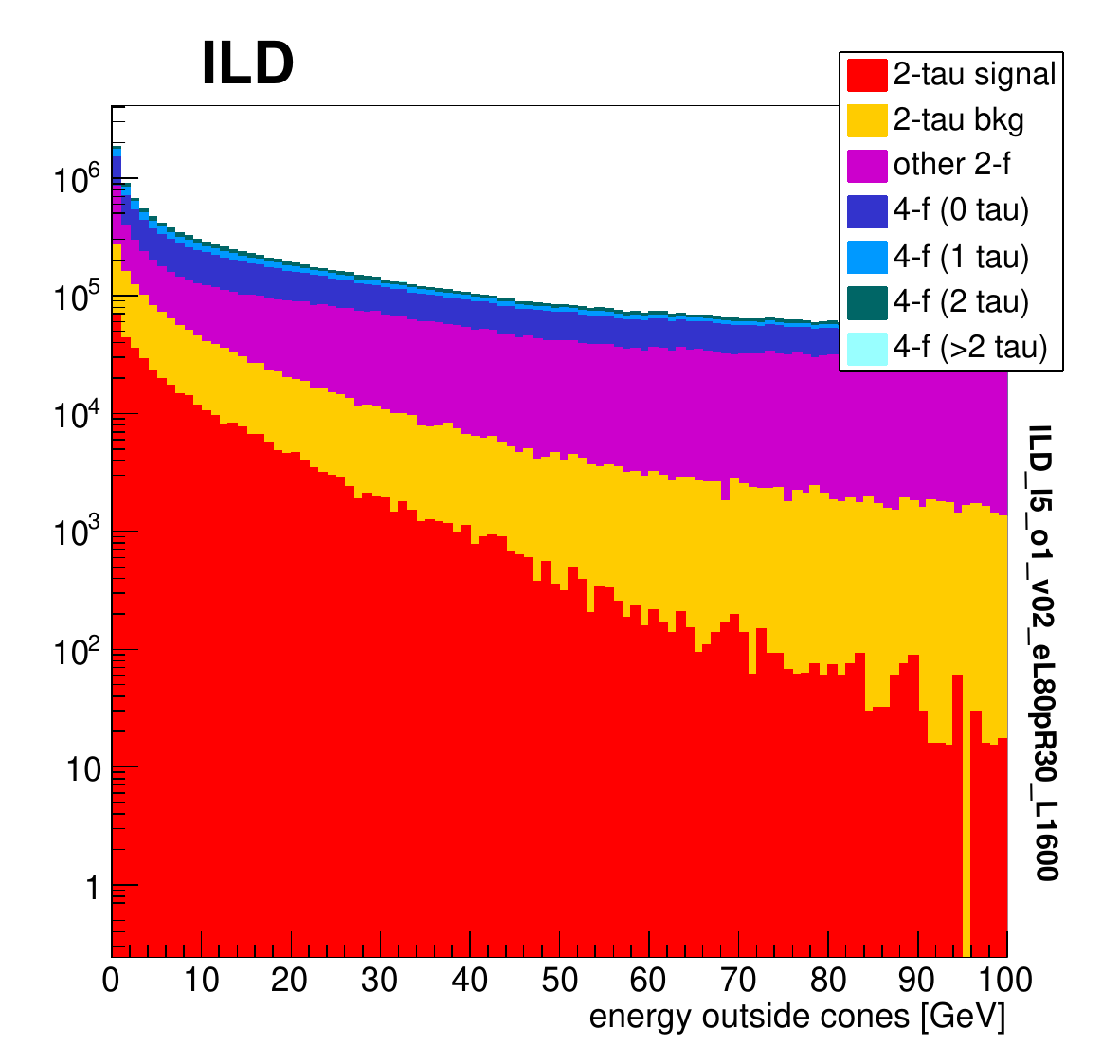} \\
\includegraphics[width=0.45\textwidth]{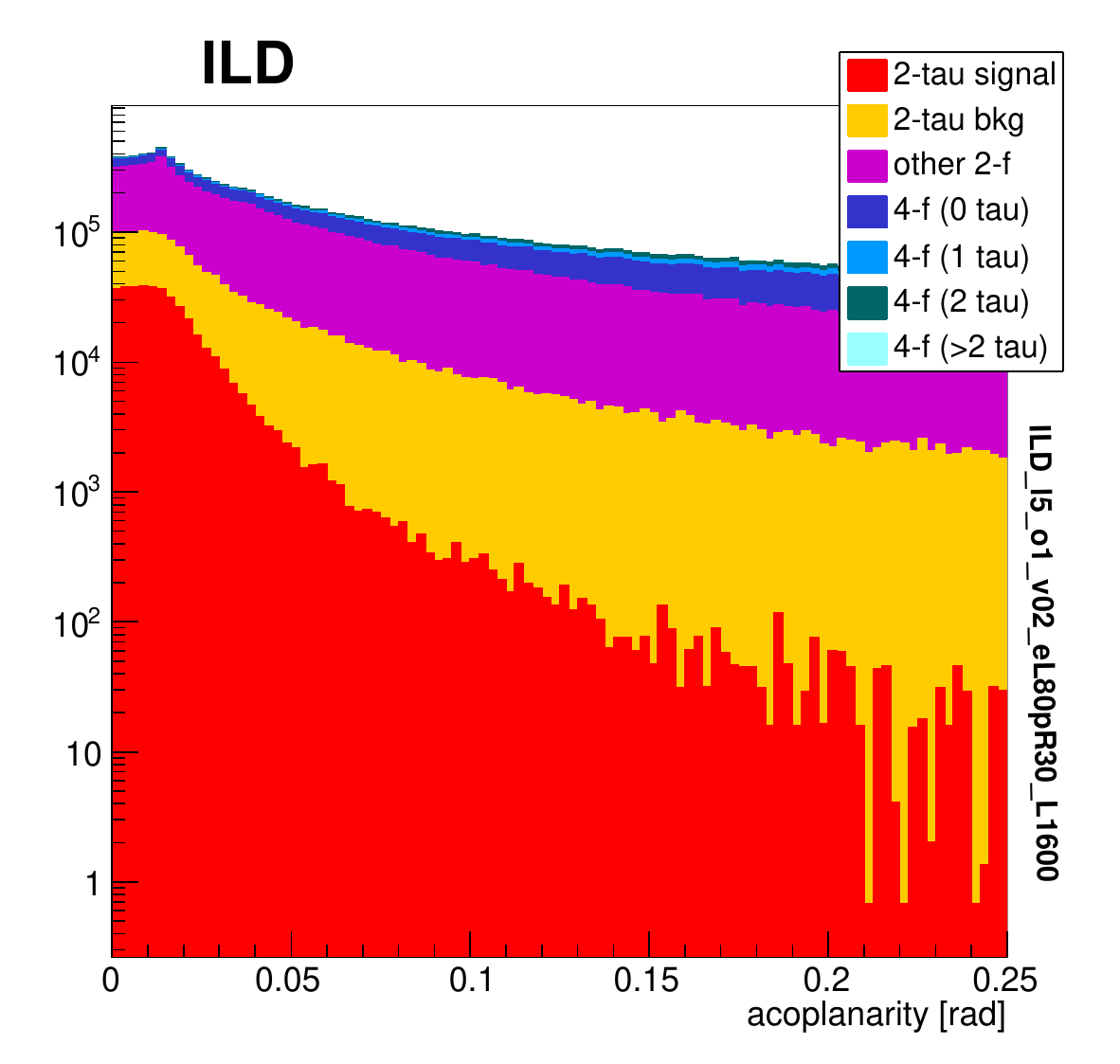} 
\includegraphics[width=0.45\textwidth]{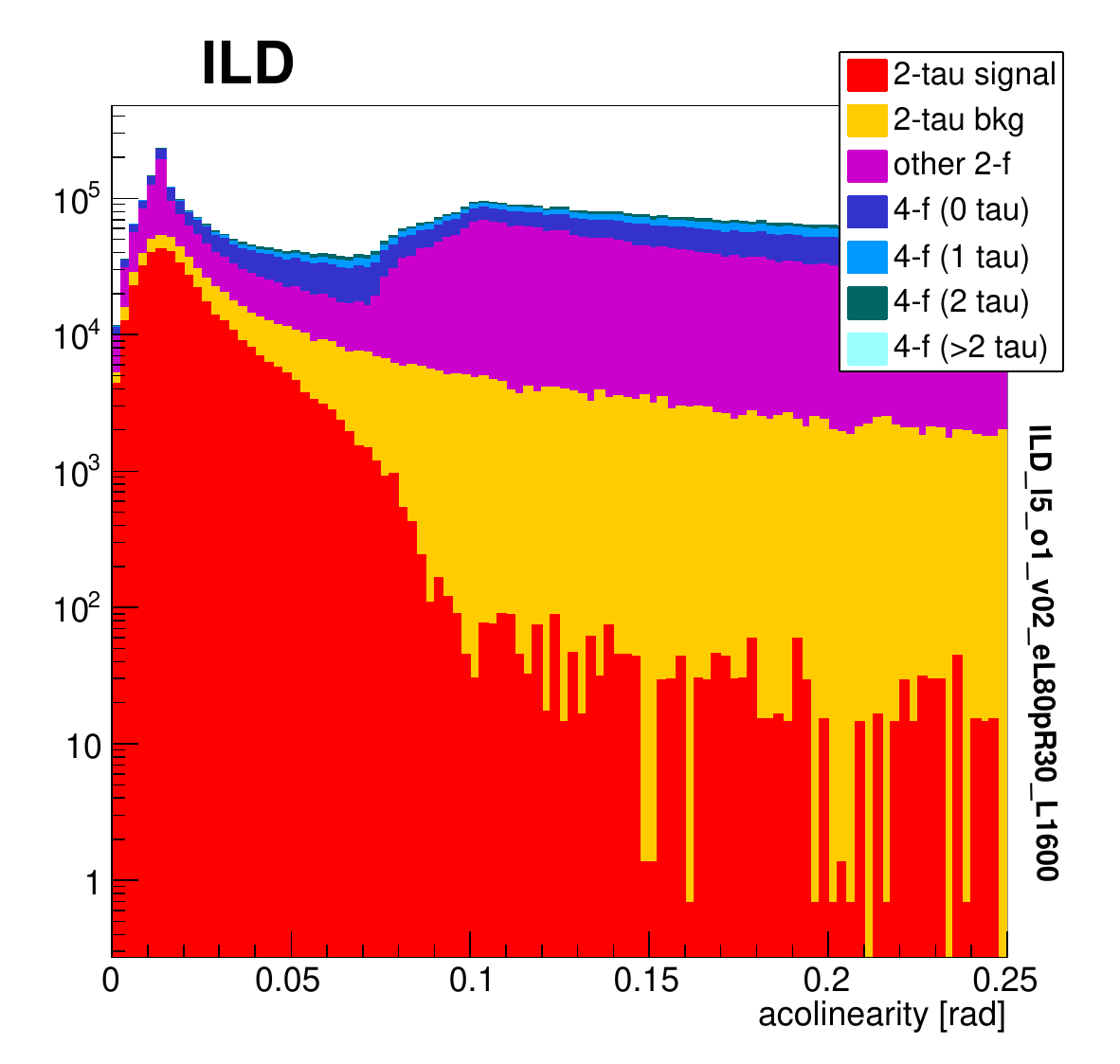} \\
\includegraphics[width=0.45\textwidth]{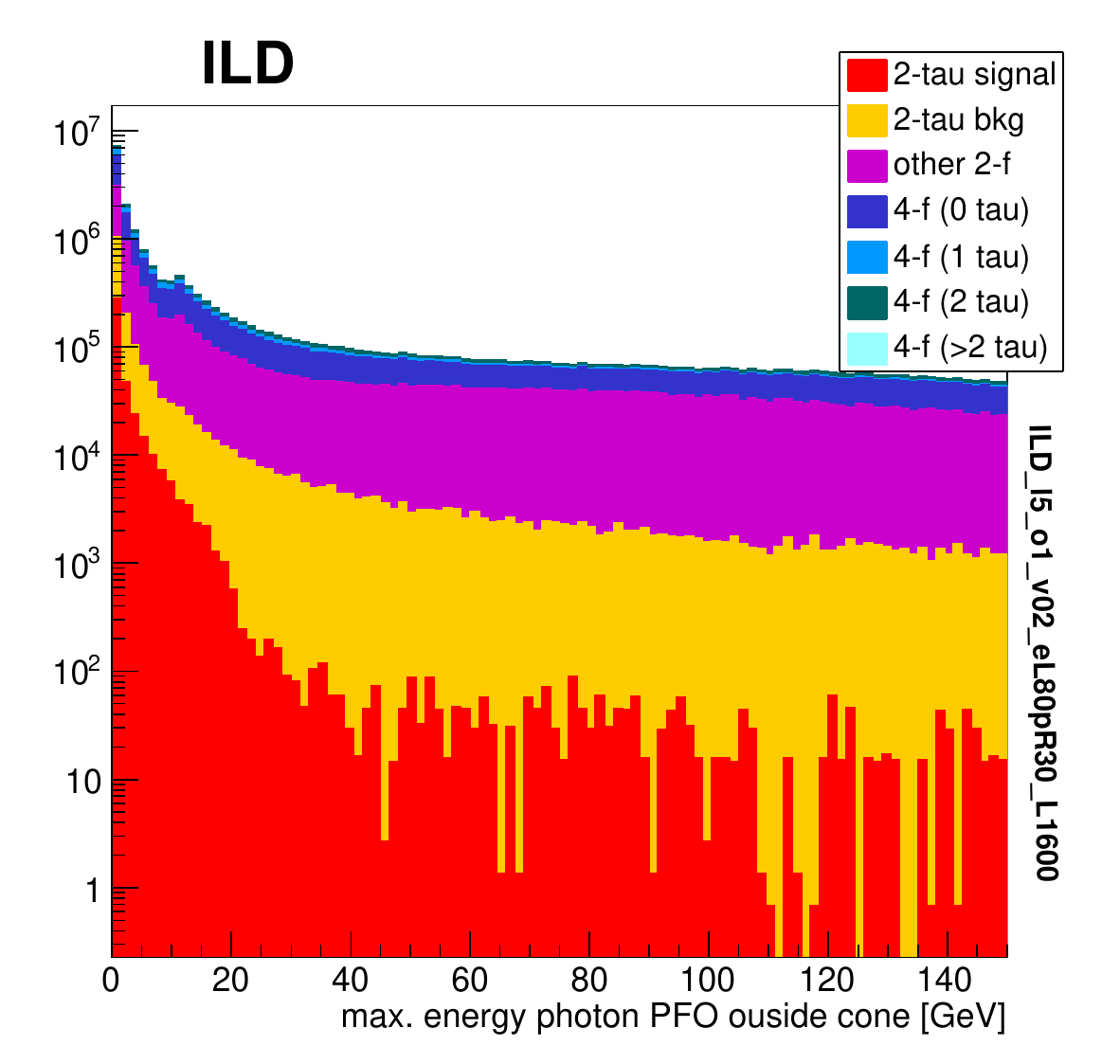}
\includegraphics[width=0.45\textwidth]{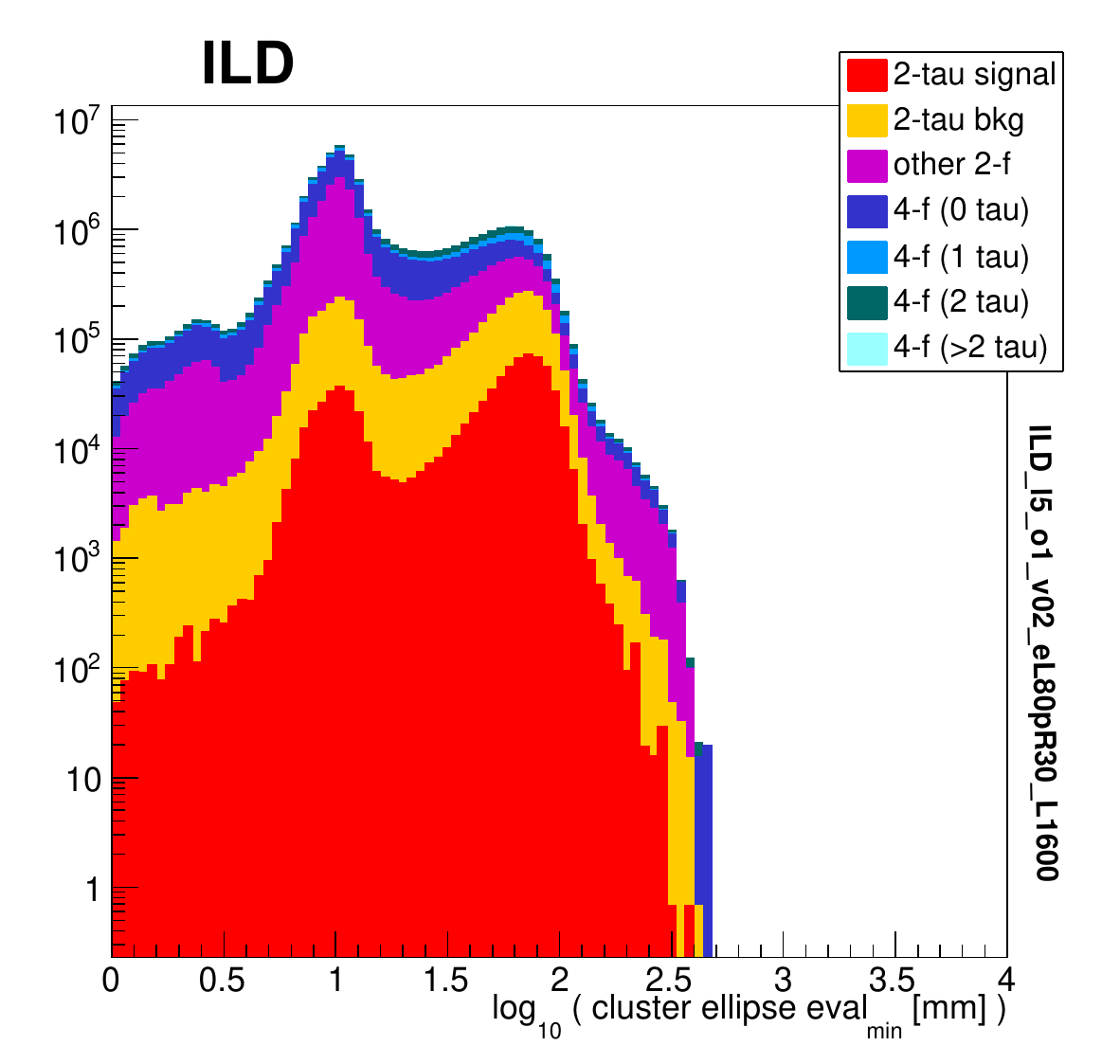}
\caption{
Distributions of some event observables used in the selection after preselection: 
(top) the smaller of the seed PFOs' energies and sum of the energy of all PFOs lying outside the cones;
(middle) acoplanarity and acolinearity of the two candidate jets; 
(lower) the energy of most energetic photon PFO found outside the jet cones, and the smaller eigenvalue of each seeds' associated calorimeter cluster shape.
The ``2-tau signal'' contribution contains di-tau events with a MC invariant mass greater than 480~GeV and at least one hadronic tau decay.
Plots normalised to $1.6~\mathrm{ab^{-1}}$ of \eLpR.
}
\label{fig:varpresel1}
\end{figure}

We first apply a simple preselection, requiring that between 2 and 12 charged PFOs have been reconstructed, to remove the majority of
events with hadronic jets.
We then look for two seed directions around which to build tau jet candidates. 
We identify the highest momentum charged PFO in the event (``first seed''). 
Once this has been found, we look for the highest momentum charged PFO
which is separated from the first seed by at least $\pi/2$ in the $x$-$y$ plane ($\delta \phi$). 
This selection makes use of the property that the two taus in signal events are
emitted back-to-back in the $x-y$ plane in the case of collinear (or no) ISR. 
If no second seed is found, the event is rejected.
We then look in narrow cones (opening angle 0.1 rad) around these two seed directions. PFOs within these cones are associated to tau jet candidates.
The calorimeter cluster associated to each seed particle is modeled as an ellipsoid, whose eigenvalues (the lengths of its axes) are used in the selection.
Distributions of some of these obervables are shown in Fig.~\ref{fig:varpresel1}.
We apply the following selection:
\begin{itemize}
\item energy of the second seed PFO less than 200 GeV [to remove di-lepton events];
\item sum of the energy [$p_T$] of PFOs lying outside the two cones less than 40 [20] GeV [remove hadronic events];
\item acoplanarity between candidate jet directions less than 0.05 rad [remove fully leptonic $WW$ events];
\item acolinearity between candidate jet directions less than 0.075 rad [remove $Z$ return events];
\item no photon-like PFO (as tagged by PandoraPFA) with energy larger than 10 GeV located outside the two cones [remove events with seen ISR];
\item no isolated leptons identified by the IsolatedLeptonTagging processor [remove dilepton events, fully leptonic tau decays];
\item the smallest and largest eigenvalues of the shower ellipsoid must respectively lie in the range $3.2 \to 63\ \mathrm{mm}$ and $6.3 \to 100\ \mathrm{mm}$.
\end{itemize}


\begin{figure}
\centering
\includegraphics[width=0.45\textwidth]{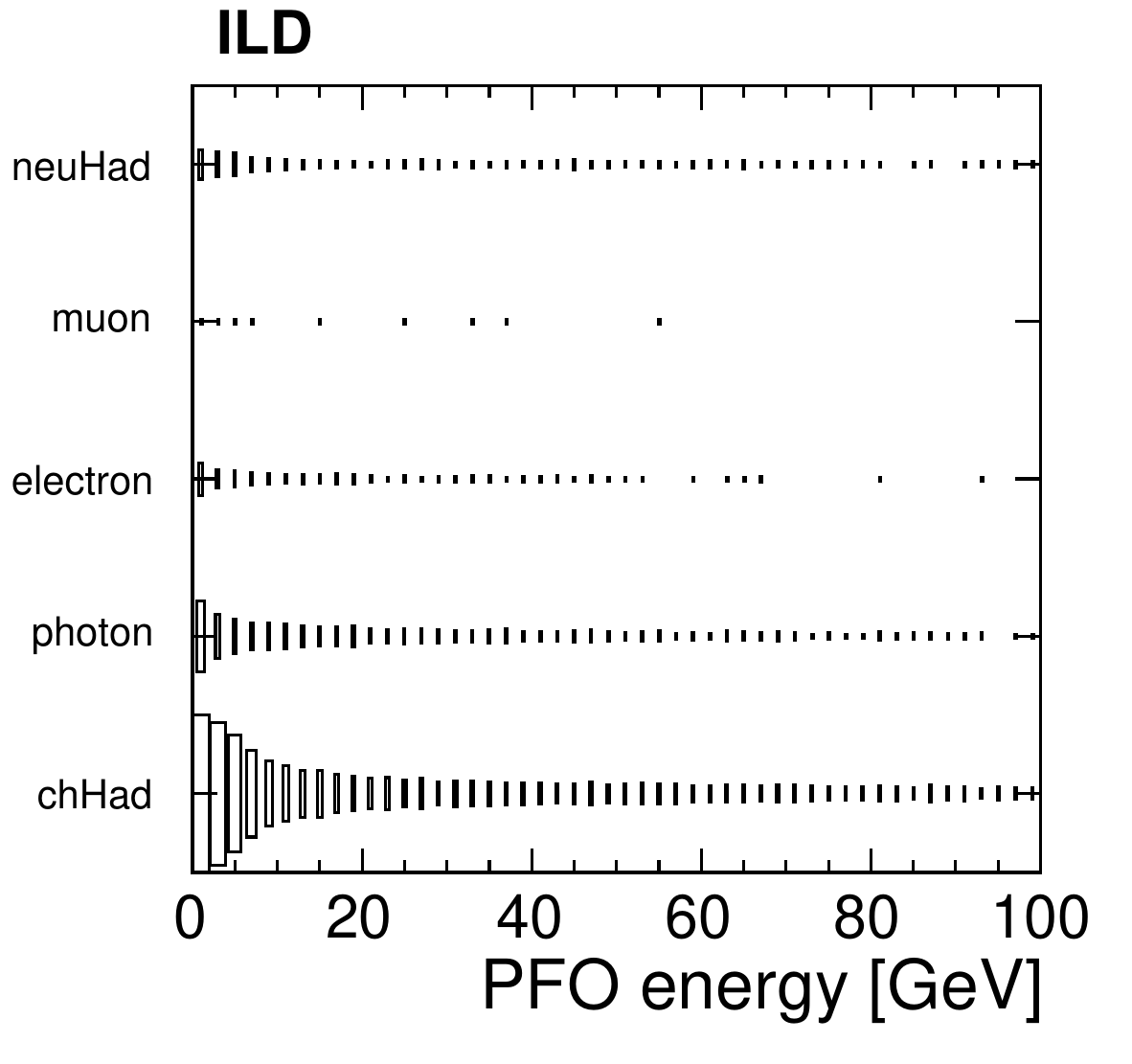} \\
\includegraphics[width=0.45\textwidth]{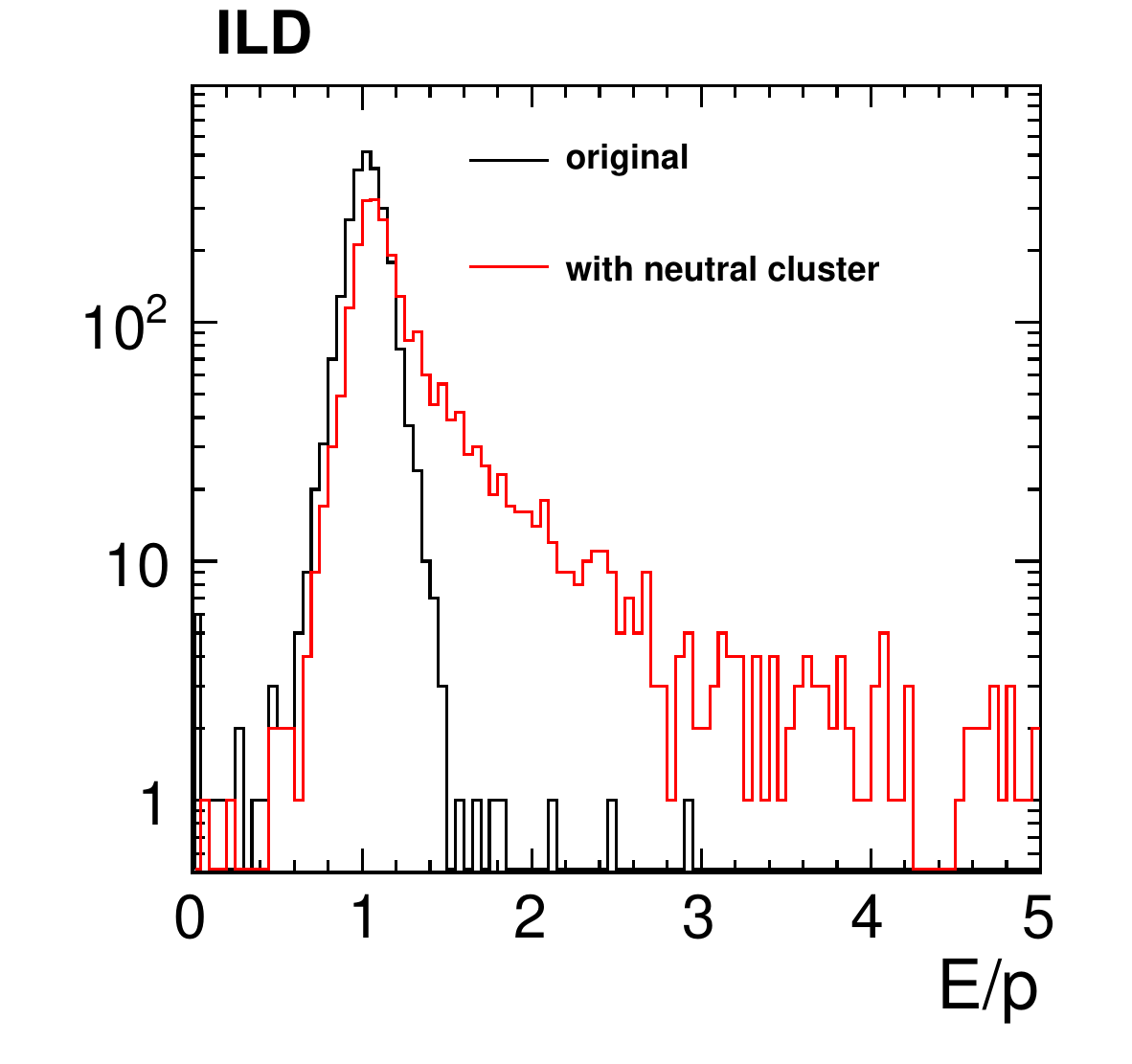}
\includegraphics[width=0.45\textwidth]{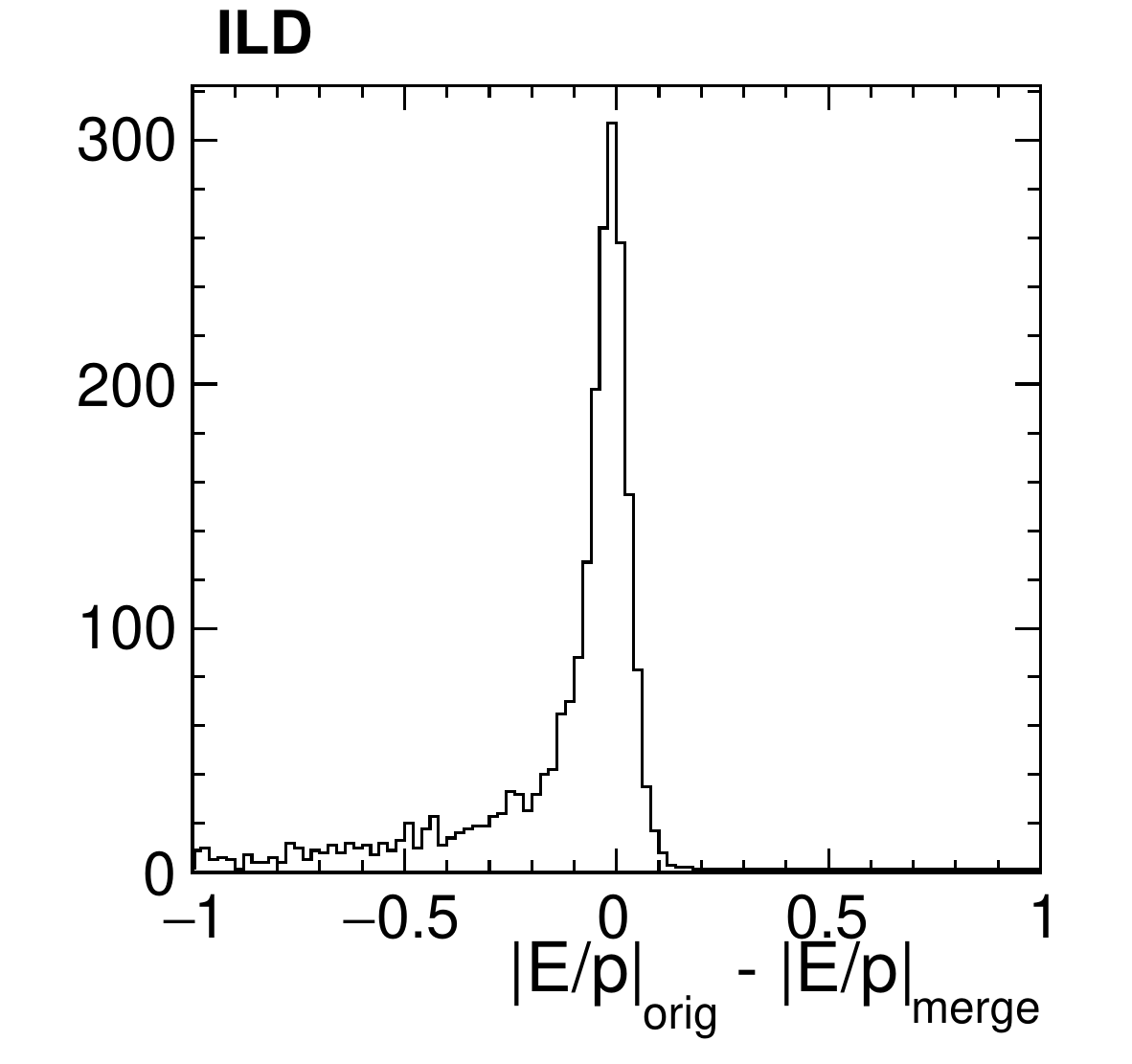}
\caption{Top: the nature of the MC particle which created PFOs flagged as neutral hadrons, as a function of PFO energy.
Lower: Events with one charged and one neutral hadron PFO;
left) the energy/momentum ratio of the charged hadron before and after adding the neutral hadron cluster energy;
right) the difference between $|\mathrm{E/p-1}|$ before and after adding the neutral hadron cluster to the charged PFO.
}
\label{fig:mistakes}
\end{figure}

We then look in more detail at the two jet cones. We observe that neutral hadrons are often reconstructed within the cones. 
However, the number of long-lived neutral hadrons produced in tau decays is quite small. In fact, the observed neutral
hadron PFOs are almost always the result of the splitting of the calorimetric shower induced by a charged hadron.
We therefore remove such neutral hadron PFOs from the event.
%
%
In Fig.~\ref{fig:mistakes}, we look at the parent particle of those clusters identified as neutral hadrons:
the majority are indeed due to fragmentation of the charged hadron. In the same figure we compare the E/p of the original charged hadron PFO, and what we obtain
when the neutral hadron cluster energy is added: the E/p distribution is degraded when the neutral hadron is added: this helps to understand why
the reconstruction algorithm chooses to split off part of the hadronic shower into a distinct PFO.

\begin{figure}
\centering
\includegraphics[width=0.45\textwidth]{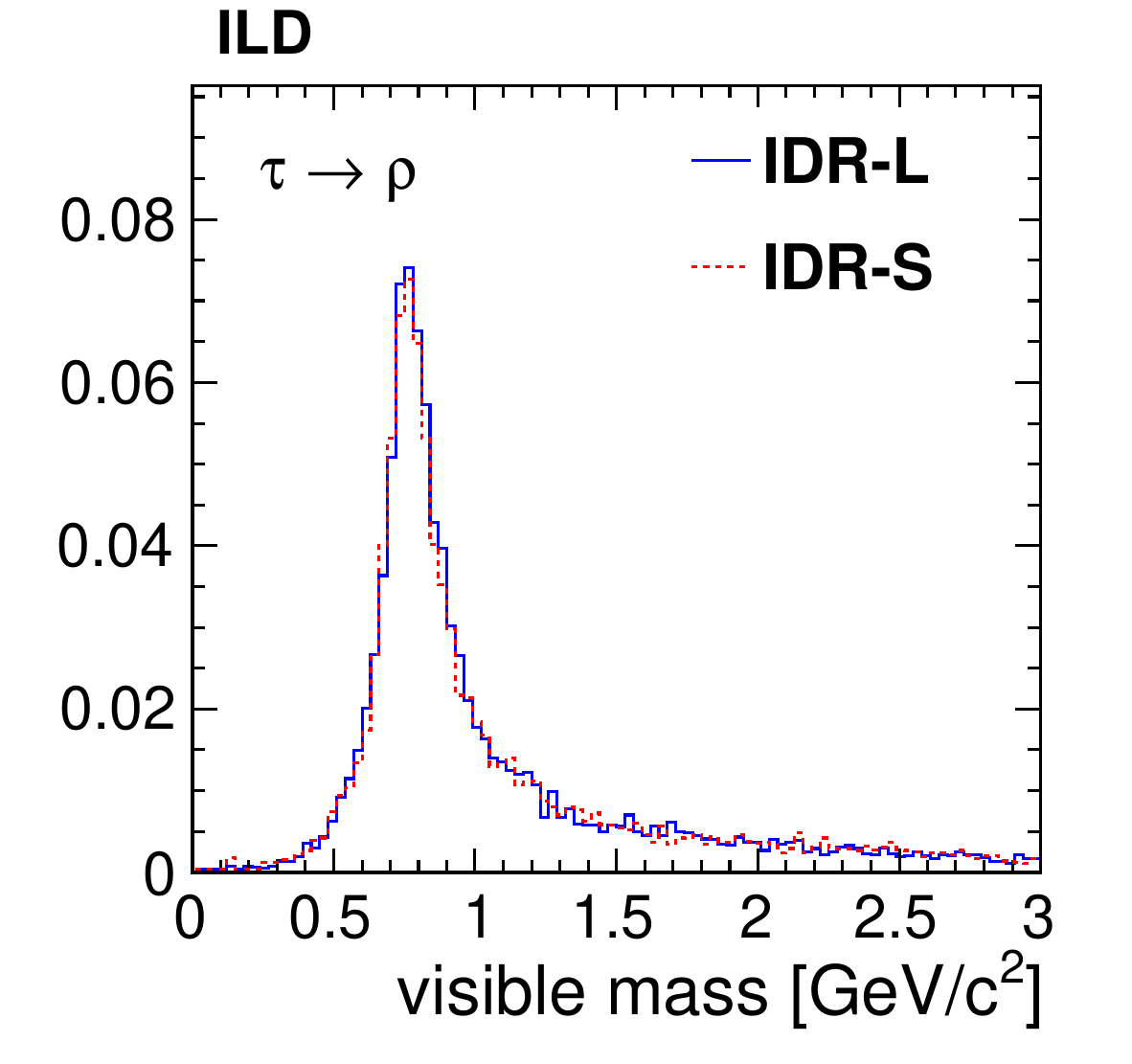}
\includegraphics[width=0.45\textwidth]{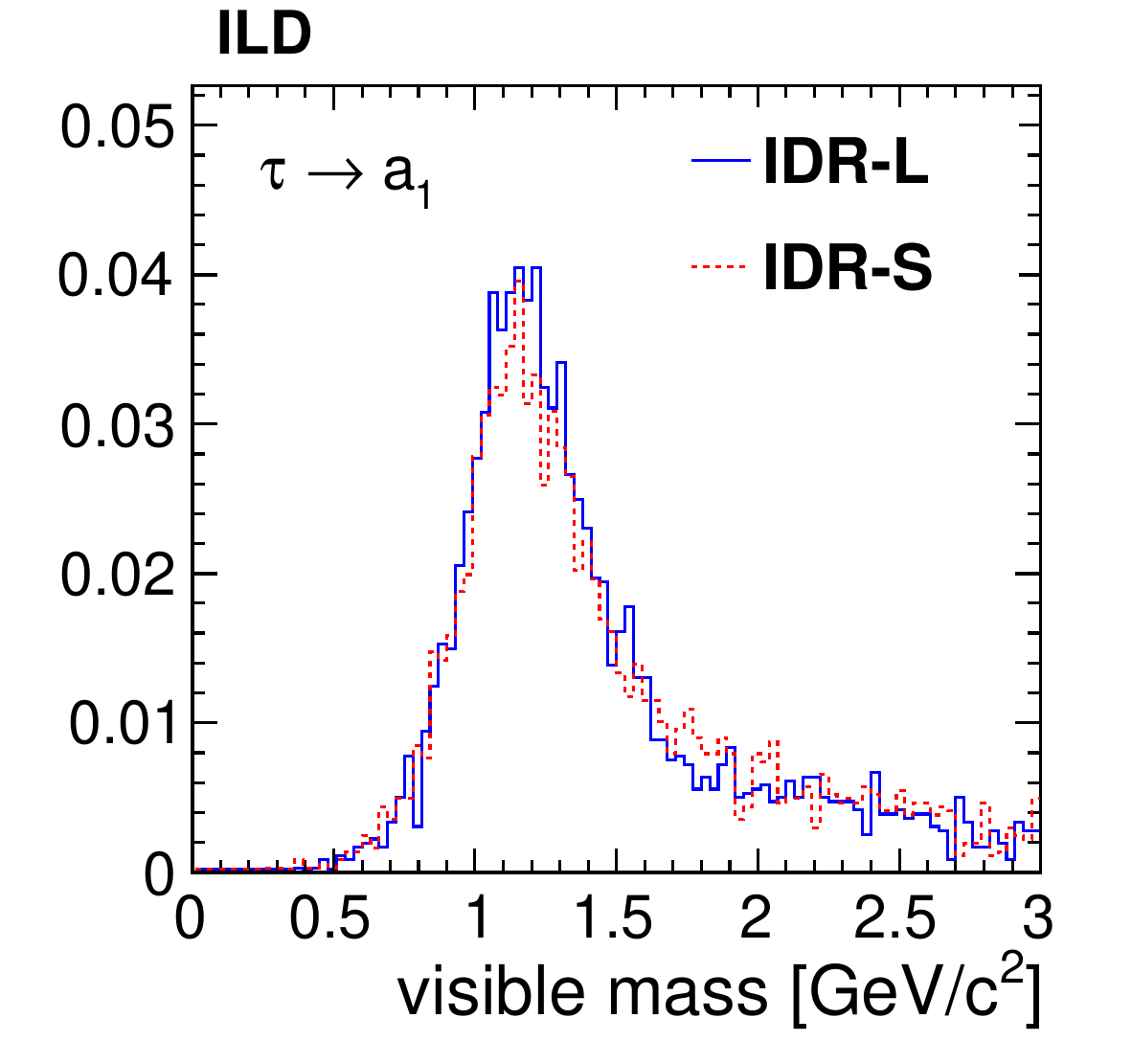}
\includegraphics[width=0.45\textwidth]{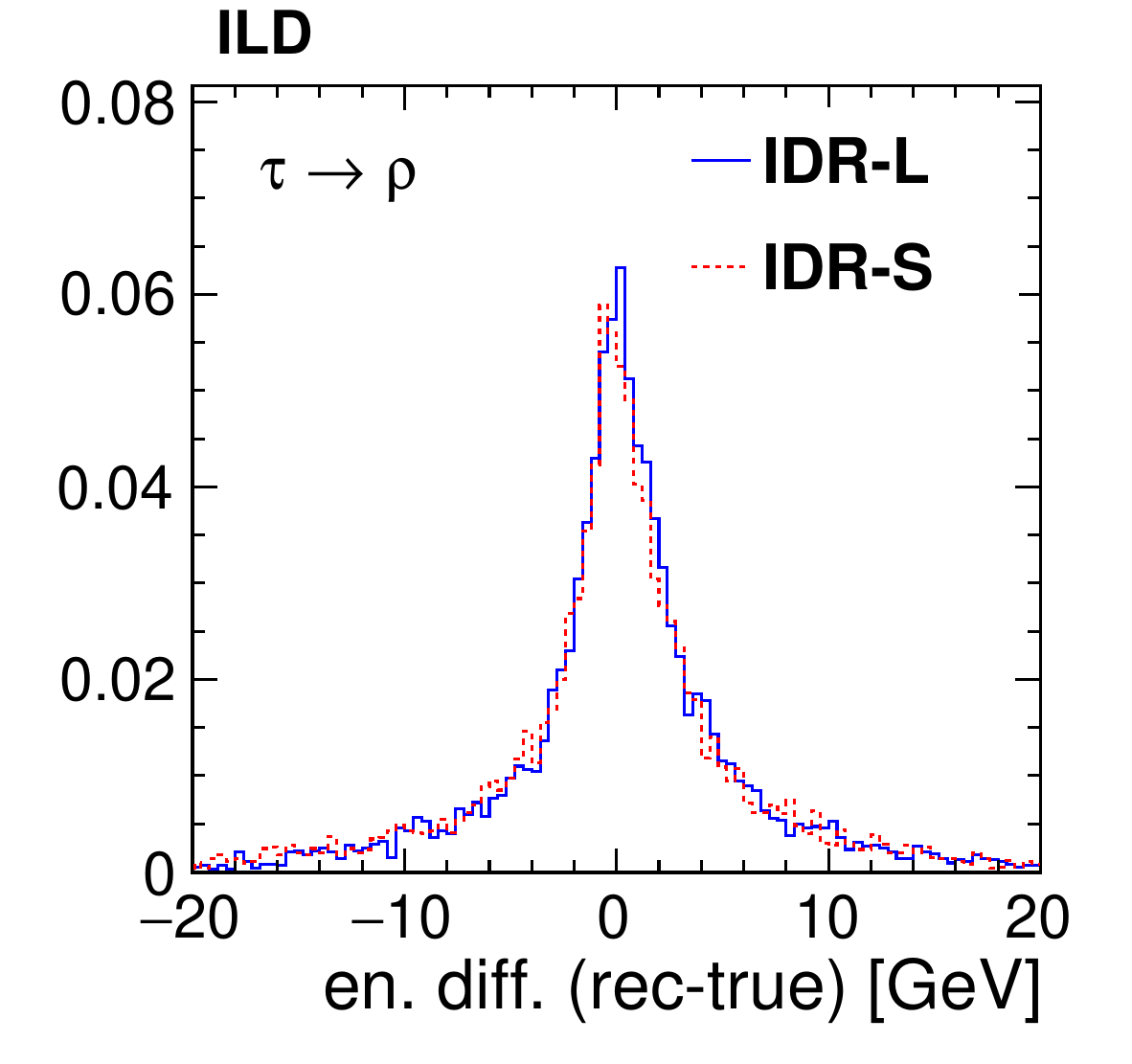}
\includegraphics[width=0.45\textwidth]{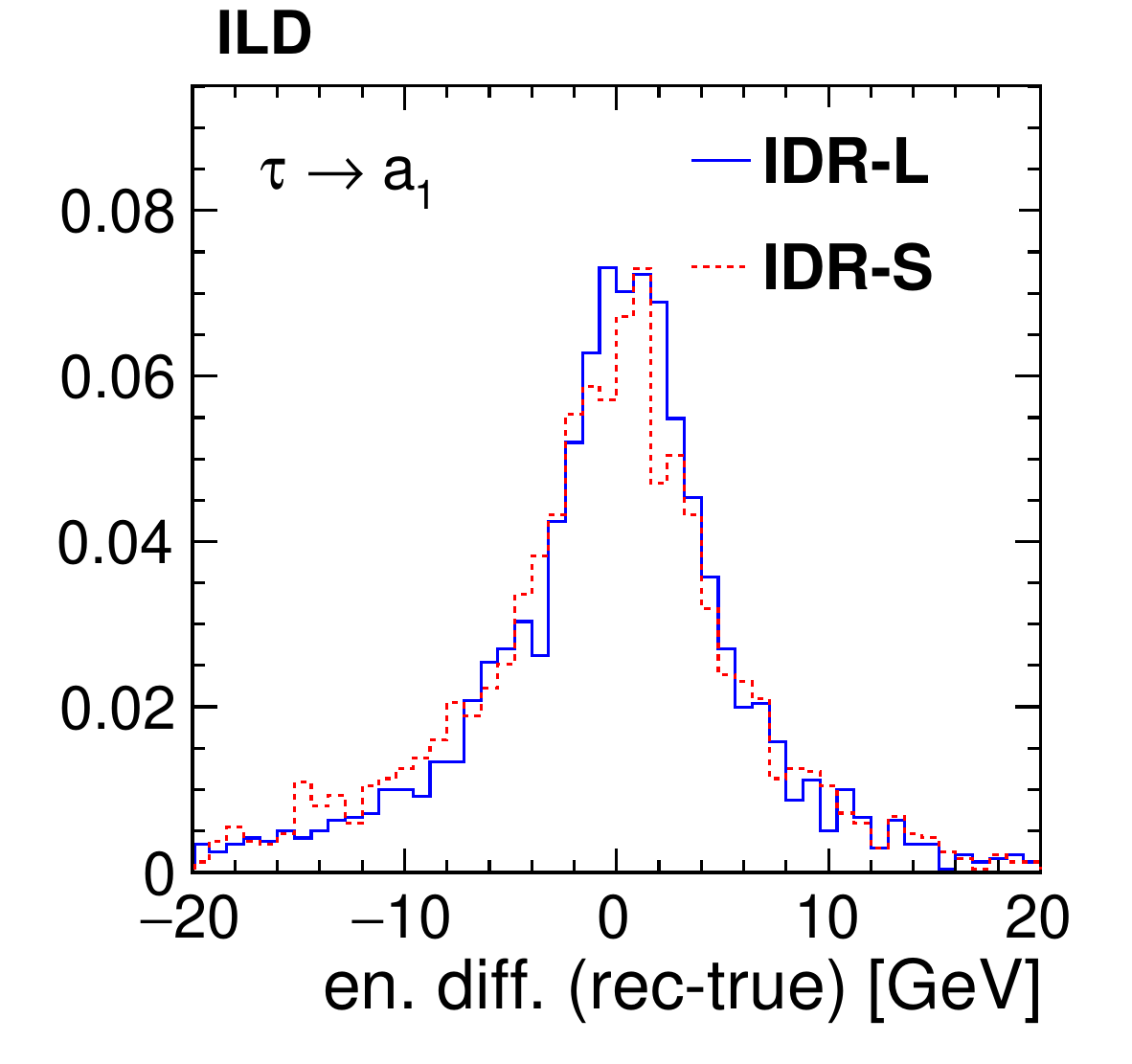}
\caption{Signal-only comparisons after general event selection: (Top) visible invariant mass and (lower) the difference between the true and reconstructed visible energy, 
in single-prong $\rho$ and $a_1$ decays.}
\label{fig:simplecomp2}
\end{figure}

In addition, if the total charge of the jet is zero, we remove the charged particle furthest from the jet's initial seed direction.
After removing these PFOs from consideration, we calculate the invariant mass of the jet. 
The distribution of this invariant mass is shown in Fig.~\ref{fig:simplecomp2}, together with distributions demonstrating how well the visible tau jet energy is reconstructed.
We require that the jets' visible invariant mass is less than 1.77~GeV ($\sim m_\tau$), and that the product of the two jets' charges is $-1$.


The selection efficiency and remaining background events at the various stages of this selection are shown in Table~\ref{tab:signalSel}.
The overall efficiency to select high invariant mass tau-pair events in which at least one tau has decayed hadronically is around 60\%. There is a $\sim 11\%$ contribution from
lower invariant mass tau-pair events in the signal sample (mostly from just below the 480~GeV threshold), and a $\sim 4\%$ contribution from remaining 
non tau-pair events in the \eLpR\ polarisation, mostly from 4-fermion (4f) processes. This 4f contribution is reduced to $\sim 1.5\%$ in the \eRpL\ polarisation scenario.

\begin{table}
\centering

\begin{tabular}{|l|c|r|r|r|rrr|}
\hline
final state            & \multicolumn{4}{c|}{$ee\to\tau\tau$} & $ee\to\mu\mu$ & other 2f & 4f \\
\hline
$m_{\tau\tau}$         & \multicolumn{2}{c|}{$>480$} & $[250,480]$ & remain-  & & & \\
\# had $\tau$ decays   & \multicolumn{2}{c|}{$\ge 1$}& $\ge 1$     & der   & & & \\
\hline
\hline
\eLpR, $1.6~\mathrm{ab^{-1}}$ & efficiency [\%] & \multicolumn{6}{c|}{expected events/1000} \\ 
\hline

original               &       100.0 &       456.9 &       463.3 &       128.3 &      4387.4 &       36801.1 &       51915.0 \\    
preselected            &        90.0 &       411.3 &       419.8 &       118.5 &      3763.0 &        7540.4 &       16236.3 \\    
two seeds              &        89.8 &       410.1 &       418.2 &       118.1 &      3529.5 &        7205.4 &       14338.7 \\    
out-of-cone activity   &        85.7 &       391.5 &       317.1 &       101.1 &      1812.6 &        1707.3 &        5886.9 \\    
acolinearity           &        83.5 &       381.6 &       304.3 &        95.1 &      1242.4 &        1116.3 &         713.0 \\    
acoplanarity           &        77.5 &       354.2 &        65.6 &        58.0 &        37.1 &           6.3 &         123.1 \\    
ISR veto               &        75.7 &       345.7 &        58.6 &        55.6 &        35.3 &           6.0 &         117.1 \\    
isolated lepton veto   &        75.6 &       345.2 &        58.5 &        50.8 &        20.5 &           5.6 &          73.5 \\    
seed cluster shape     &        70.1 &       320.1 &        53.5 &         1.3 &         2.9 &           5.0 &          24.5 \\    
candidate jet mass     &        61.9 &       282.8 &        39.0 &         1.3 &         1.8 &           0.6 &          17.2 \\    
jets' charge           &        60.1 &       274.6 &        37.3 &         1.2 &         1.3 &           0.1 &          14.3 \\    
\hline
\hline
\eRpL, $1.6~\mathrm{ab^{-1}}$  &        59.6 &       226.9 &        29.5 &         1.3 &         1.1 &           0.2 &           3.9 \\
\hline
\eLpL, $0.4~\mathrm{ab^{-1}}$  &        60.0 &        41.1 &         5.6 &         0.2 &         0.2 &           0.0 &           2.2 \\
\hline
\eRpR, $0.4~\mathrm{ab^{-1}}$  &        59.7 &        35.7 &         4.7 &         0.2 &         0.2 &           0.0 &           1.1 \\
\hline
\end{tabular}
\caption{Selection efficiencies and expected event numbers at different stages of the selection (see text for details). 
Results are shown for the large detector IDR-L, with full details for the \eLpR\ polarisation, and final results for the other polarisations.
}
\label{tab:signalSel}
\end{table}

\clearpage

\section{Tau decay mode selection}

\begin{figure}
\centering
\includegraphics[width=0.45\textwidth]{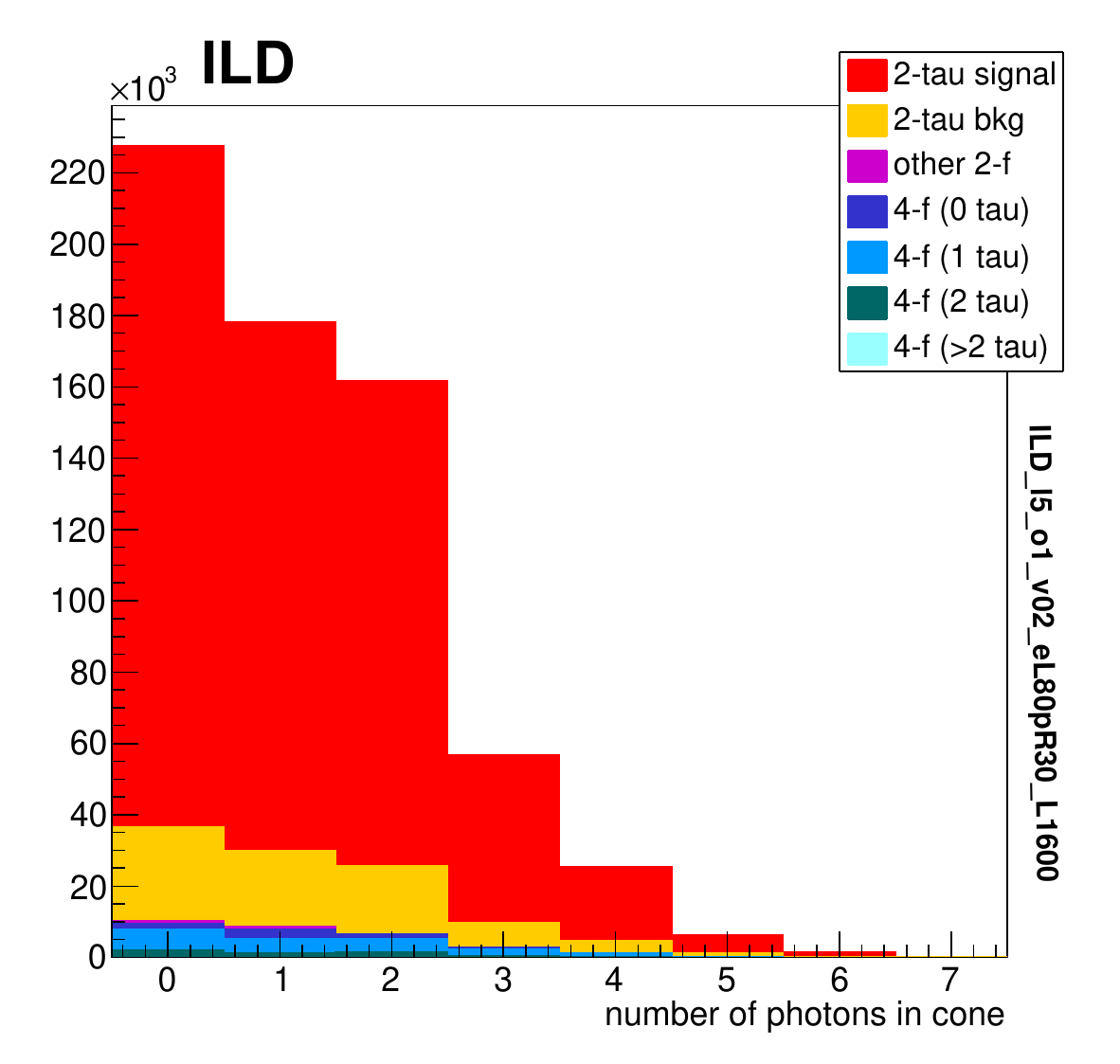} \\
\includegraphics[width=0.45\textwidth]{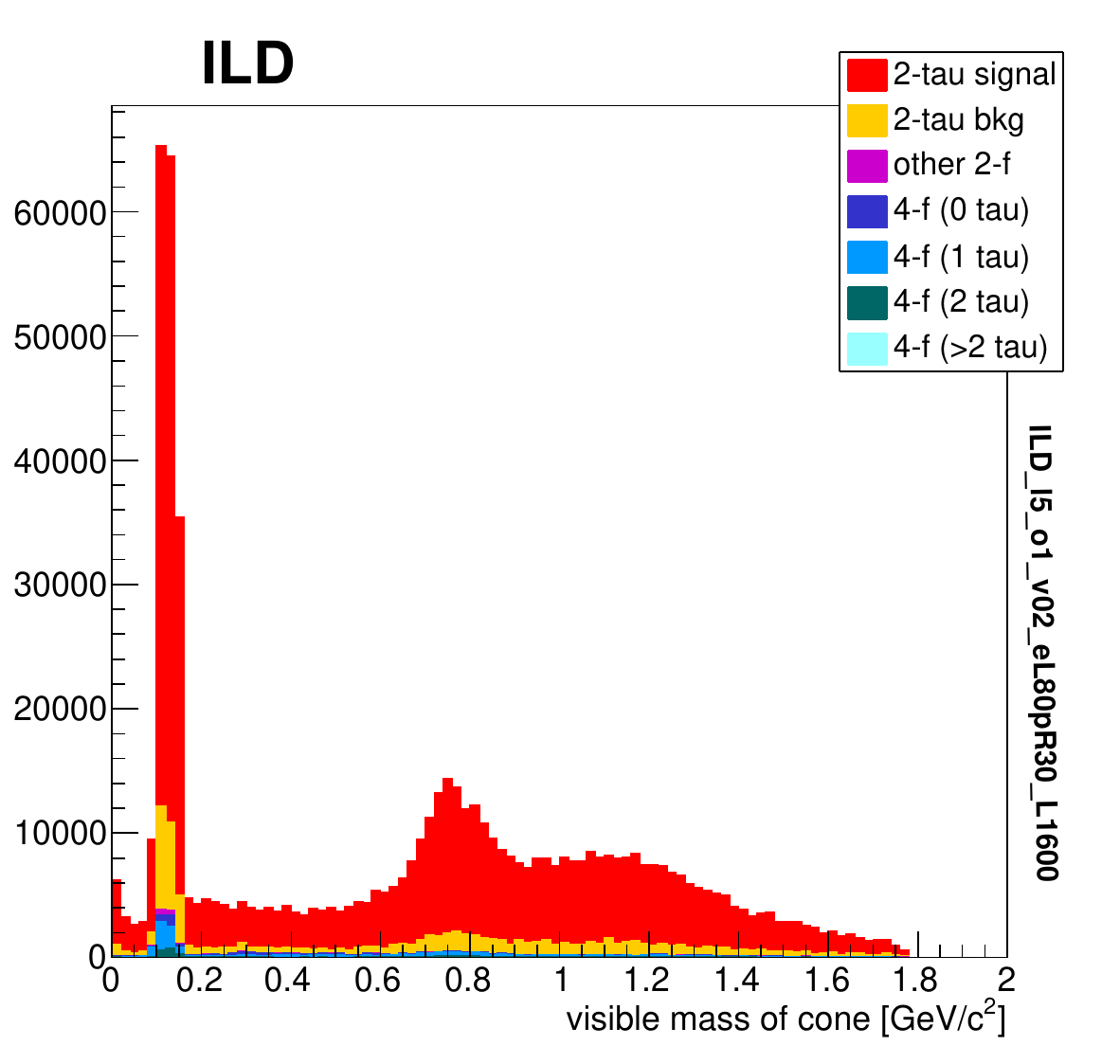}
\includegraphics[width=0.45\textwidth]{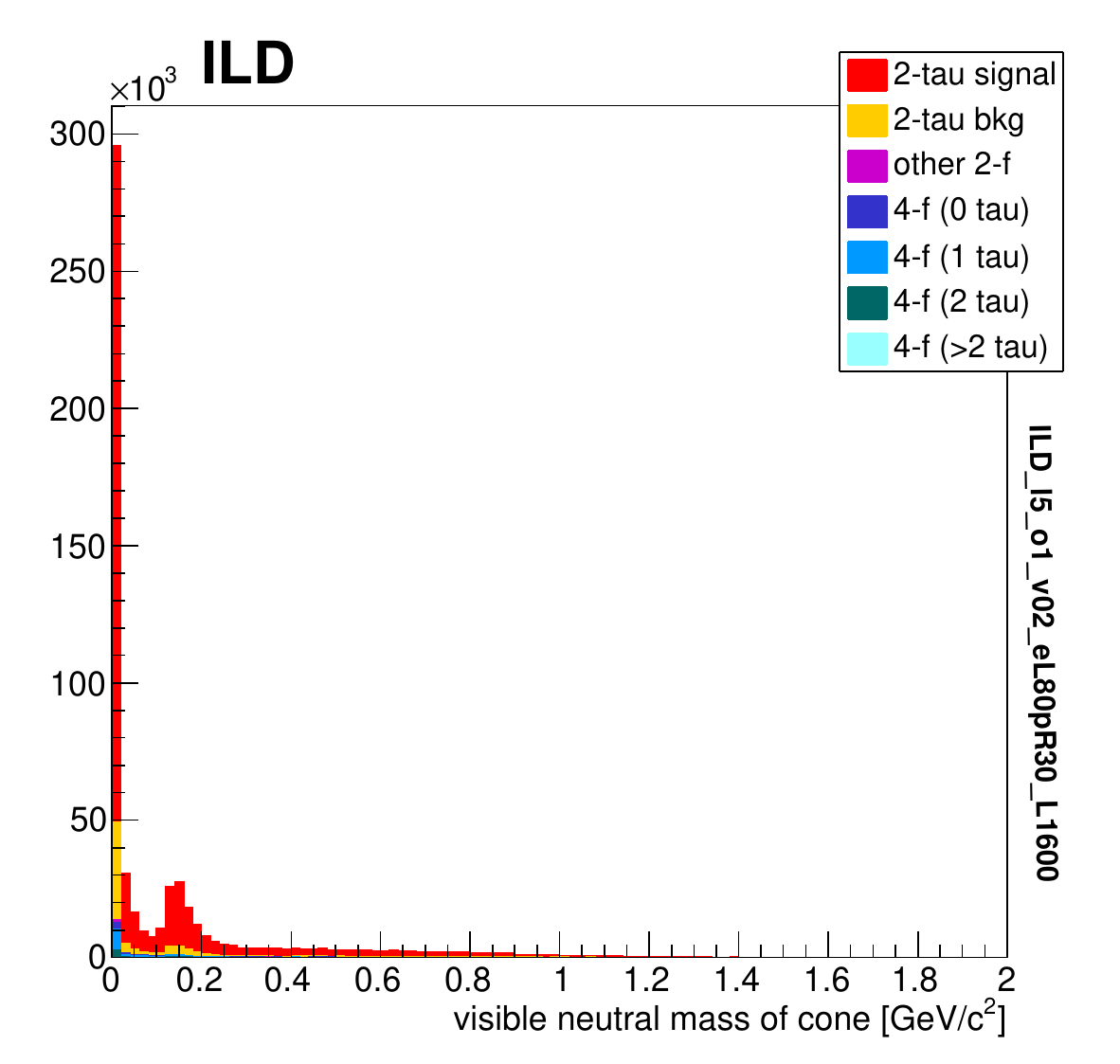}
\caption{Distributions after the general event selection, before decay mode identification. 
The number of photon PFOs found per candidate jet, and the invariant mass of (left) all and (right) all neutral PFOs inside the jet.
IDR-L model, normalised to $1.6~\mathrm{ab^{-1}}$ of \eLpR.
}
\label{fig:varpresel3}
\end{figure}

Distributions of the number of reconstructed photons and visible invariant mass (both total and neutral) of tau jets in selected events is shown in Fig.~\ref{fig:varpresel3}.
These observables will be used to distinguish the different tau lepton decay modes.
%

\begin{figure}
\centering
\includegraphics[width=0.45\textwidth]{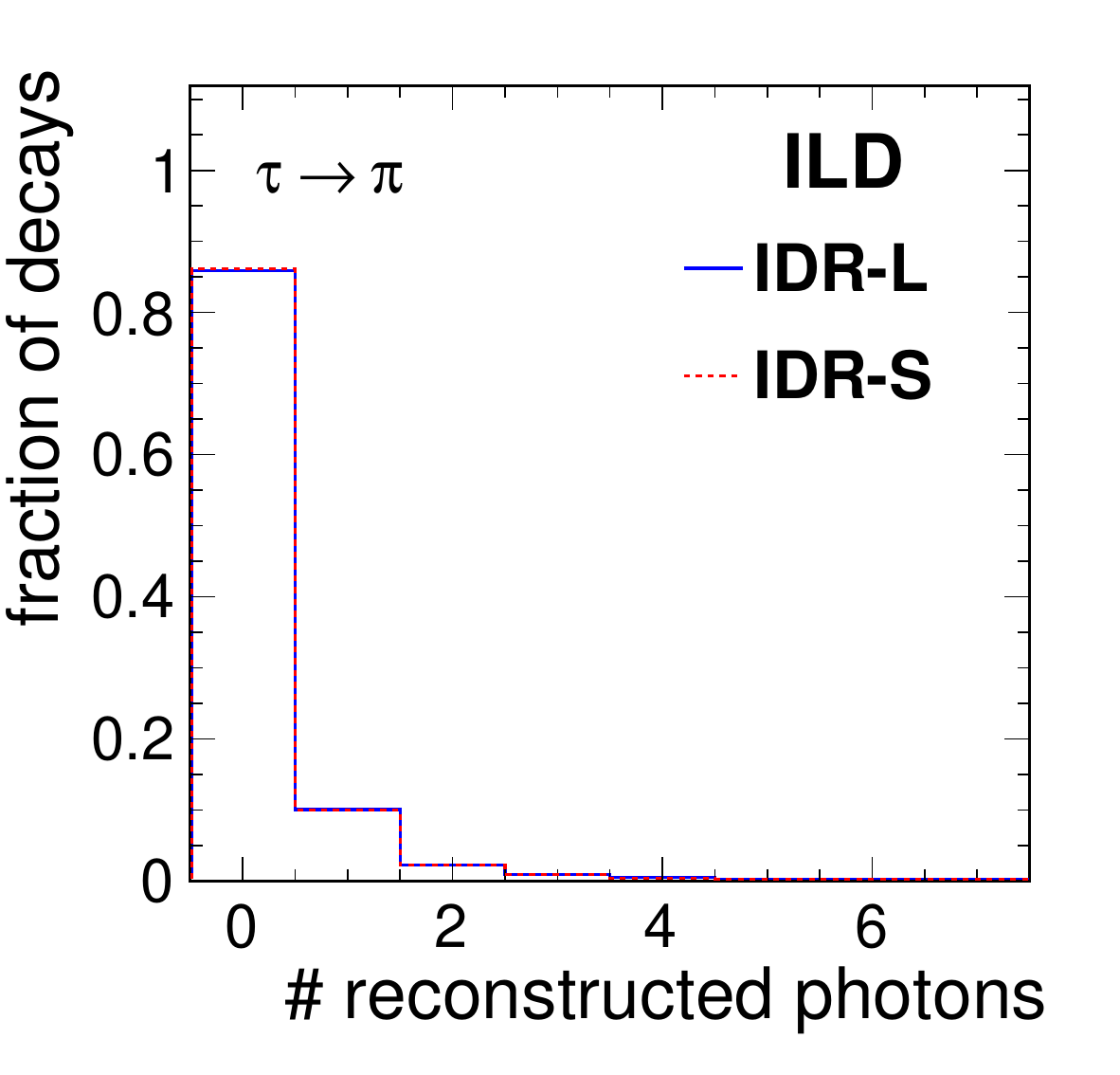}
\includegraphics[width=0.45\textwidth]{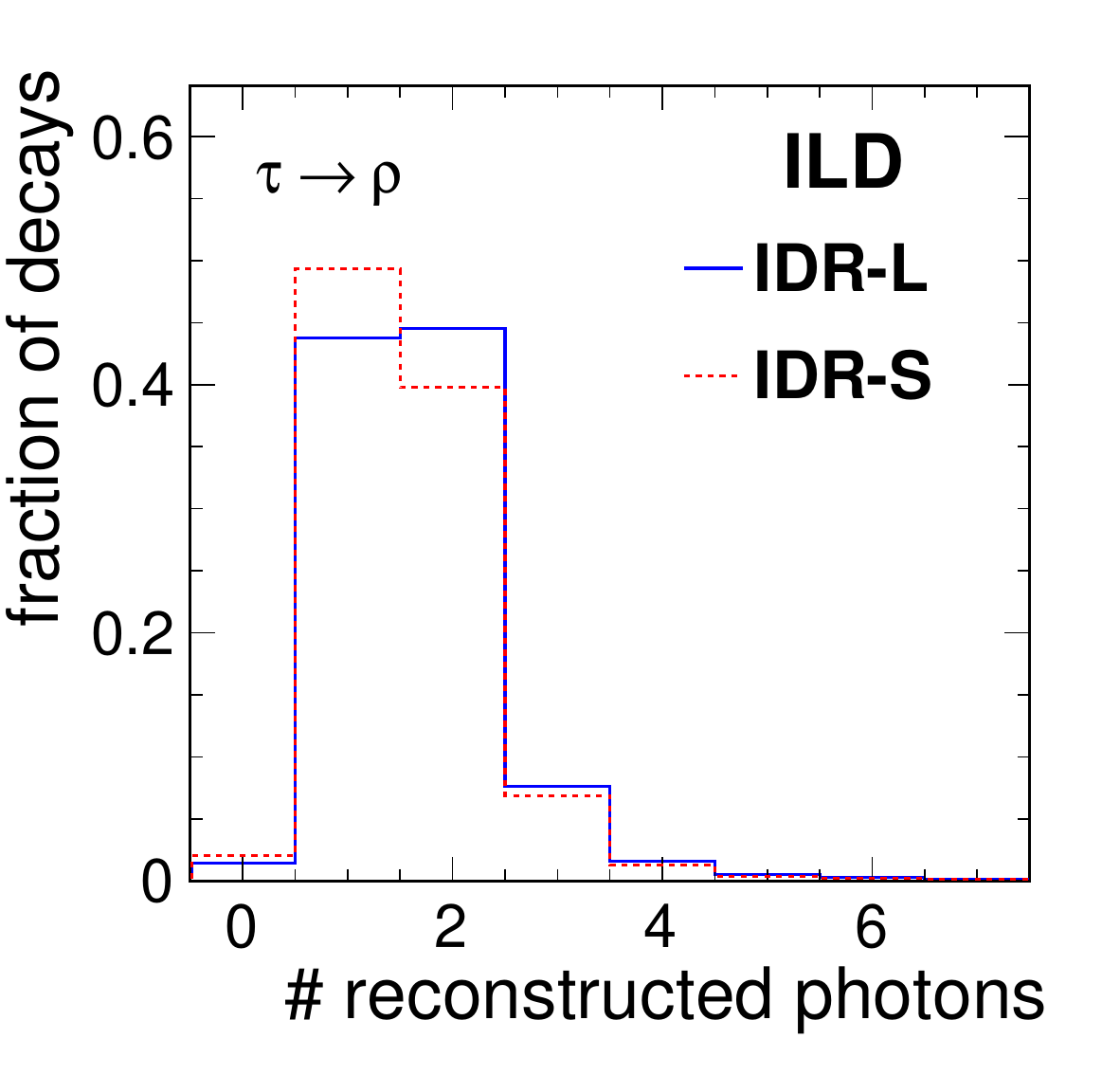}
\includegraphics[width=0.45\textwidth]{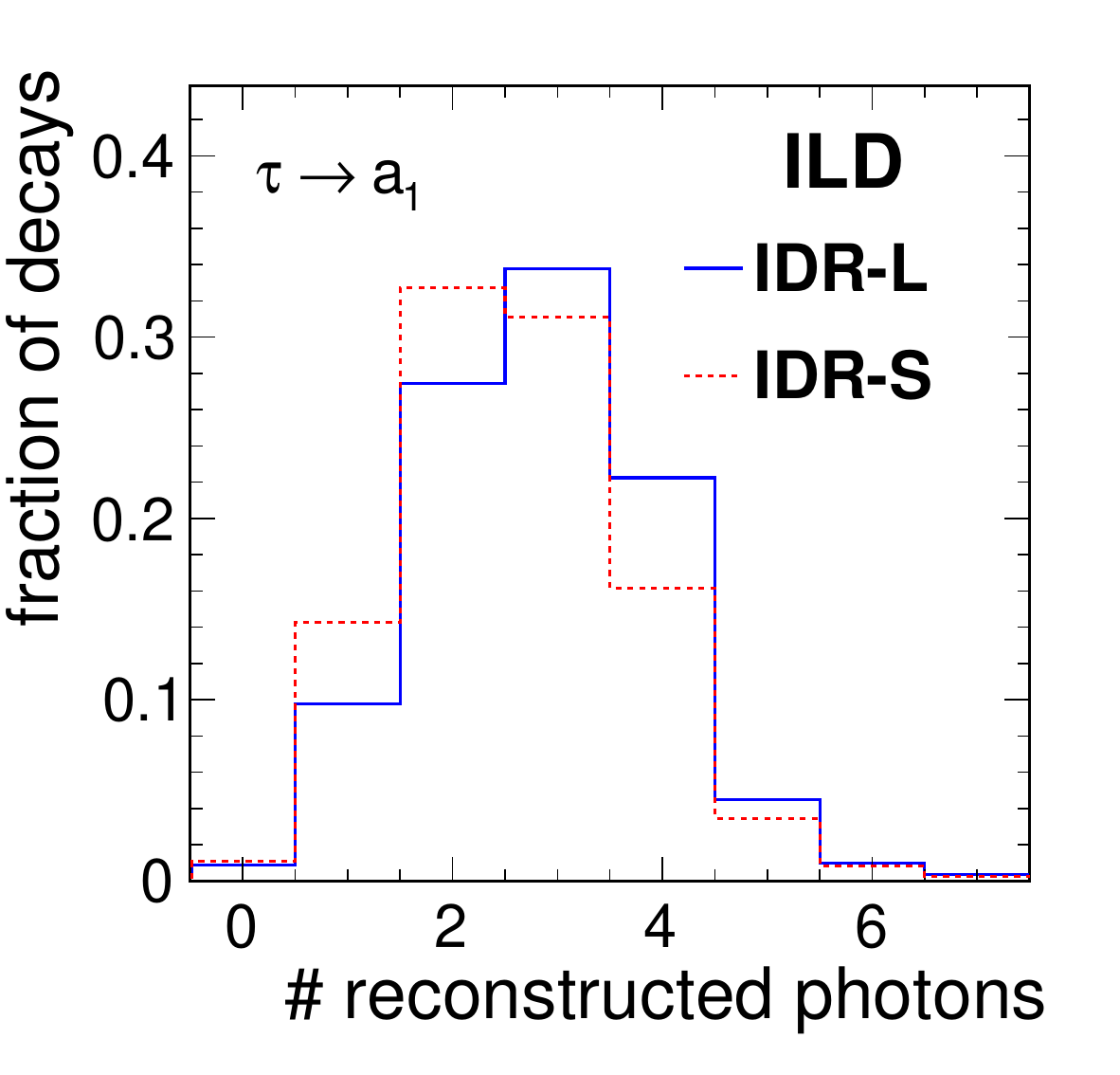}
\caption{Signal-only comparisons after general event selection: the number of reco photons in candidate jets matched to taus decaying in the single pion, rho, and single-prong ${a_1}$ modes.
The expected number of photons produced in these decays is respectively 0, 2, and 4 in these decay models.
}
\label{fig:simplecomp1}
\end{figure}

\begin{figure}
\centering
\includegraphics[width=0.45\textwidth]{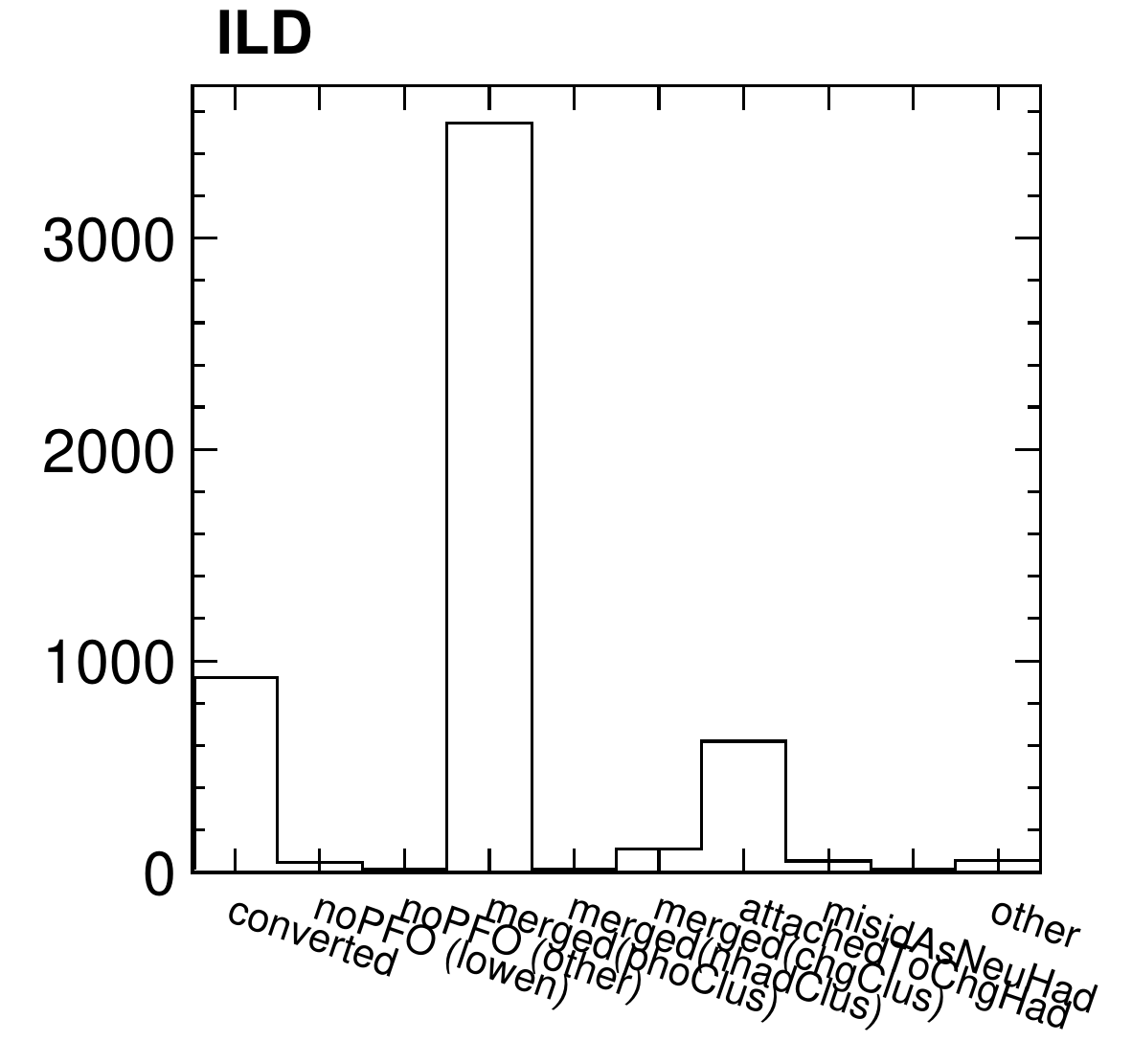}
\caption{
Reason for which only a single photon PFO was found in $\tau \to \rho$ decays (see text for details).
IDR-L detector model.
}
\label{fig:merge}
\end{figure}

\begin{figure}
\centering
\includegraphics[width=0.45\textwidth]{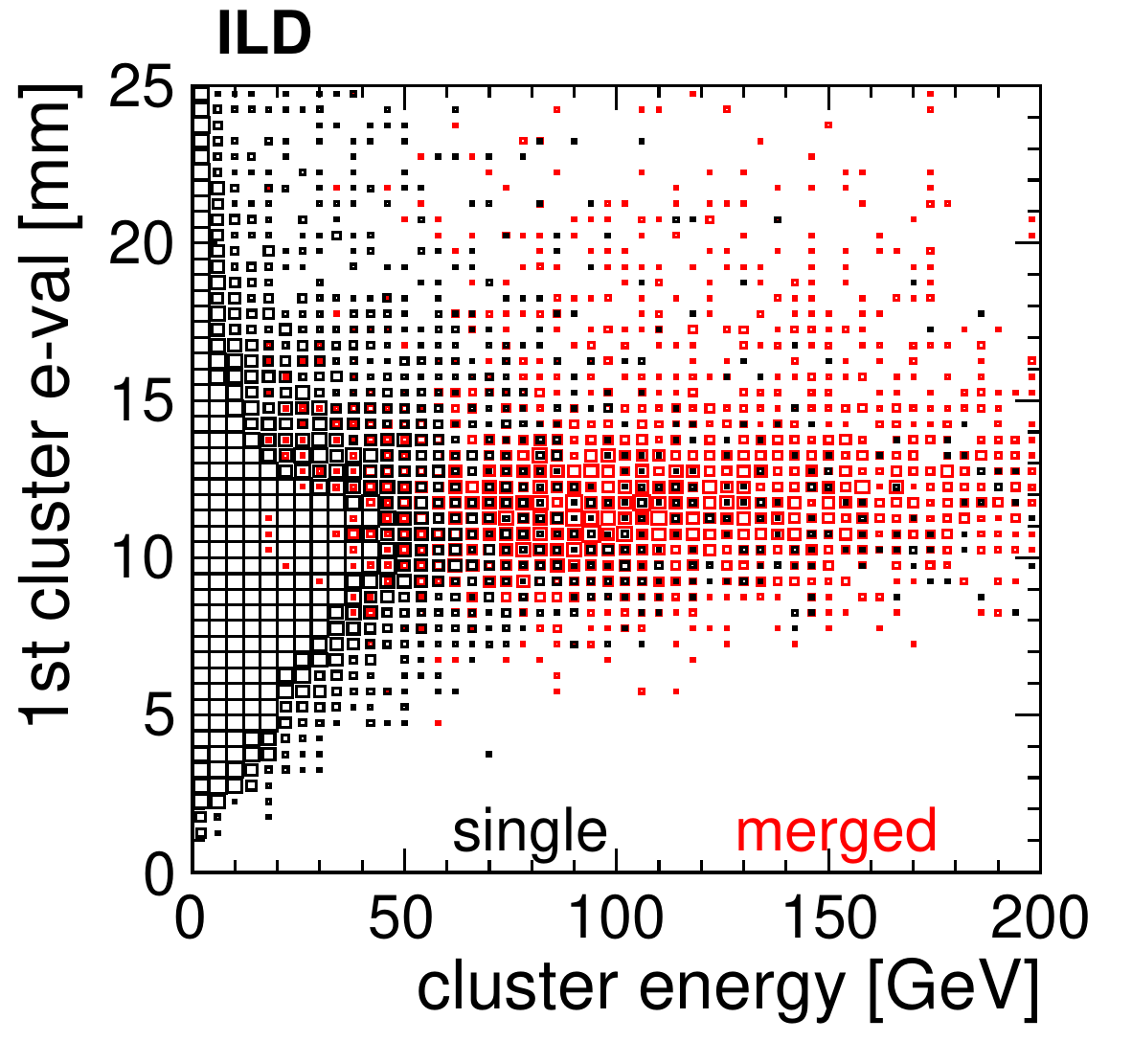}  
\includegraphics[width=0.45\textwidth]{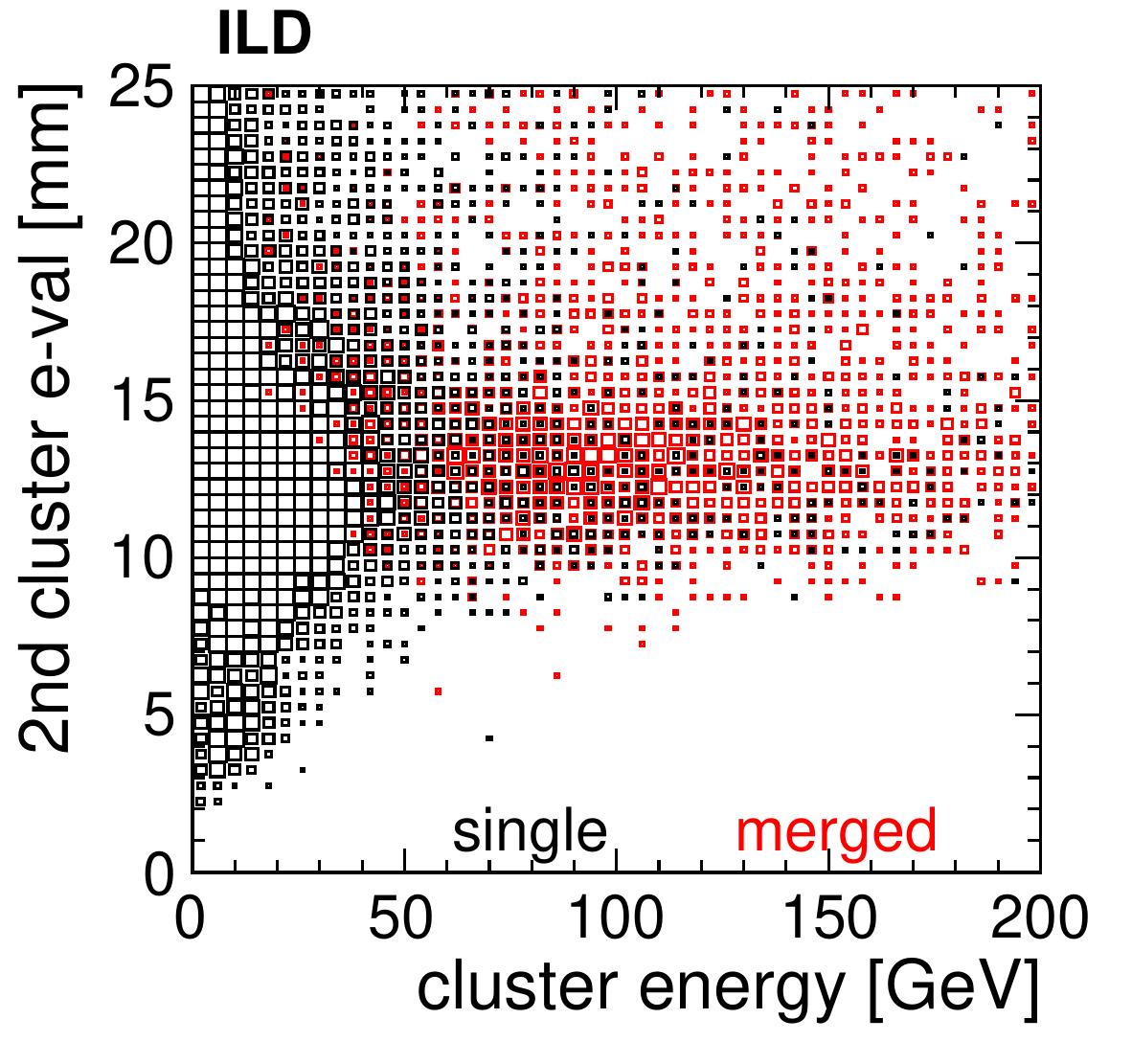}  
\includegraphics[width=0.45\textwidth]{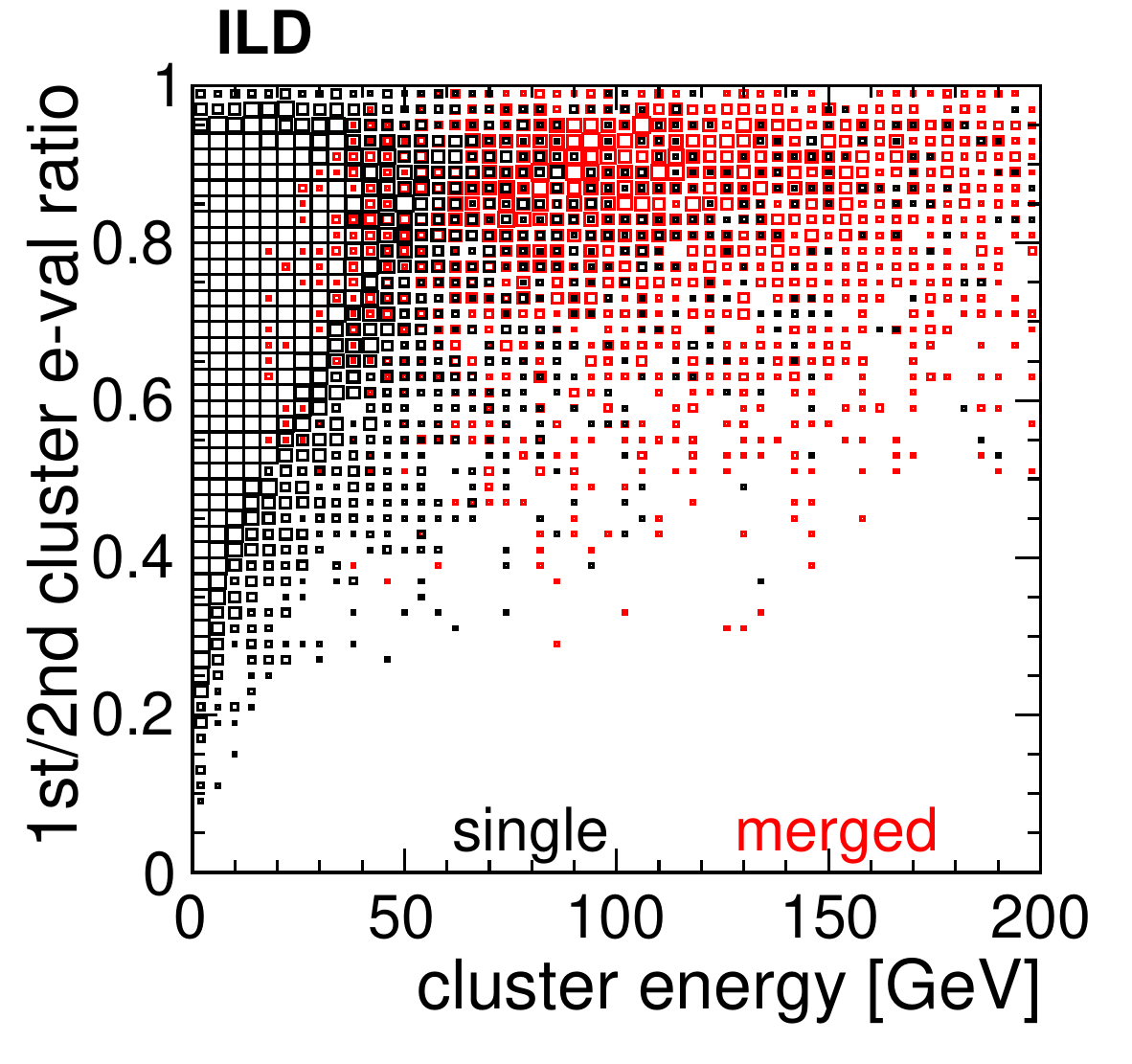}
\caption{First (smallest) and second eigenvalues and their ratio, of the ellipsoid fitted to photon-like clusters 
found in $\tau \rightarrow \rho$ decay jets. The contribution from clusters resulting from the merger of two clusters
is shown in red, and single-photon clusters in black. IDR-L detector model.
}
\label{fig:clustermerge}
\end{figure}

Figure~\ref{fig:simplecomp1} shows the number of photon-like PFOs identified within the jet cone, in the case that
the cone is matched (in angle) to various tau decays at the MC level. In the case of $\rho$ decays, in around half of tau jets
only a single photon cluster is reconstructed.
Figure~\ref{fig:merge} shows the reasons that only a single photon cluster is sometimes found in $\rho$ decays:
\begin{itemize}
\item ``converted'': at least one photon converted in the tracking region (and was not identified as such in the event reconstruction);
\item ``noPFO (lowen)'': no PFO associated with the photon was found (and at least one photon energy $<$ 300~MeV);
\item ``noPFO (other)'': no PFO associated with the photon was found (and both photon energies $>$ 300 MeV);
\item ``merged (phoClus)'': the two photons were attached to the same photon PFO;
\item ``merged (nhadClus)'': the two photons were attached to the same neutral hadron PFO;
\item ``merged (chgClus)'': the two photons were attached to the same charged hadron PFO;
\item ``attachedToChgHad'': one photon was attached to a charged hadron PFO; and 
\item ``misidAsNeuHad'': one photon was reconstructed as a neutral hadron PFO.
\end{itemize}
The most common reason for mis-counting the number of photons is that the two photons
have been merged into a single photon-like cluster. To investigate whether the shape of 
the resulting cluster can be used to identify such ``merged photon'' clusters, we show in 
Fig.~\ref{fig:clustermerge} the smaller two eigenvalues of the ellipsoid which describes
the shape of the calorimetric cluster, for clusters which originate in a single photon, and those which
are the result of a di-photon merger. There is no clear difference between the two polulations,
apart from rather different distributions of the clusters' energy.


\begin{figure}
\centering
\includegraphics[width=0.45\textwidth]{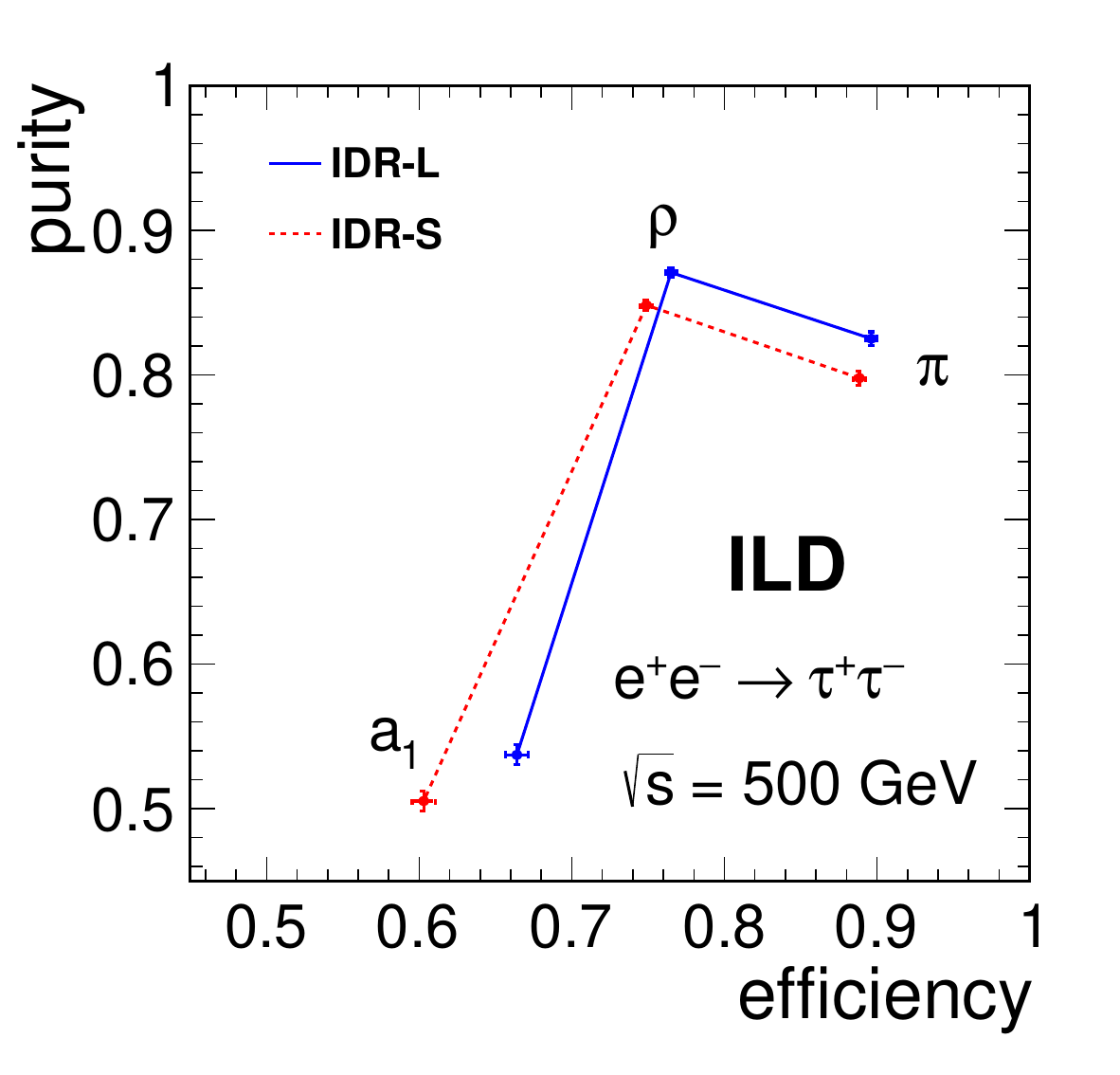}
\includegraphics[width=0.45\textwidth]{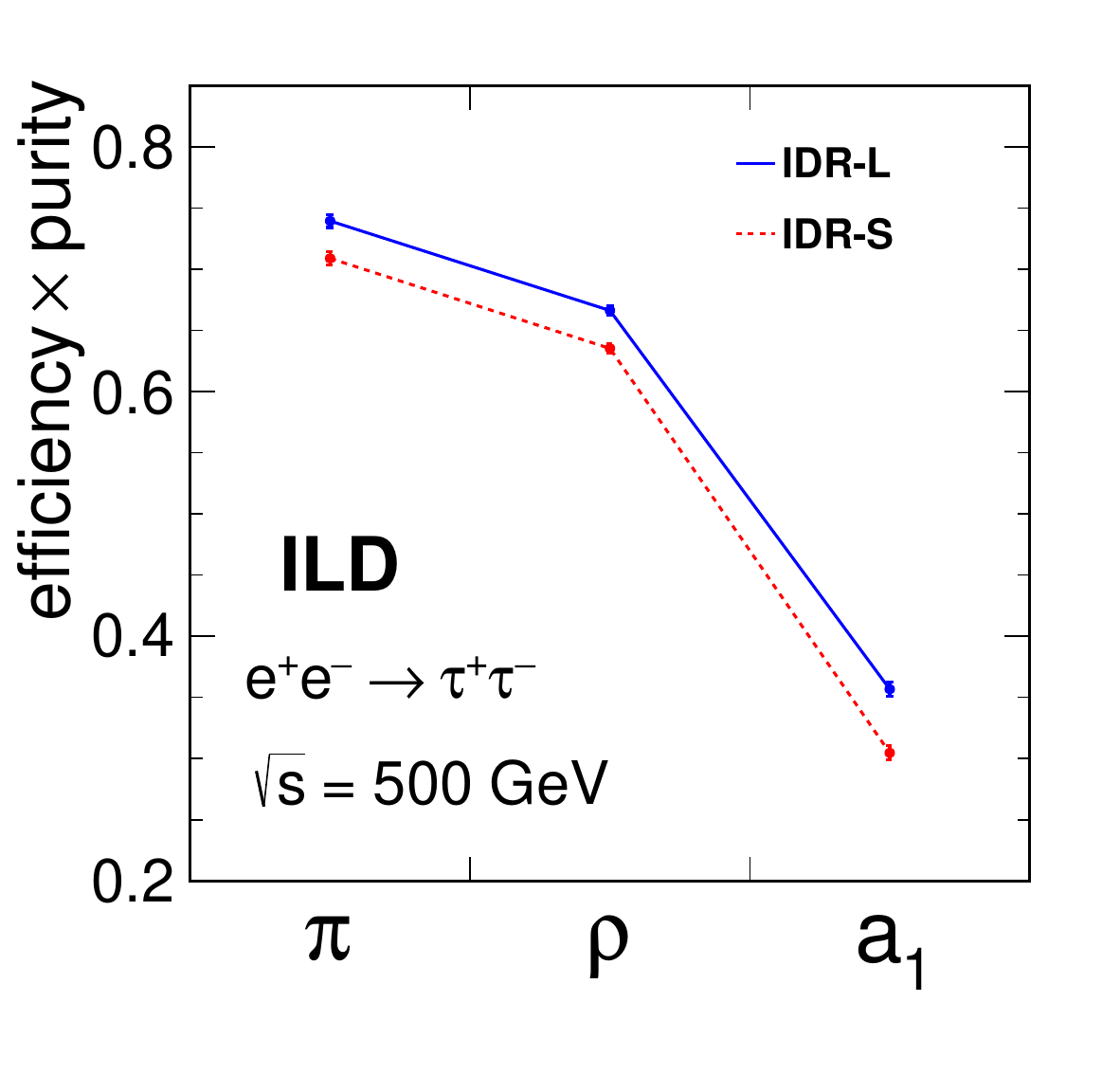}
\caption{Separation of single prong tau decay modes: the efficiency and purity (left) and their product (right) of the decay mode selection described in the text.
The purity definition includes only other high mass di-tau events as background.
}
\label{fig:decaymode}
\end{figure}

To decide whether a jet originates from \tpn\ or \trn, we first require that it contains a single charged PFO.
A cut-based selection is based on 3 observables of particles in the trimmed candidate jet: 
\begin{itemize}
\item number of identified photon PFOs;
\item total invariant mass of all visible particles; and 
\item total invariant mass of all neutral visible particles.
\end{itemize}
The same selection criteria are used in both detectors models. The performance of this identification in both models is shown in Table~\ref{tab:decayMode} 
and summarised graphically in Fig.~\ref{fig:decaymode}.


\begin{table}
\centering
\begin{tabular}{|l|rrrr|r|}
\hline
 & \multicolumn{4}{c|}{true MC decay} & \\ 
 &  \tpn  &  \trn  &  \tAn  & $\tau^\pm \to$ other & purity \\
\hline
\hline
 & \multicolumn{5}{c|}{IDR-L} \\
\hline
%
%
%
selected as \tpn    & $    89.27 \pm     0.38 $ & $        2.06 \pm     0.12 $ & $        0.87 \pm     0.13 $ & $        9.22 \pm     0.29 $ & $       82.11 \pm     0.45 $ \\ 
selected as \trn    & $     6.47 \pm     0.30 $ & $       75.21 \pm     0.36 $ & $       13.32 \pm     0.48 $ & $        5.81 \pm     0.23 $ & $       86.79 \pm     0.30 $ \\ 
selected as \tAn    & $     2.20 \pm     0.18 $ & $       13.03 \pm     0.28 $ & $       64.32 \pm     0.68 $ & $        6.74 \pm     0.25 $ & $       53.86 \pm     0.65 $ \\ 
\hline
\hline
 & \multicolumn{5}{c|}{IDR-S} \\
\hline
%
%
%
selected as \tpn   & $    88.28 \pm     0.40 $ & $        3.16 \pm     0.15 $ & $        1.11 \pm     0.15 $ & $        9.93 \pm     0.30 $ & $       79.20 \pm     0.47 $ \\ 
selected as \trn   & $     7.56 \pm     0.33 $ & $       73.45 \pm     0.37 $ & $       17.14 \pm     0.54 $ & $        5.92 \pm     0.23 $ & $       84.64 \pm     0.32 $ \\ 
selected as \tAn   & $     2.18 \pm     0.18 $ & $       13.82 \pm     0.29 $ & $       58.64 \pm     0.70 $ & $        6.33 \pm     0.24 $ & $       50.65 \pm     0.66 $ \\ 
\hline
\end{tabular}
\caption{Selected 1-prong tau candidates in signal events: decay mode identification efficiency in large and small models (unpolarised sample),
and the purity considering only backgrounds from other high mass di-tau events.}
\label{tab:decayMode}
\end{table}

As shown in both the table and the plot, the performance of the large detector model is somewhat better than the small one, with slightly better
efficiency and/or purity when selecting these three decay modes.

\begin{figure}
\centering
\includegraphics[width=0.8\textwidth]{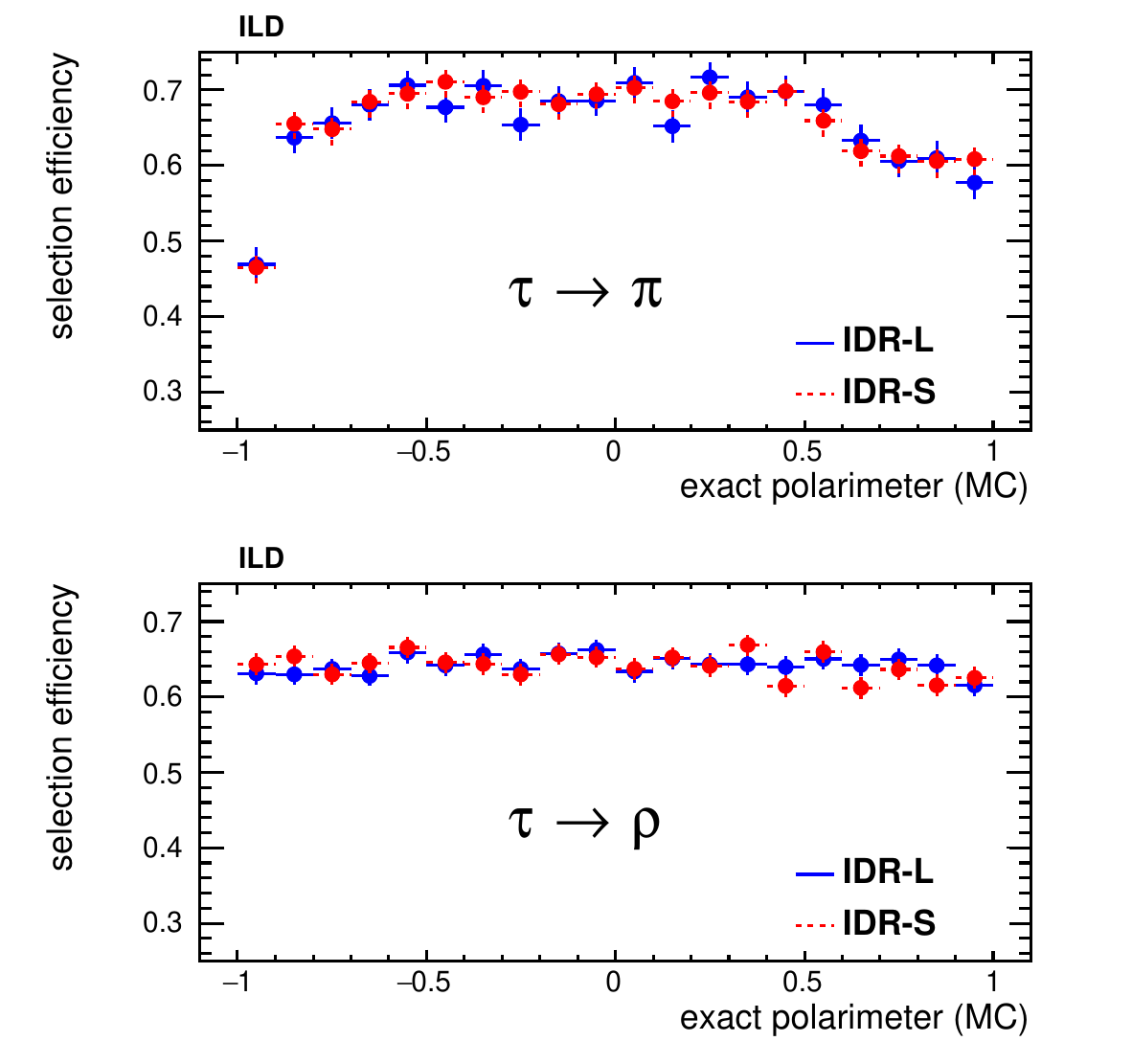}
\caption{
Selection and reconstruction efficiency for taus in the two considered decay modes, as a function of the MC ``optimal'' polarimeter.
}
\label{fig:effPol}
\end{figure}

The efficiency to select events and correctly identify tau decay modes may have some dependence on the helicity of the taus involved.
Such a dependence might introduce a bias on the extraction of the tau polarisation, if this is not corrected for.
The dependence on the selection and reconstruction efficiency on the optimal polarimeters, calculated using MC truth information, is shown in Fig.~\ref{fig:effPol}.
Some dependence is seen, most notably around $-1$ for $\pi^\pm$ decays: this can be understood since it corresponds to a
very soft charged pion.

\clearpage

\section {Polarimeter estimation}

\begin{figure}
\centering
\includegraphics[width=0.45\textwidth]{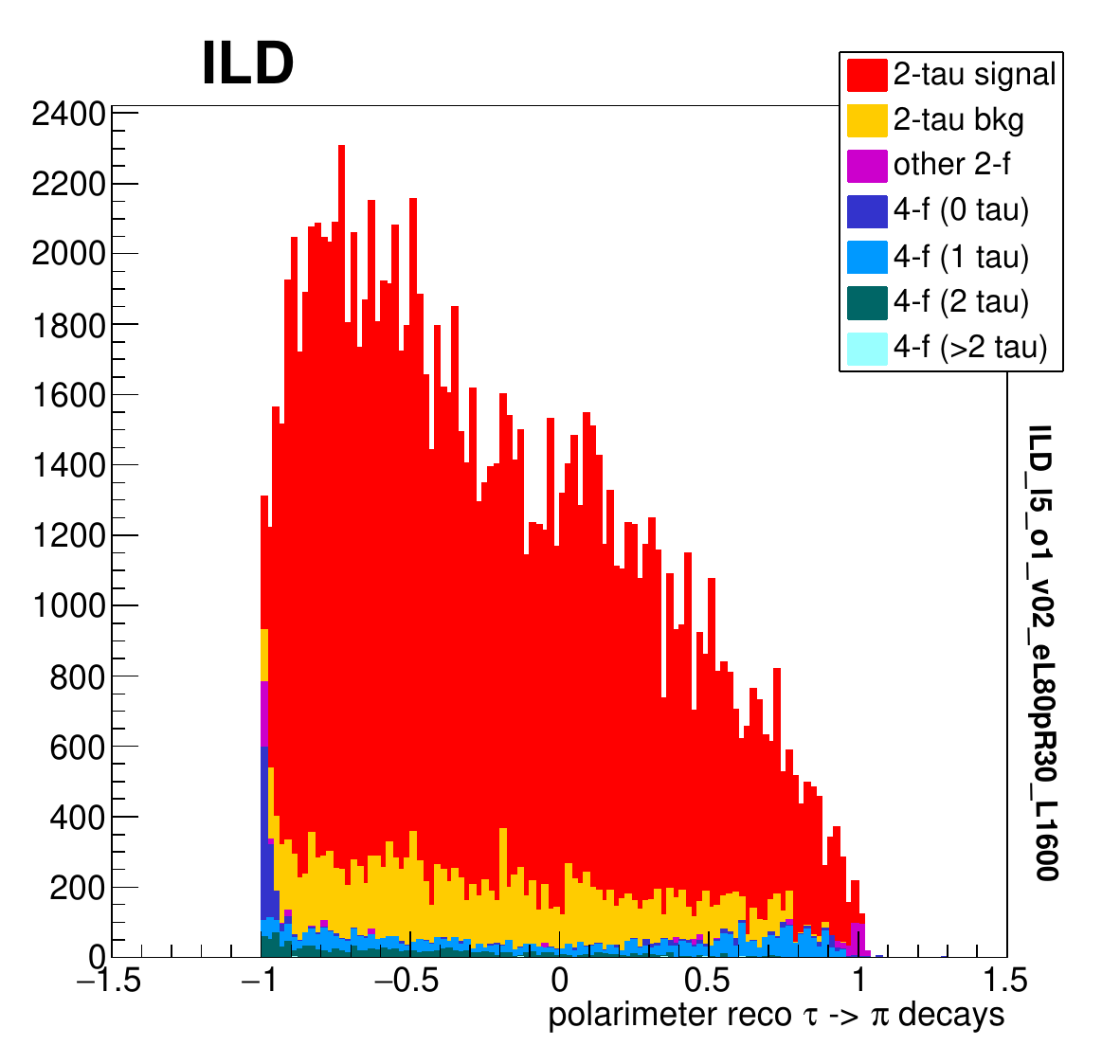}
\includegraphics[width=0.45\textwidth]{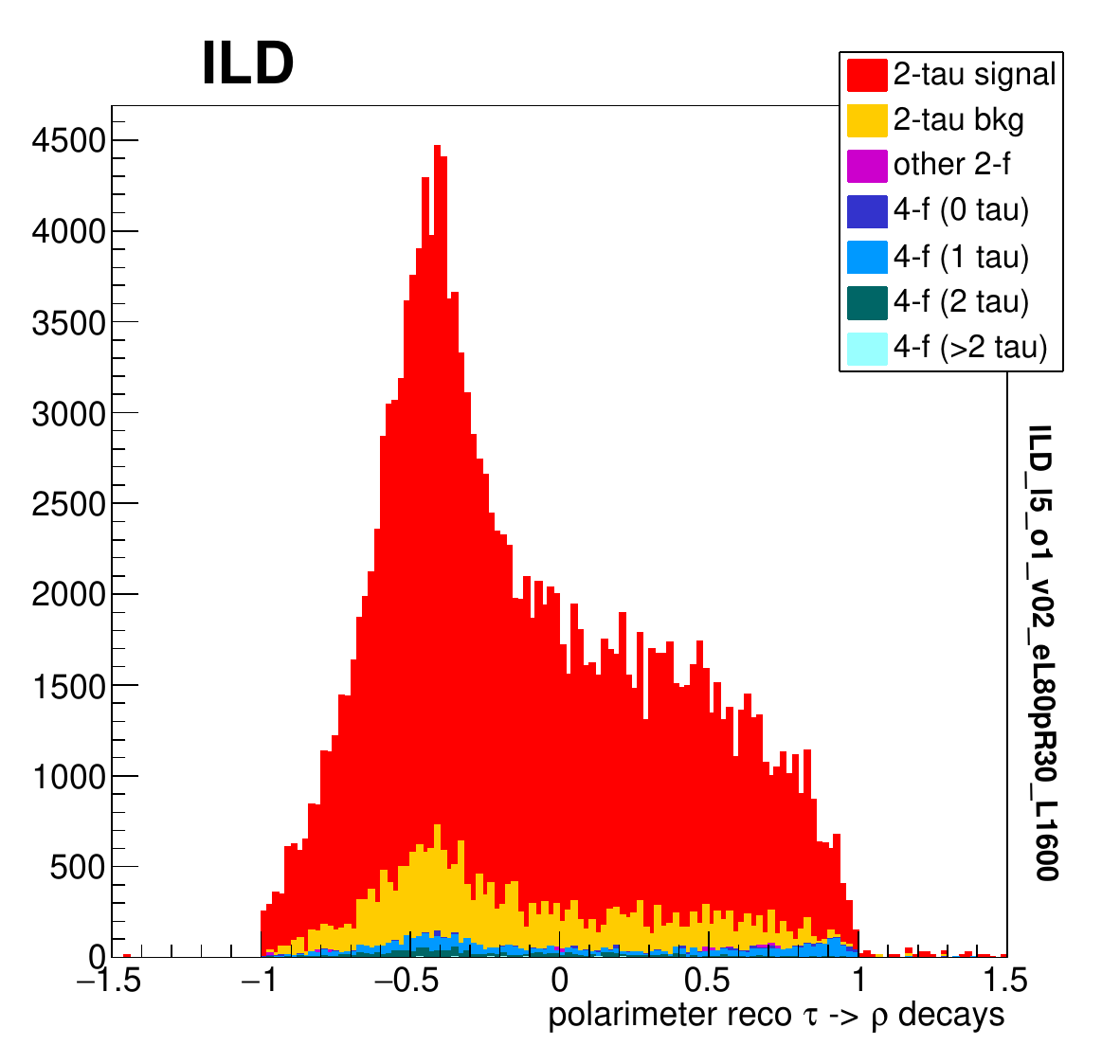}
\caption{Distributions after event selection and tau decay mode identification. Reconstructed ``approximate'' polarimeters in jets identified as $\tau \to \pi, \rho$ decays.
Plots normalised to $1.6~\mathrm{ab^{-1}}$ of \eLpR.
}
\label{fig:varsel2}
\end{figure}

We here discuss how to extract polarisation information from measurements of the tau decay products.
In the present analysis, we adopt an approach which makes use of only the
visible 4-vectors of the charged (and potentially neutral) pions produced in the tau decay, as described in \cite{duflot}.
We use just \tpn\ and \trn\ decays. In the case of \tpn, we use just the fraction of the beam energy carried by the pion, while
$\rho$ decays make use of a more complicated function of the measured charged and neutral pion momenta, as described in the above reference and reproduced in an appendix to this note.

\begin{figure}
\centering
\includegraphics[width=0.45\textwidth]{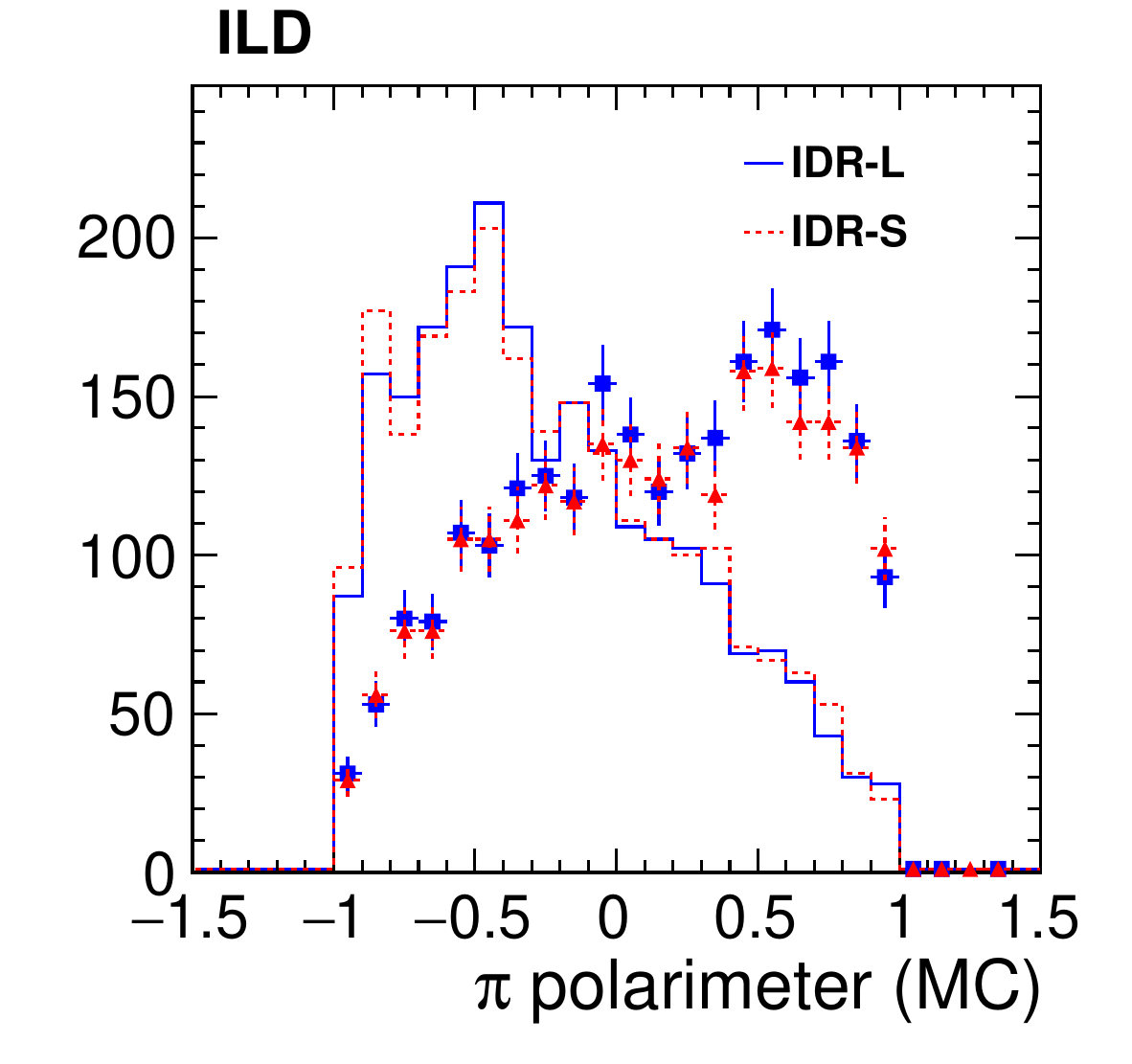}
\includegraphics[width=0.45\textwidth]{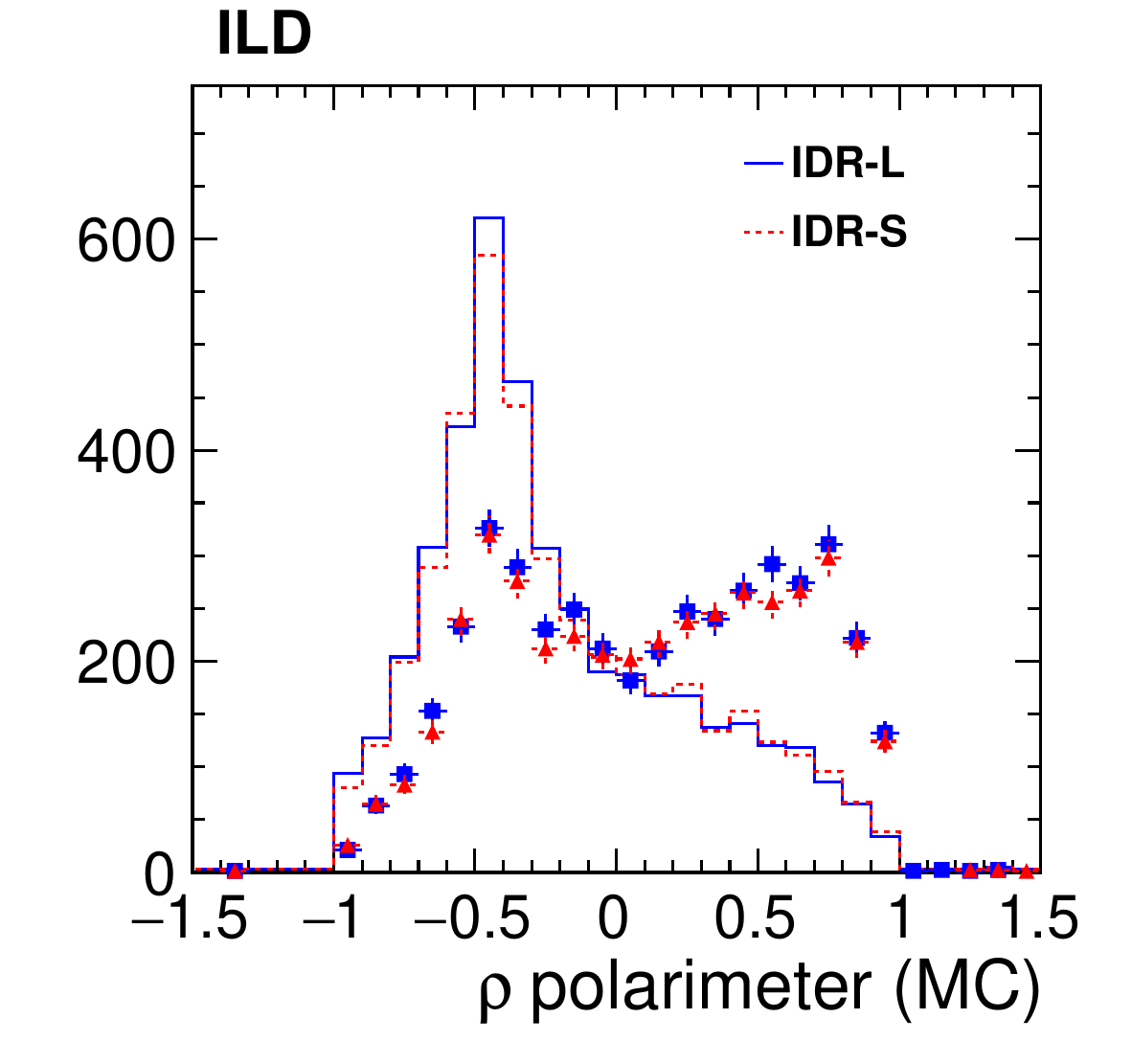} \\
\includegraphics[width=0.45\textwidth]{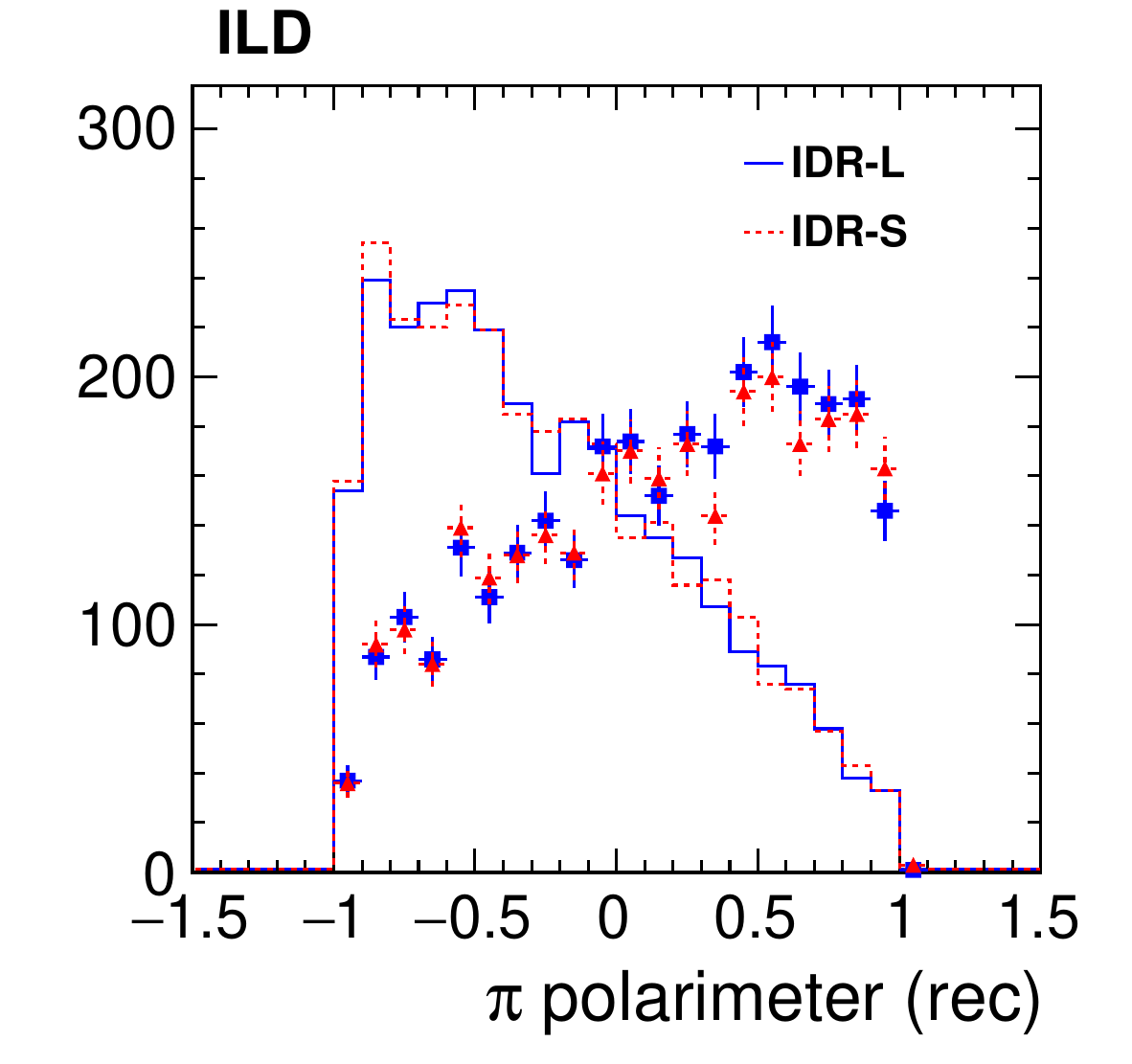}
\includegraphics[width=0.45\textwidth]{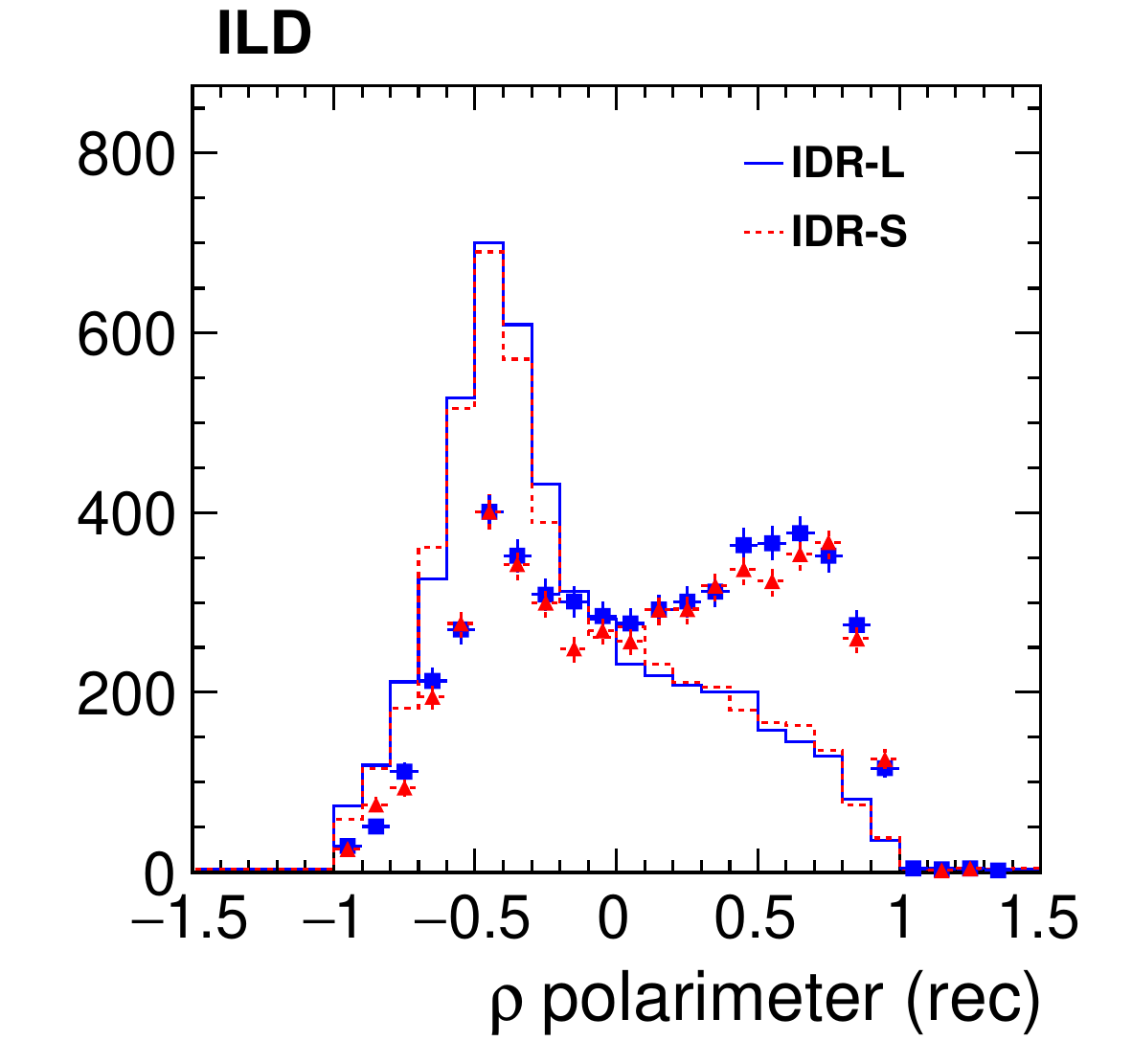} \\
\includegraphics[width=0.45\textwidth]{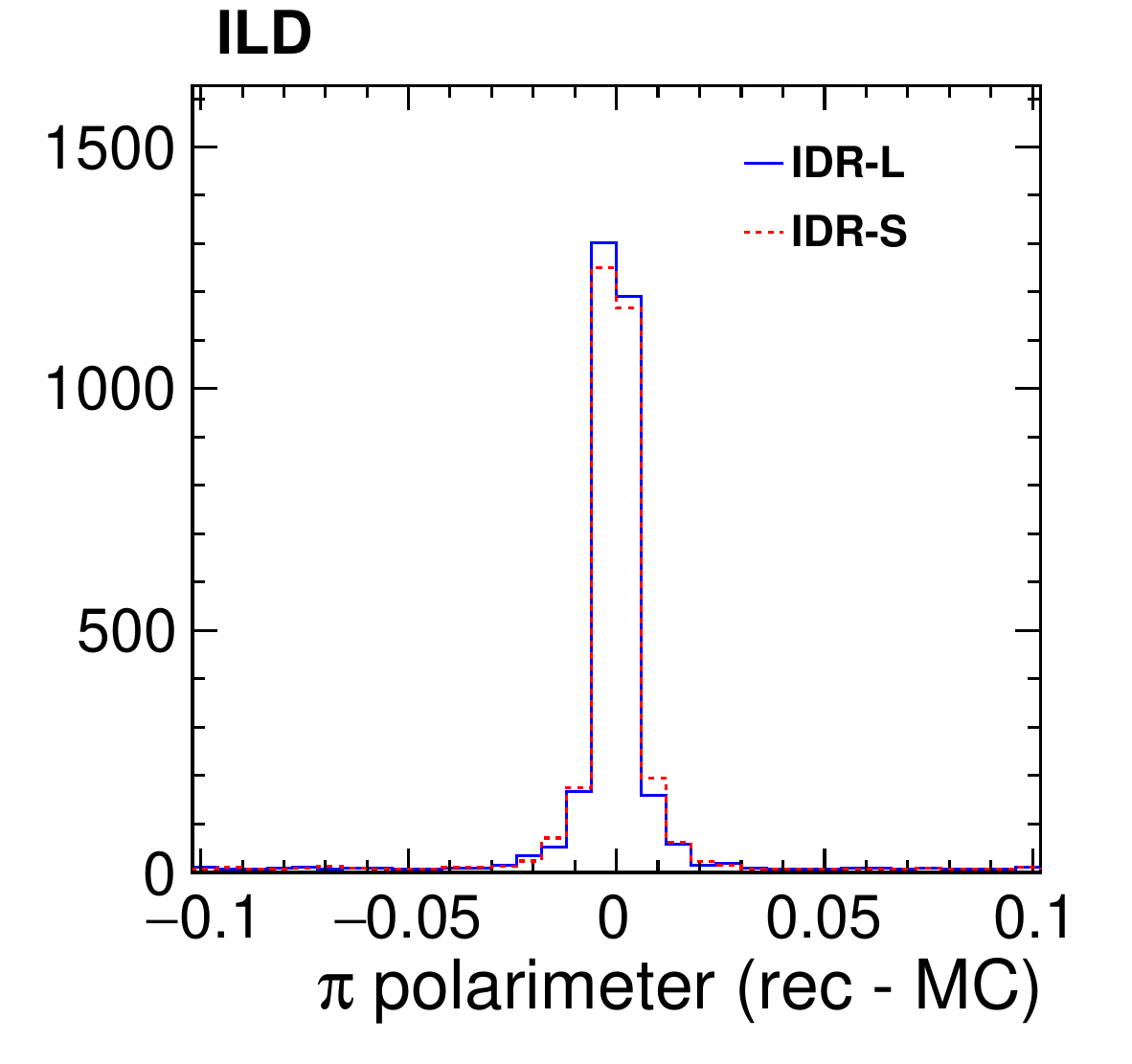}
\includegraphics[width=0.45\textwidth]{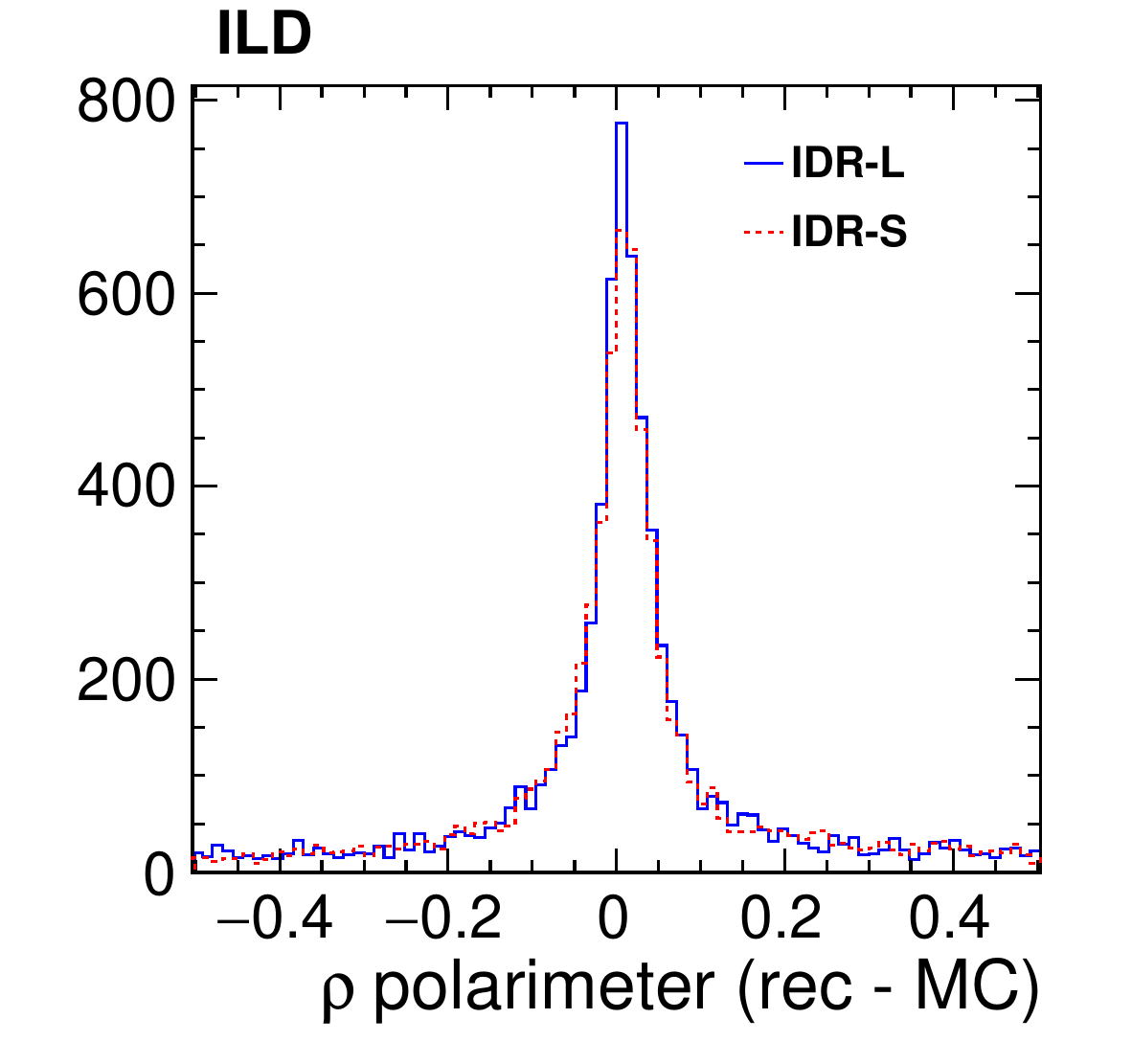}
\caption{``Approximate'' polarimeters in selected and correctly identified tau jets. 
Top: distributions of the MC polarimeters for jets in selected signal events:
the line histograms are for 100\% \PUREeLpR\ beam polarisation, and the markers with error bars are for 100\% \PUREeRpL\ polarisation.
Middle: the same for the reconstructed polarimeters.
Bottom: the tau-by-tau difference between polarimeters calculated using reconstructed and MC truth particle momenta.
}
\label{fig:recoPols}
\end{figure}

Figure~\ref{fig:varsel2} shows the reconstructed polarimeters in tau jet candidates identified as these decays in selected signal and background events.
In Fig.~\ref{fig:recoPols} we show the signal-only reconstructed polarimeters in the two detector models and for 100\% polarisation scenarios, and their difference to the MC truth,
for tau jets whose decay has been correctly identified, in selected events.
The shapes of the polarimeter distributions clearly show differences between the polarisation scenarios.
The difference between the polarimeters extracted using reconstructed and true particle momenta show that the 
pion polarimeters can be much more precisely extracted than for $\rho$ decays, and that the large model ILD-L gives slightly better resolution than the smaller one ILD-S. 


\begin{figure}
\centering
\includegraphics[width=0.45\textwidth]{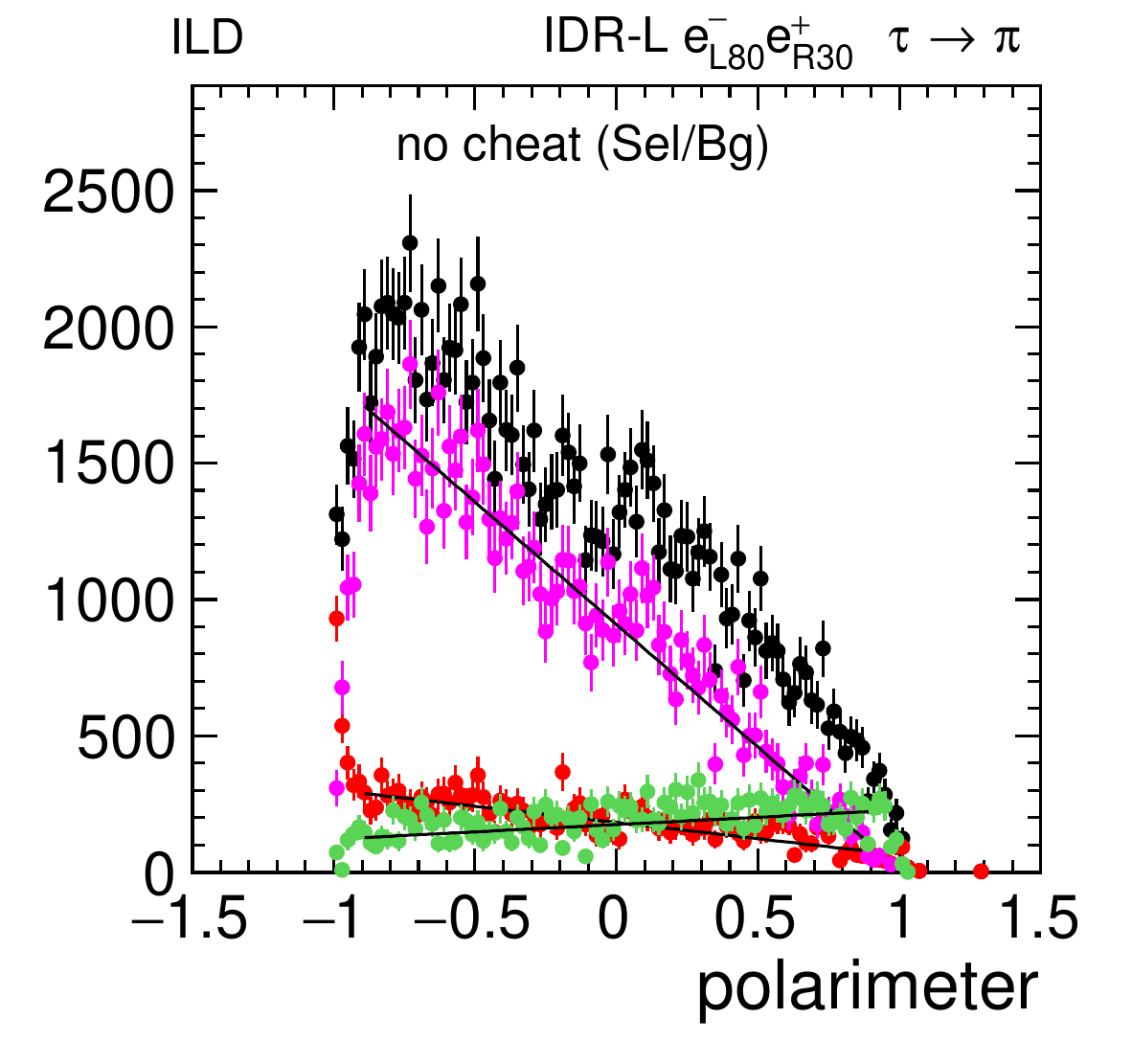}
\includegraphics[width=0.45\textwidth]{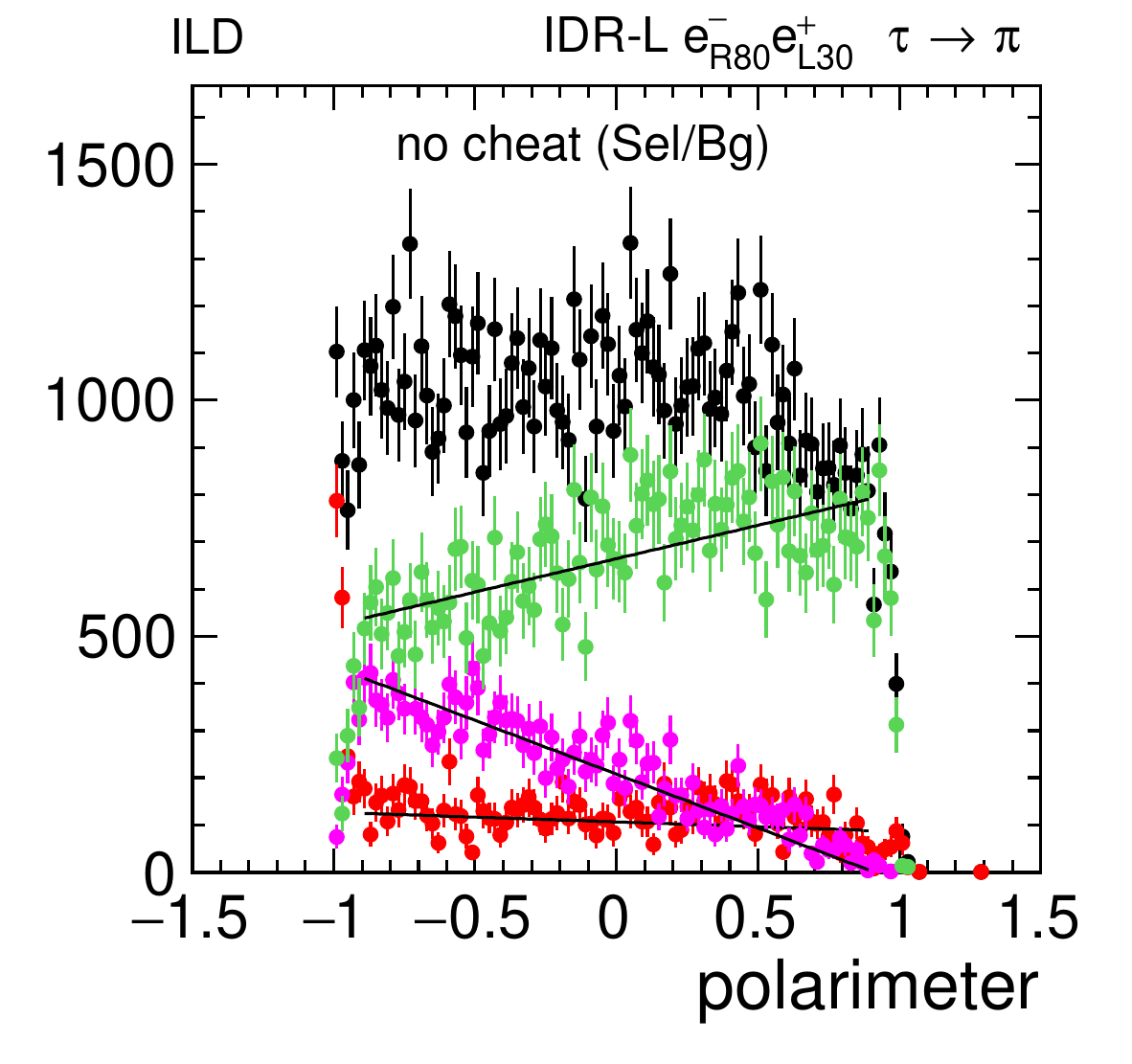} \\
\includegraphics[width=0.45\textwidth]{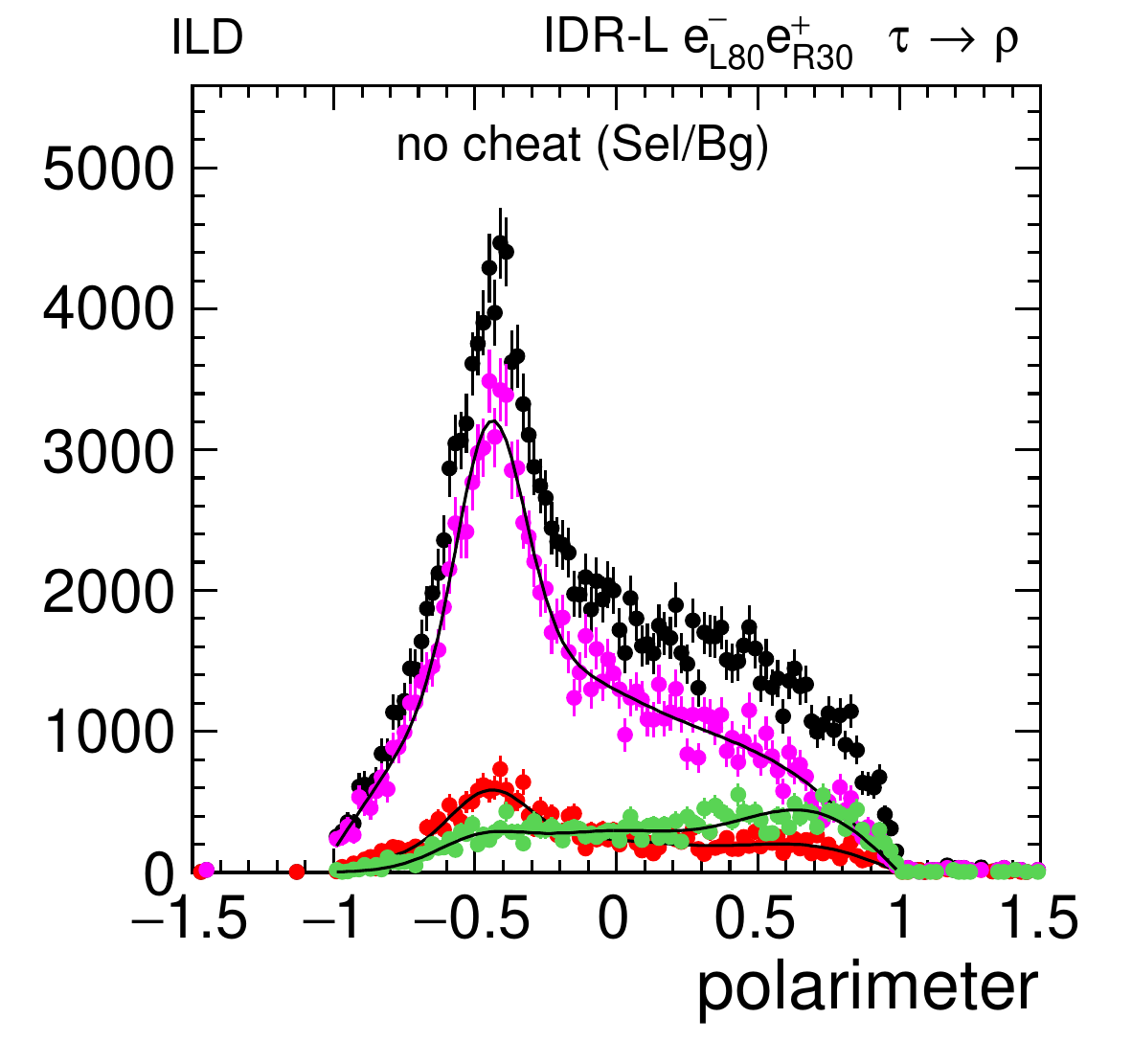}
\includegraphics[width=0.45\textwidth]{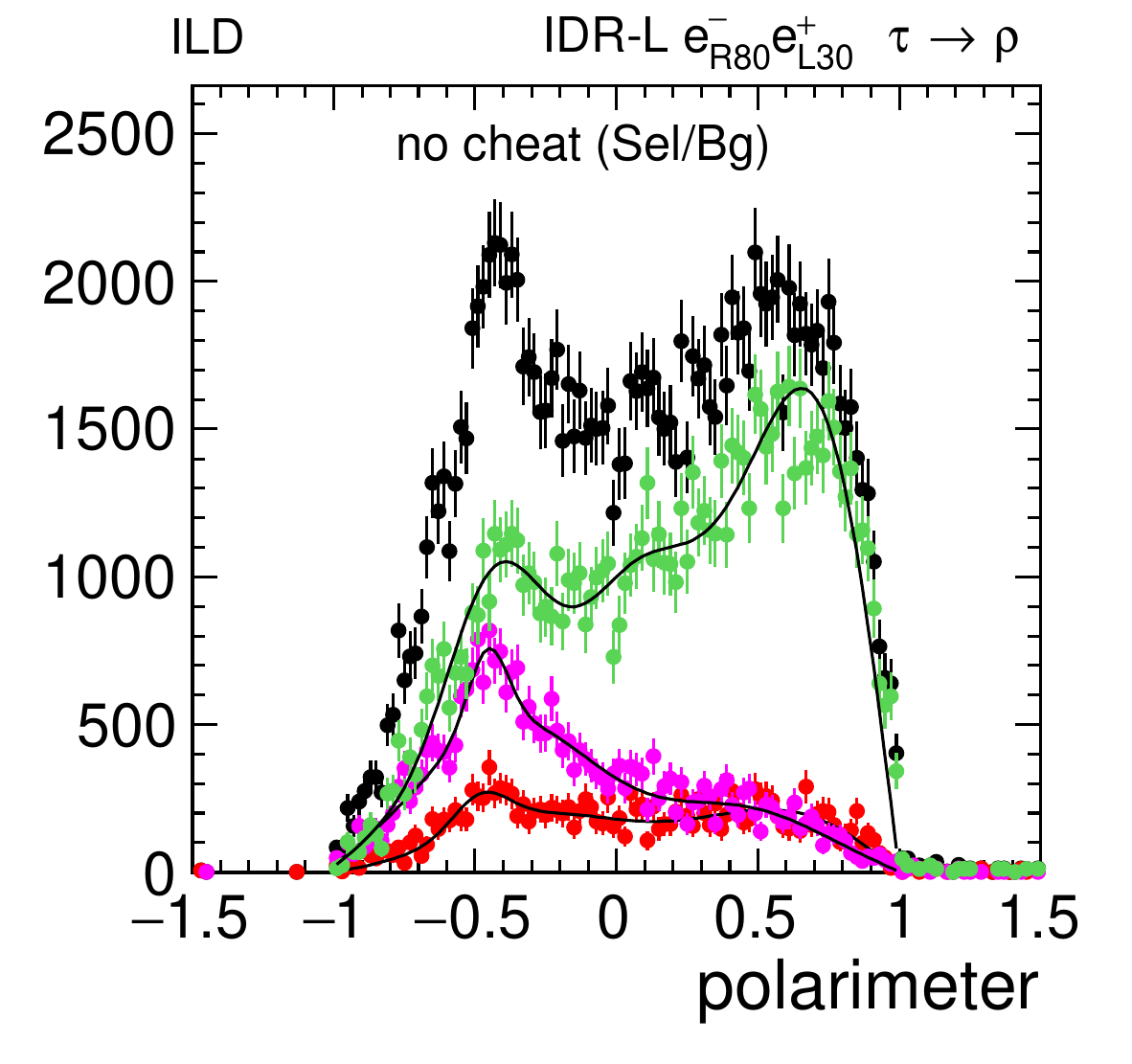}
\caption{Reconstructed polarimeter templates for \tpn\ (top) and \trn\ (bottom) decays in the IDR-L detector model, scaled to the expected integrated luminosity in the \eLpR\ and \eRpL\ polarisation scenarios. 
Black=total, pink=negative helicity signal, green=positive helicity signal, red=background contriutions. 
Error bars are due to finite MC statistics, and 
lines representing the fitted functions use to describe the individual contributions are superimposed.
}
\label{fig:fitTemplates-pirho}
\end{figure}

The distributions, normalised to the expected integrated luminosity and 80\%/30\% beam polarisation, 
are shown in Fig.~\ref{fig:fitTemplates-pirho} for \eLpR\ and \eRpL\ polarisations in the two considered tau decay modes.
The total distribution is split into contributions from positive and negative helicity taus from the signal process, and one for selected background processes.
These distributions were fit to approriate functions, which were used to obtain smoothed input templates, to minimise the effect of 
statistical fluctuations in the MC datasets used. For \tpn\ and exact \trn\ polarimeters a simple
1st degree polynomial was used, while a more complex 8--parameter function was used to descibe the distribution of the approximate polarimeter in \trn\ decays.
Where appropriate, a slightly limited range was considered, excluding regions near $\pm 1$.


%

\begin{figure}
\centering
\includegraphics[width=0.45\textwidth]{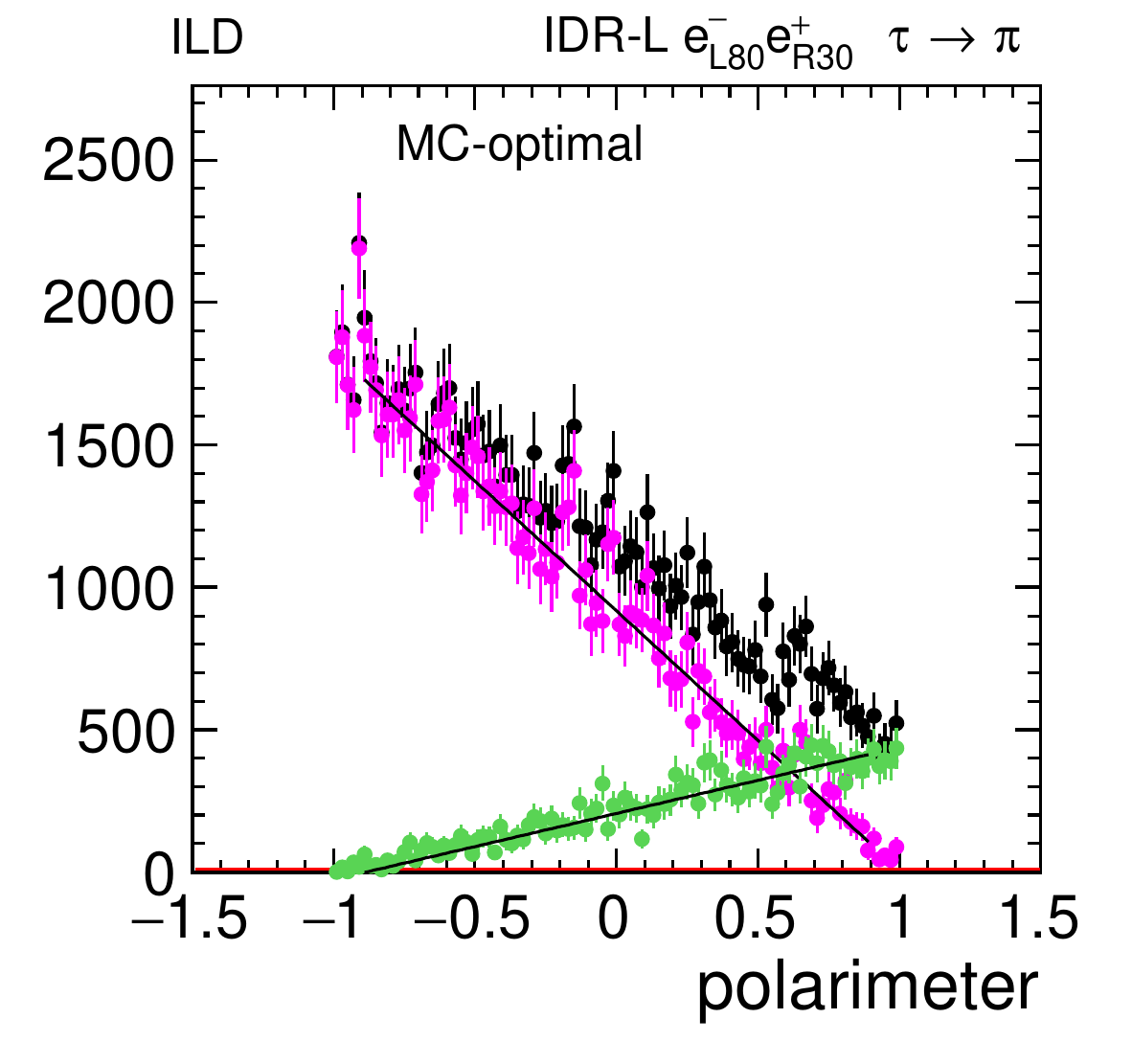}
\includegraphics[width=0.45\textwidth]{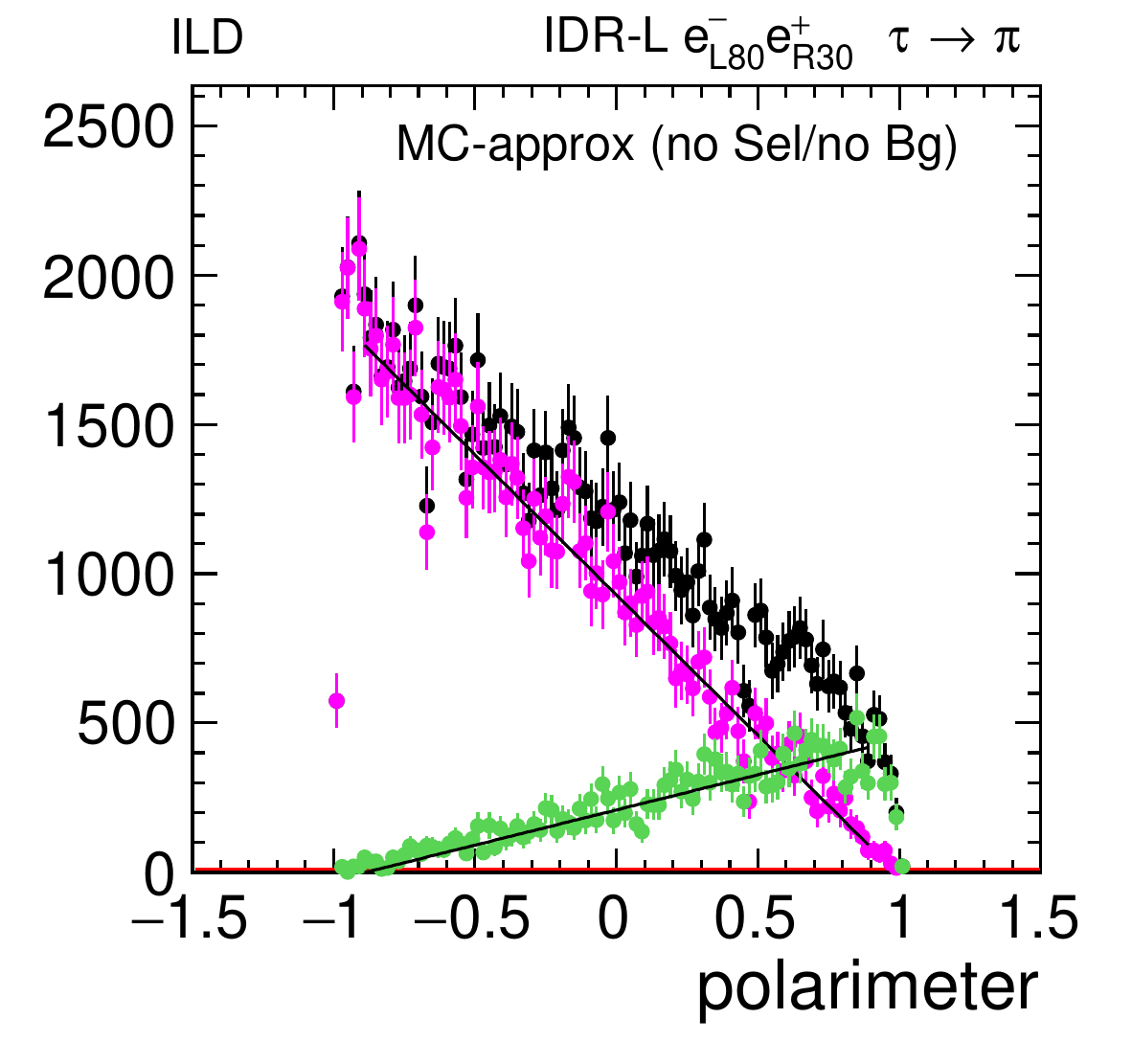}
\includegraphics[width=0.45\textwidth]{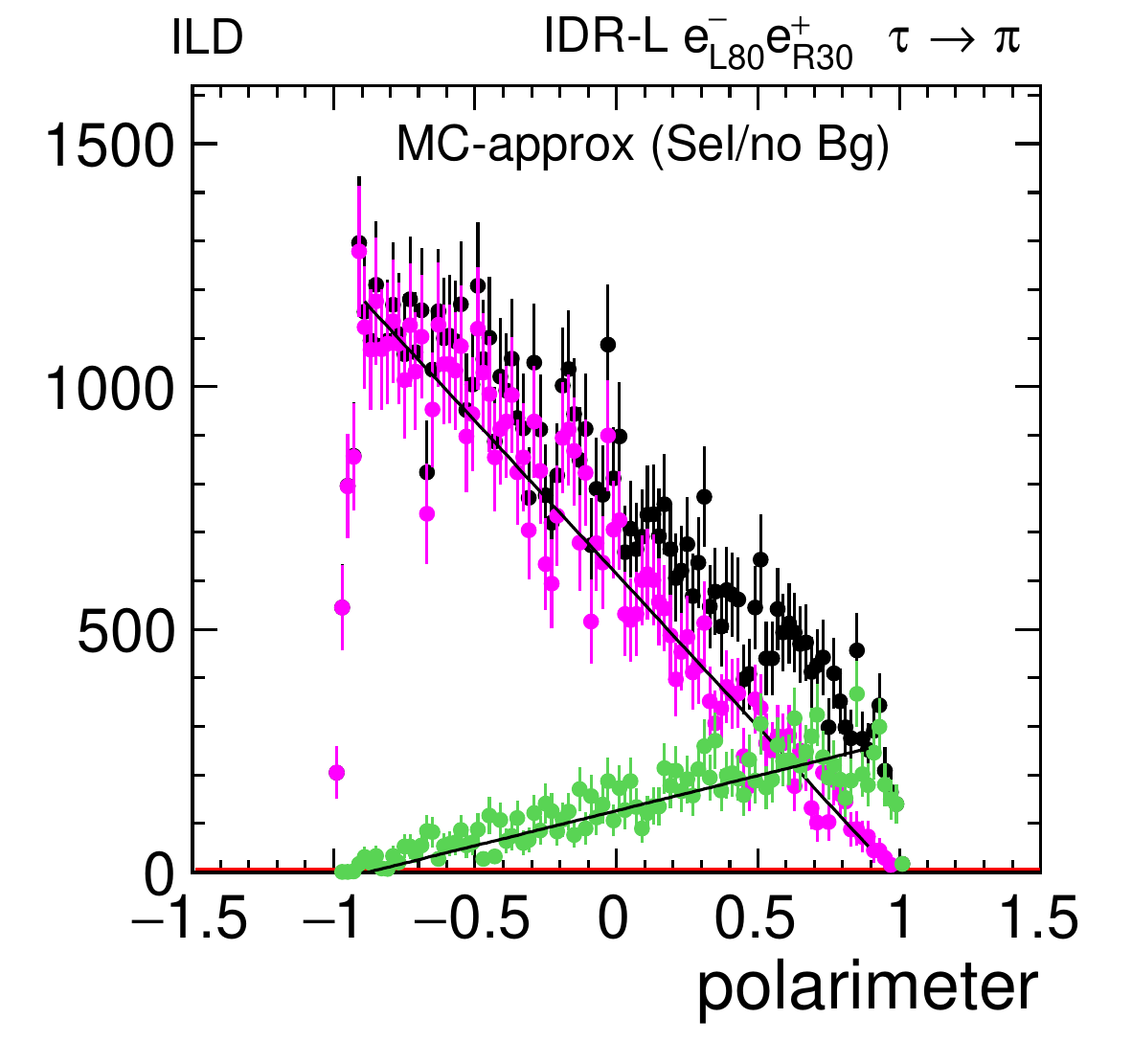}
\includegraphics[width=0.45\textwidth]{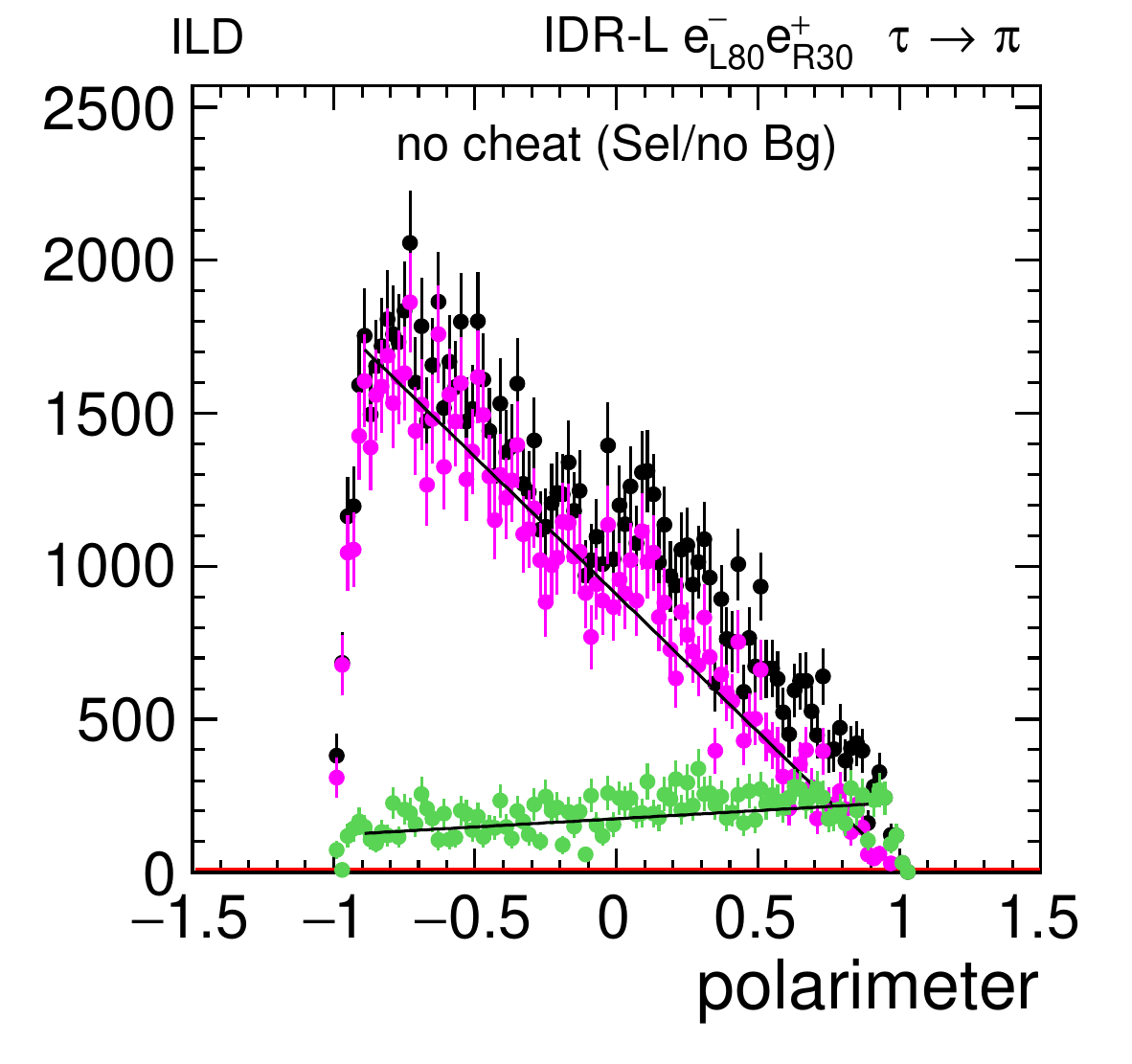}
\includegraphics[width=0.45\textwidth]{figs_updated/noCheatSEL_L_lr_pi_templates-eps-converted-to.pdf}
\caption{Polarimeter templates for \tpn\ decays at the different levels of ``cheating'', for IDR-L in the \eLpR\ scenario.
Black=total, pink=negative helicity signal, green=positive helicity signal, red=background contriutions. 
Error bars are due to finite MC statistics, and 
lines representing the fitted functions use to describe the individual contributions are superimposed.
}

\label{fig:fitTemplates-TTT}
\end{figure}

\begin{figure}
\centering
\includegraphics[width=0.45\textwidth]{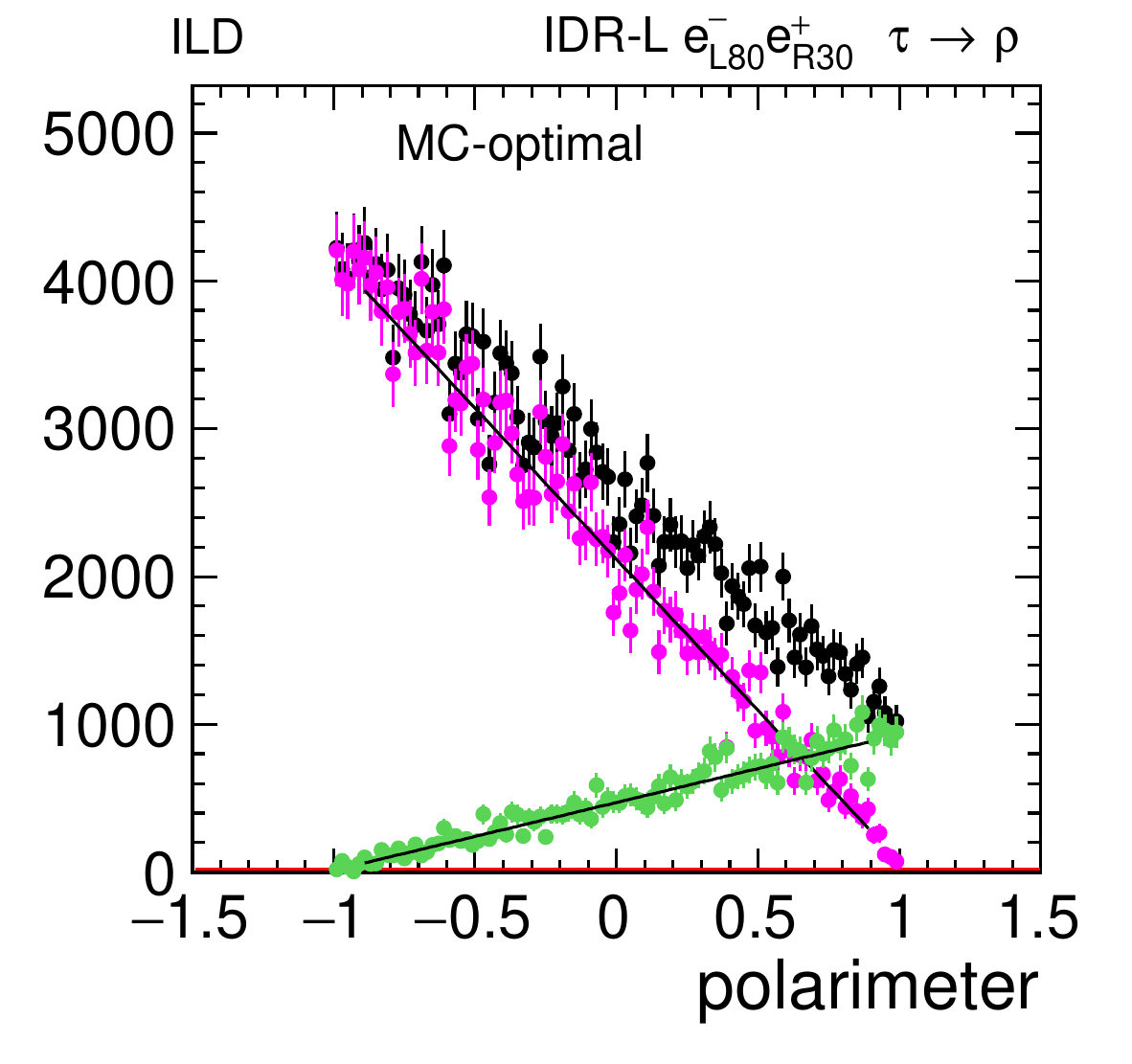}
\includegraphics[width=0.45\textwidth]{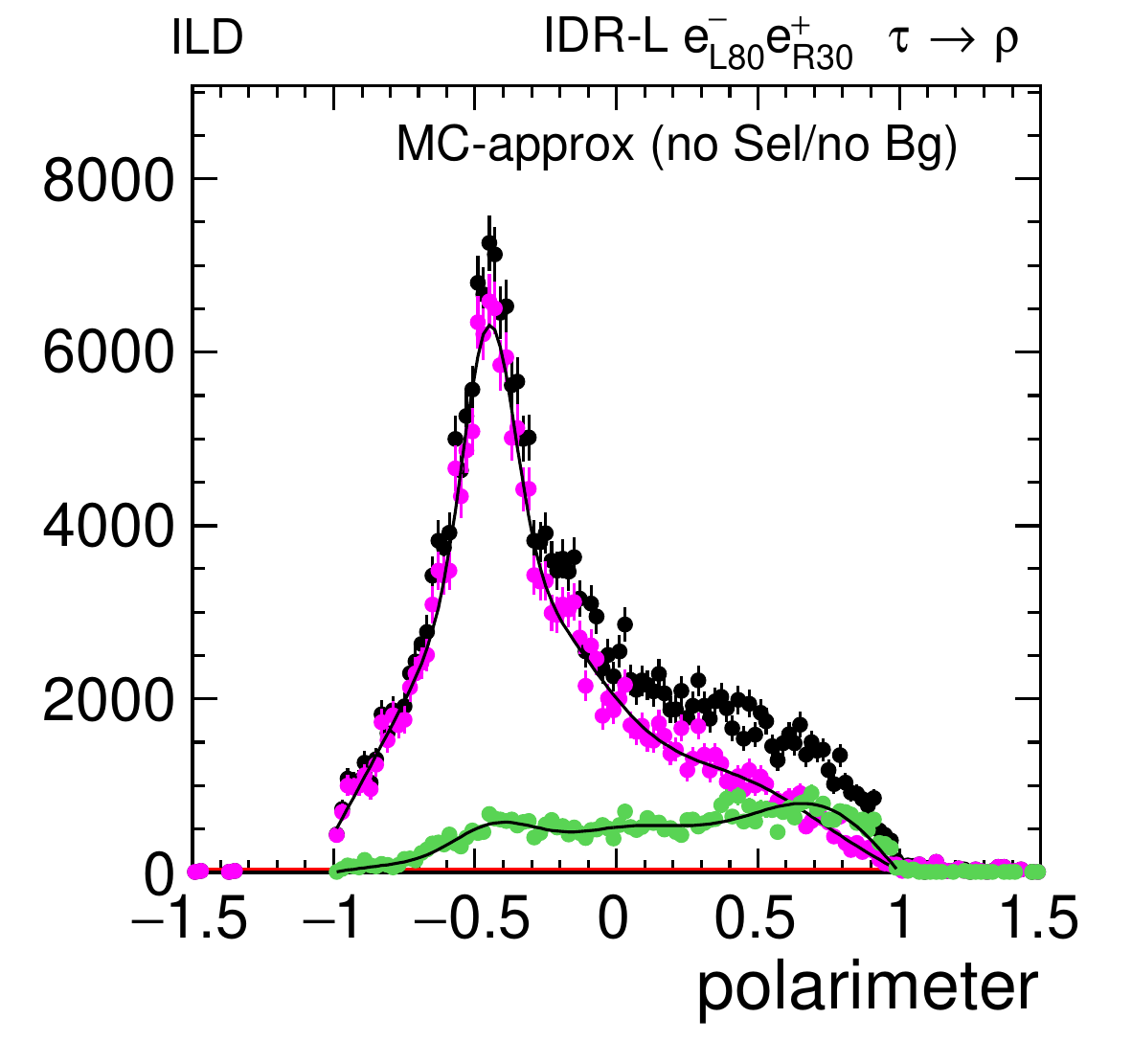}
\includegraphics[width=0.45\textwidth]{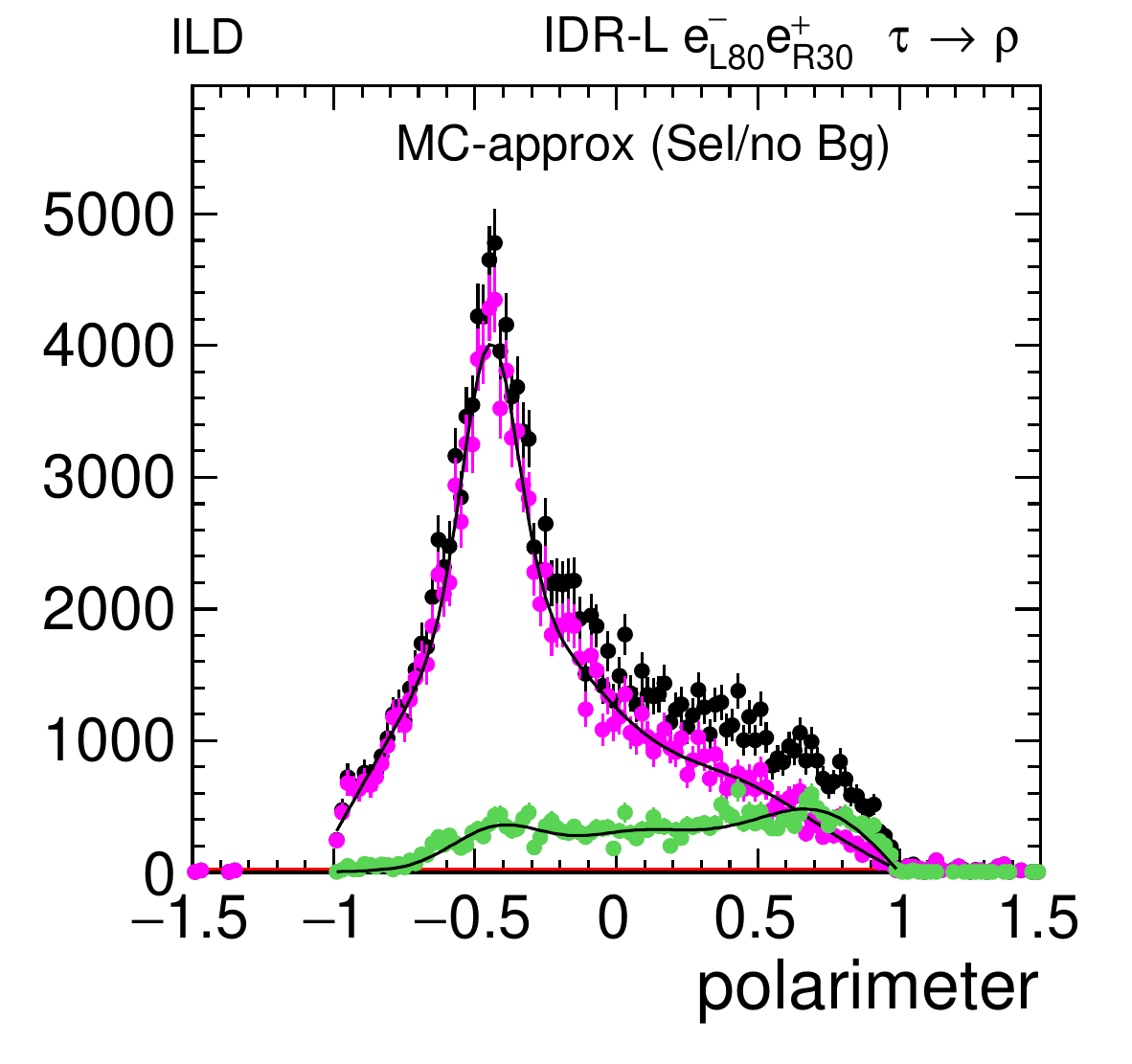}
\includegraphics[width=0.45\textwidth]{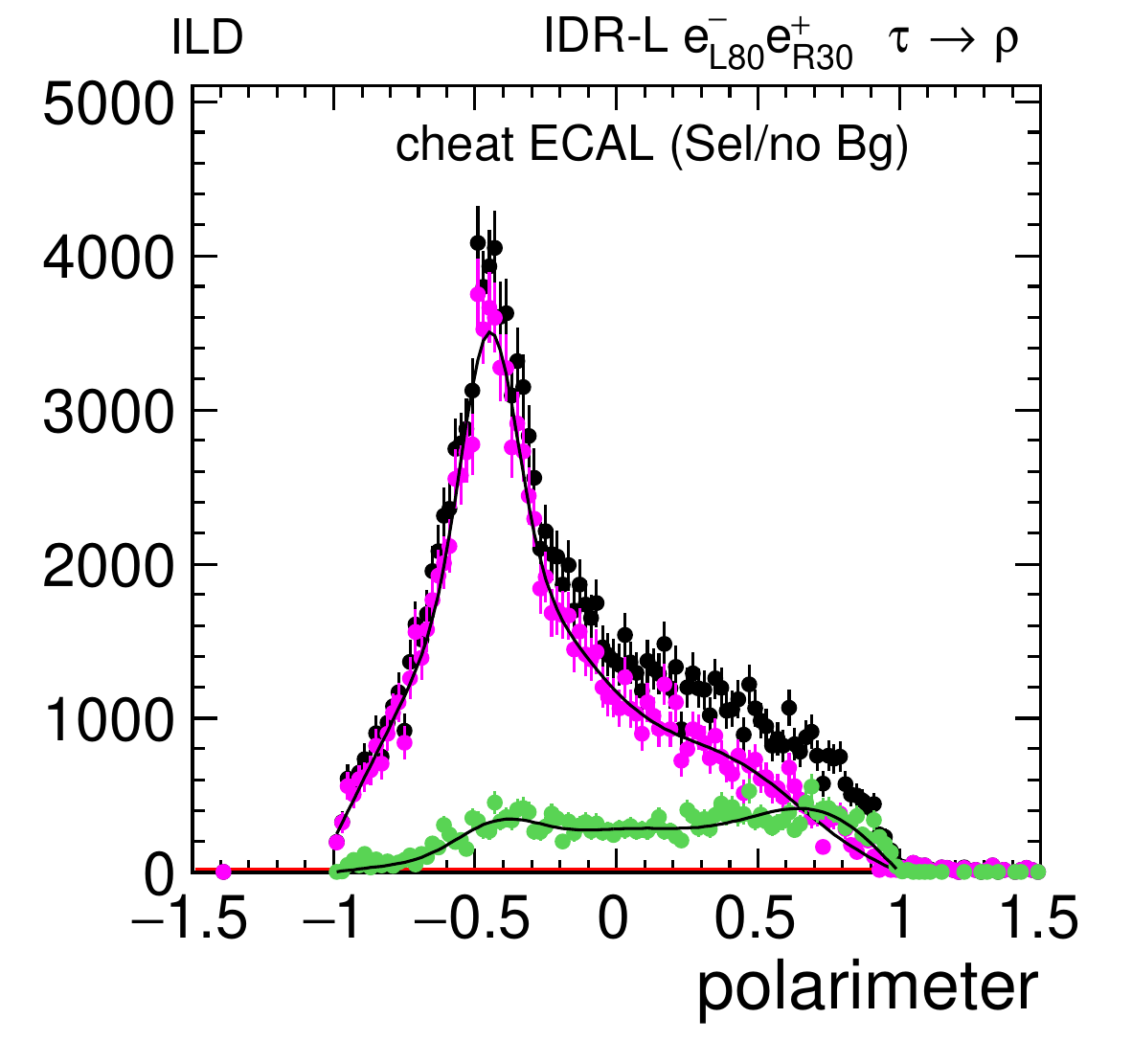}
\includegraphics[width=0.45\textwidth]{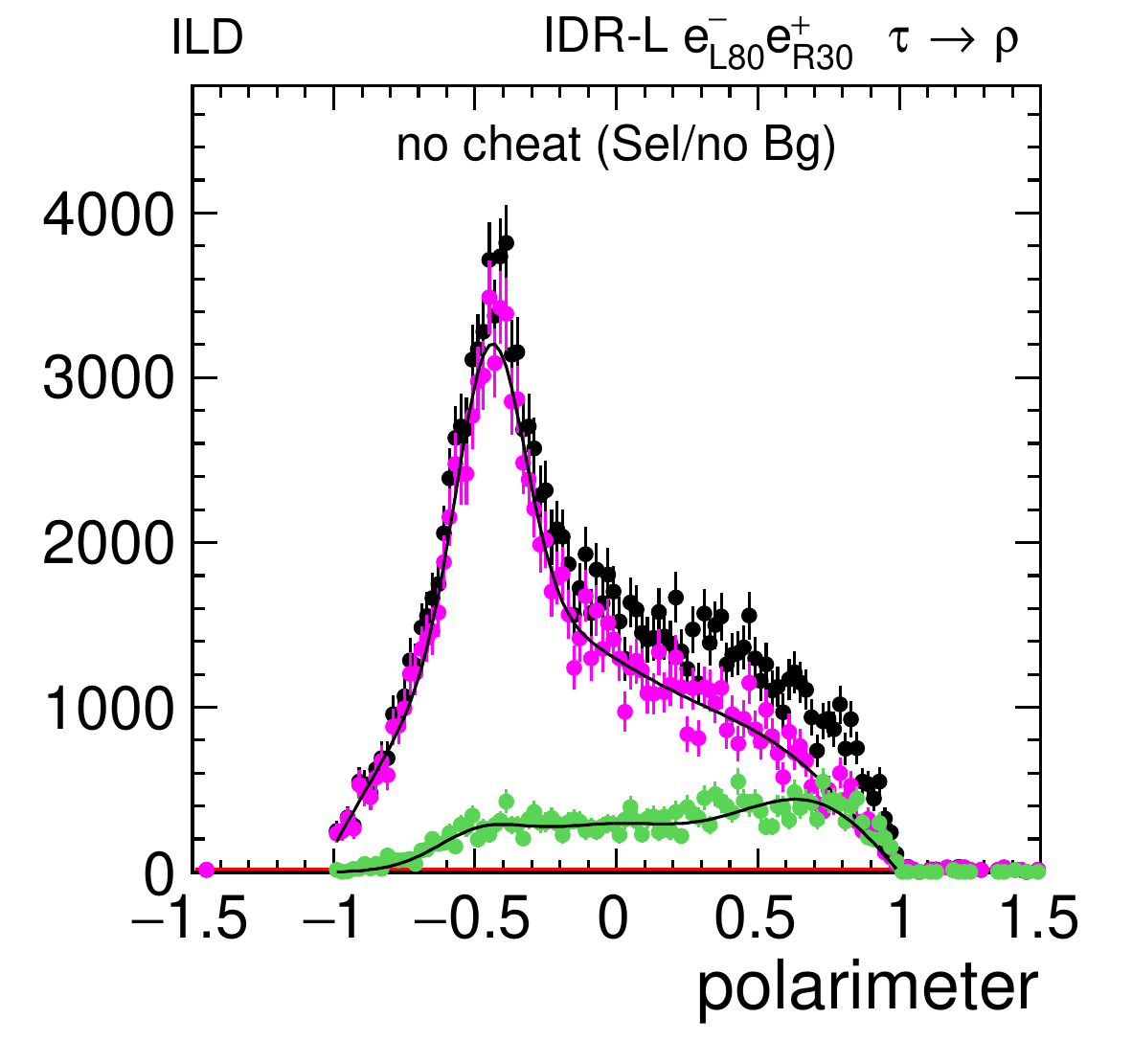}
\includegraphics[width=0.45\textwidth]{figs_updated/noCheatSEL_L_lr_rho_templates-eps-converted-to.pdf}
\caption{Polarimeter templates for \trn\ decays at the different levels of ``cheating''. Data scaled to the \eLpR\ scenario.}
\label{fig:fitTemplates-TTTrho}
\end{figure}

The template distributions at different stages of ``cheating'', from the exact MC truth to the final selected and reconstructed stage, are shown in 
figs.~\ref{fig:fitTemplates-TTT} and \ref{fig:fitTemplates-TTTrho} respectively for the \tpn\ and \trn\ modes.

\clearpage

\section{Tau polarisation measurement}

\begin{figure}
\centering
\includegraphics[width=0.45\textwidth]{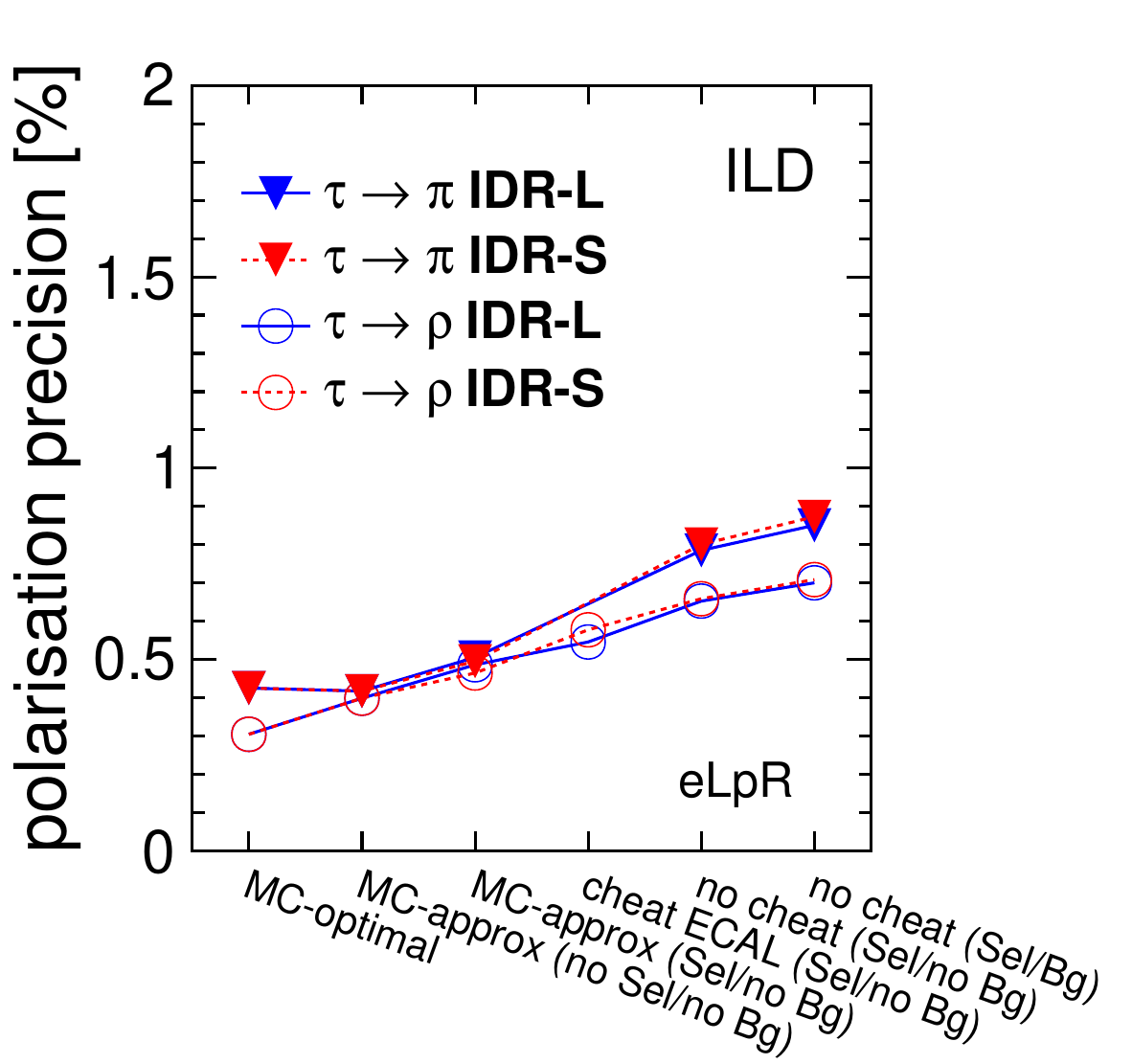}
\includegraphics[width=0.45\textwidth]{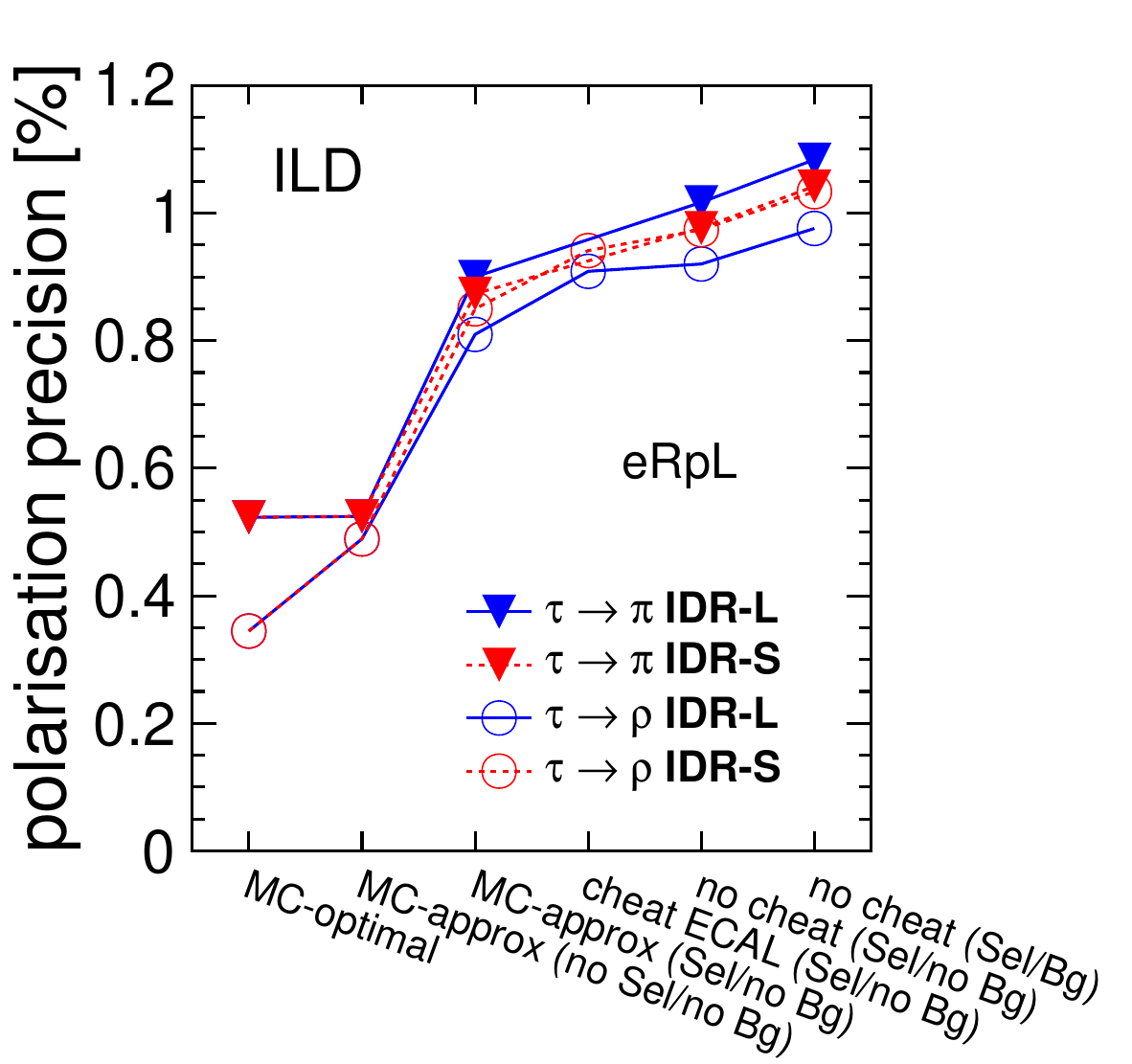} \\
\includegraphics[width=0.45\textwidth]{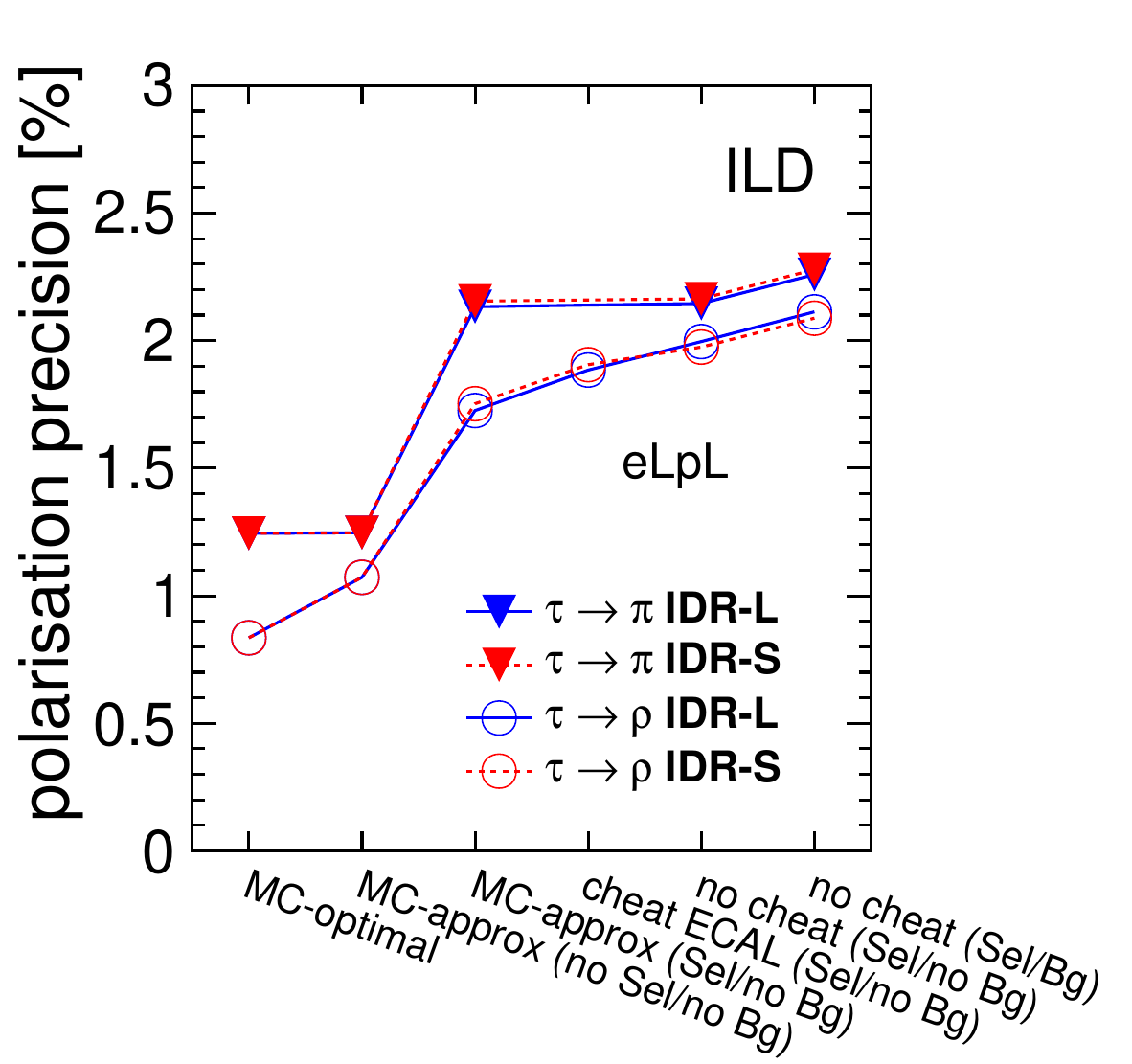}
\includegraphics[width=0.45\textwidth]{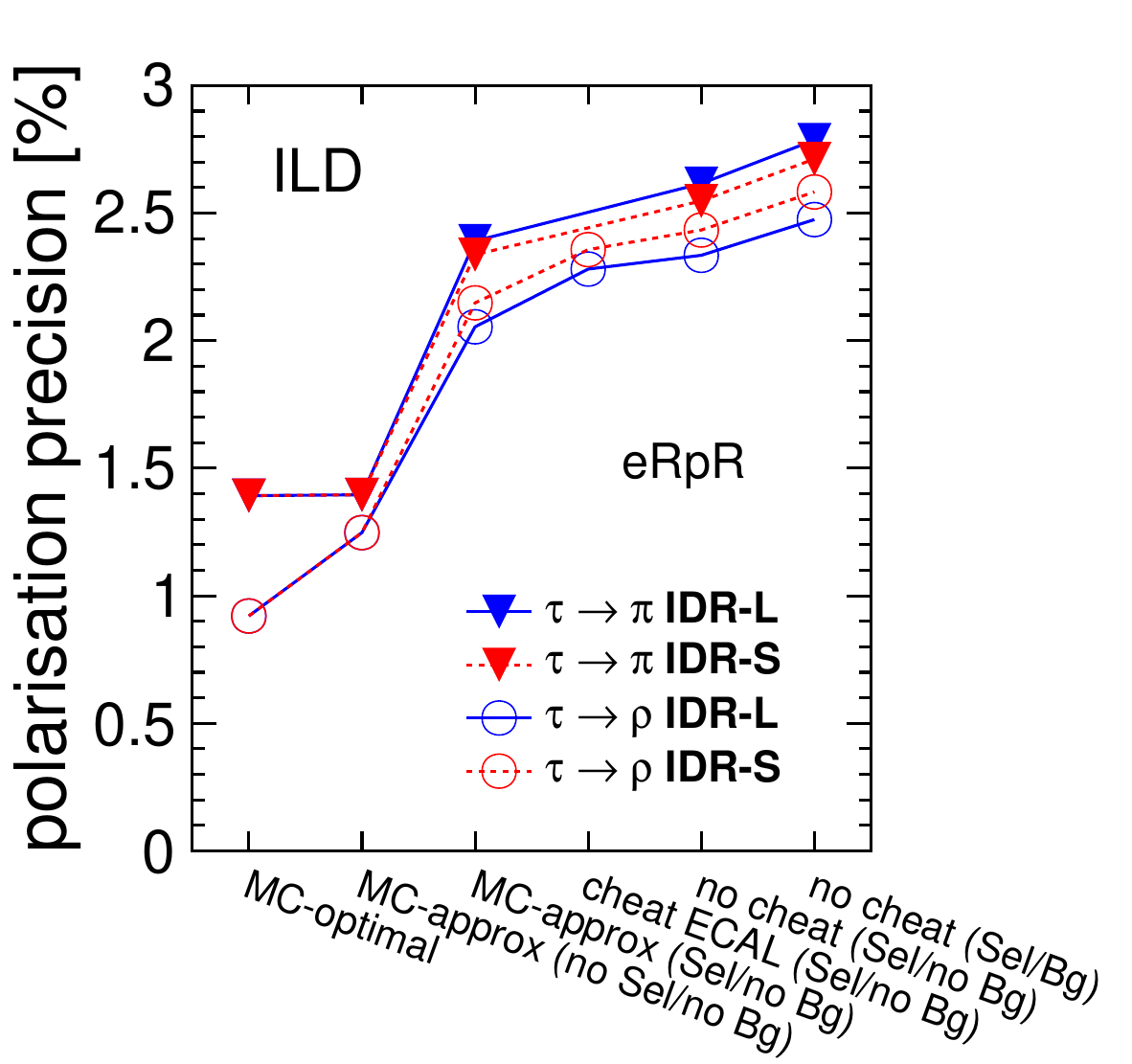}
\caption{
Estimated polarisation precision at different levels of cheating, for different decay modes, detector models, and polarisation sets.
}
\label{fig:fitResults-cheatingFIT}
\end{figure}

These templates were used to run a series of pseudo-experiments, in which event samples were randomly generated from the smoothed templates.
Each sample was then fitted to the same templates, allowing the relative contributions of the two helicity states (i.e. the tau polarisation) to vary. 
This procedure was repeated at various levels of ``cheating''. 
In the first, we use the ``optimal'' form for the polarimeters, including the neutrino momenta, and using the MC truth for 4-momenta.
In the ``approximate'' case, we use the polarimeter forms introduced in Sec.~\ref{sec:pol} and given in the appendix. 
We also investigate the effect of event selection, photon energy resolution, and selected backgrounds.
The precision on the determination of this polarisation (the mean of the distribution of parameter fit errors in the ensemble of pseudo-experiments) 
are shown in Table~\ref{tab:polFitUPDATE} and
graphically in 
Fig.~\ref{fig:fitResults-cheatingFIT}.

The \trn\ decay mode has somewhat better precision than \tpn, thanks to its larger branching ratio.
The simpler \tpn\ mode does nonetheless make a significant contribution, thanks to its more powerful polarimetry and more precise reconstruction.
The final experimental precision obtained by combining the two decay modes is around 0.6\% for the \eLpR\ and \eRpL\ portions, and around 1.5\% (1.7\%) for \eLpL\ (\eRpR).
This precision on the polarisation is around twice worse than the theoretically possible value obtained ignoring experimental effects.
The various considered experimental effects which contribute to this degradation all do so to a somewhat similar extent. 

There is no clear advantage for either of the two detector models. 
Although some differences were seen at the intermediate stages of the analysis (e.g. in the identification of the tau lepton decay modes),
only very small differences are seen in the final sensitivity to the tau polarisation.

\begin{table}
\scriptsize
\centering

\begin{tabular}{|l||rr|rr|rr|rr|rr|}
\hline
Cheat           & optimal-MC & \multicolumn{3}{|c|}{APPROX}  & \multicolumn{2}{|c|}{ECAL}  &  \multicolumn{4}{|c|}{NONE} \\
\hline
evtSel          & \multicolumn{2} {|c|}{NO} & \multicolumn{8} {|c|}{YES} \\
\hline
BG              & \multicolumn{8} {|c|}{NO} & \multicolumn{2} {|c|}{YES} \\
\hline
detector        & \multicolumn{2} {|c|}{} & IDR-L & IDR-S & IDR-L & IDR-S & IDR-L & IDR-S & IDR-L & IDR-S  \\

\hline  
\hline

\multicolumn{11}{|c|}{} \\
\multicolumn{11}{|c|}{\tpn} \\
\multicolumn{11}{|c|}{} \\
\hline
 & \multicolumn{10}{|l|}{Number of tau jet candidates} \\
\hline
\eLpR  & $  114963 $ & $  114943 $ & $   73964 $ & $   73855 $ & & & $  101035 $ & $  102846 $ & $  118702 $ & $  121804 $  \\ 
\eRpL  & $   96002 $ & $   95911 $ & $   61519 $ & $   60632 $ & & & $   80695 $ & $   80815 $ & $   91878 $ & $   92649 $  \\ 
\eLpL  & $   17240 $ & $   17235 $ & $   11087 $ & $   11055 $ & & & $   15079 $ & $   15323 $ & $   17687 $ & $   18120 $  \\ 
\eRpR  & $   15085 $ & $   15073 $ & $    9673 $ & $    9552 $ & & & $   12767 $ & $   12819 $ & $   14643 $ & $   14801 $  \\ 
\hline
& \multicolumn{10}{|l|}{Mean statistical error on tau polarisation \% } \\
\hline
\eLpR  & $ 0.43 $ & $ 0.42 $ & $ 0.51 $ & $ 0.50 $ & & & $ 0.79 $ & $ 0.80 $ & $ 0.85 $ & $ 0.87 $  \\ 
\eRpL  & $ 0.50 $ & $ 0.50 $ & $ 0.61 $ & $ 0.63 $ & & & $ 1.03 $ & $ 1.02 $ & $ 1.07 $ & $ 1.07 $  \\ 
\eLpL  & $ 1.21 $ & $ 1.19 $ & $ 1.46 $ & $ 1.43 $ & & & $ 2.17 $ & $ 2.19 $ & $ 2.34 $ & $ 2.38 $  \\ 
\eRpR  & $ 1.37 $ & $ 1.35 $ & $ 1.68 $ & $ 1.71 $ & & & $ 2.67 $ & $ 2.66 $ & $ 2.81 $ & $ 2.82 $  \\ 

\hline
\hline

\multicolumn{11}{|c|}{} \\
\multicolumn{11}{|c|}{\trn} \\
\multicolumn{11}{|c|}{} \\
\hline
 & \multicolumn{10}{|l|}{Number of tau jet candidates} \\
\hline  
\eLpR  & $  262263 $ & $  251882 $ & $  158639 $ & $  160448 $ & $  145649 $ & $  146563 $ & $  150564 $ & $  151797 $ & $  175626 $ & $  177120 $  \\ 
\eRpL  & $  215809 $ & $  211529 $ & $  133288 $ & $  131849 $ & $  122022 $ & $  121749 $ & $  124653 $ & $  124952 $ & $  141555 $ & $  142186 $  \\ 
\eLpL  & $   39266 $ & $   37796 $ & $   23806 $ & $   24019 $ & $   21849 $ & $   21966 $ & $   22557 $ & $   22728 $ & $   26267 $ & $   26486 $  \\ 
\eRpR  & $   33987 $ & $   33211 $ & $   20925 $ & $   20769 $ & $   19165 $ & $   19146 $ & $   19613 $ & $   19677 $ & $   22388 $ & $   22507 $  \\ 
\hline  
& \multicolumn{10}{|l|}{Mean statistical error on tau polarisation \% } \\
\hline  
\eLpR  & $ 0.30 $ & $ 0.40 $ & $ 0.49 $ & $ 0.46 $ & $ 0.55 $ & $ 0.58 $ & $ 0.65 $ & $ 0.66 $ & $ 0.70 $ & $ 0.71 $  \\ 
\eRpL  & $ 0.35 $ & $ 0.48 $ & $ 0.61 $ & $ 0.61 $ & $ 0.71 $ & $ 0.72 $ & $ 0.78 $ & $ 0.78 $ & $ 0.83 $ & $ 0.82 $  \\ 
\eLpL  & $ 0.83 $ & $ 1.07 $ & $ 1.34 $ & $ 1.33 $ & $ 1.49 $ & $ 1.55 $ & $ 1.73 $ & $ 1.75 $ & $ 1.85 $ & $ 1.88 $  \\ 
\eRpR  & $ 0.92 $ & $ 1.25 $ & $ 1.58 $ & $ 1.59 $ & $ 1.82 $ & $ 1.83 $ & $ 1.98 $ & $ 2.04 $ & $ 2.10 $ & $ 2.16 $  \\ 
\hline  
\end{tabular}
\caption{
Number of tau jet candidates and the estimated precisions on the polarisation measurement in \tpn\ (upper) and \trn\ (lower) decays, at various different levels of cheating.
}
\label{tab:polFitUPDATE}
\end{table}

\clearpage

\section{Conclusion}

The reconstruction and selection of high mass pairs of tau leptons at ILC-500 was investigated.
Polarimeters were reconstructed in the \tpn\ and \trn\ decay modes, and used to estimate the tau polarisation.
The final experimental sensitivity to the tau polarisation around 0.5\% in the majority \eLpR\ and \eRpL\
portions of the forseen integrated luminosity, and around 1.5\% for the \eLpL\ and \eRpR\ portions, where less
integrated luminosity was assumed. 
Contributions to the experimental sensitivity arise
from various sources (selection ineffeciency, backgrounds, tau decay mode identification, photon energy resolution) at rather similar levels.

The performance of two detector models, IDR-L and IDR-S, was compared. Although the larger IDR-L model performed somewhat better
at reconstructing the number of photons and at identifying tau decay modes, the final precision on the tau polarisation
measurement of the two models is very similar.

\section*{Acknowledgments}

We thank our ILD colleagues, in particular M.~Berggren, for helpful discussions on the analysis and this manuscript.
We would like to thank the LCC generator working group and the ILD software working group for providing the simulation and reconstruction tools and 
producing the Monte Carlo samples used in this study.
This work has benefited from computing services provided by the ILC Virtual Organization, supported by the national resource providers of the 
EGI Federation and the Open Science GRID.

\appendix

\section{Explicit expressions for polarimeters}

\label{appendix:1}

We here reproduce the expressions given in \cite{duflot}, and used in the present analysis, for polarimeters in \tpn\ and \trn\ decays.

\subsection{\tpn}

The polarimeter $\omega_\pi$ can then be written as
\begin{equation}
\omega_\pi = 2 x - 1
\end{equation}
where $x = E_\pi / E_\tau $, the ratio of the pion energy to that of the $\tau$ (which is assumed to be half the centre-of-mass energy in the present case of di-tau production).

\subsection{\trn}

Define $Q^2$ as the squared invariant mass of the two-pion system, 
$\theta$ as the angle between the direction of the hadronic system and the $\tau$ momentum in the $\tau$ rest frame, 
and $\beta$ as the angle between the directions of the charged pion and the total hadronic momenta, in the hadronic rest frame.
The angle $\psi$, between the $\tau$ and (minus) the hadronic momentum, in the hadronic rest frame, which can in the case
of di-tau production at known centre-of-mass energy, be calculated as
\begin{equation}
\cos \psi = \frac{ x ( m_\tau^2 + Q^2 ) - 2 Q^2 }{ (m_\tau^2 - Q^2 ) \sqrt{ x^2 - 4 Q^2 / s } }
\end{equation}
where $x - 2 E_h / \sqrt{s} $, with $E_h$ being the energy of the hadronic system in the lab frame and $s$ the squared centre-of-mass energy.

The polarimeter $\omega_\rho$ can then be written as
\begin{equation}
\omega_\rho = 
\frac{ 
  (-2+ \frac{m_\tau^2}{Q^2} + 2 (1 + \frac{m_\tau^2}{Q^2})\frac{3 \cos \psi - 1}{2} \frac{3 \cos^2 \beta - 1}{2}  ) \cos \theta + 3 \sqrt{ \frac{m_\tau^2}{Q^2}} \frac{3 \cos^2 \beta - 1}{2} \sin 2\psi \sin\theta  
}
{
2 + \frac{m_\tau^2}{Q^2} - 2  (1 - \frac{m_\tau^2}{Q^2}) \frac{3 \cos \psi - 1}{2} \frac{3 \cos^2 \beta - 1}{2}  
}
\end{equation}
(corresponding to eq.~3.11 of \cite{duflot}).

\section{More sophisticated reconstruction methods which don't (yet) work very well}

In the case of single pion decay, the optimal polarimeter is very simple: the ratio of the pion energy to the beam energy (assuming that the taus are exactly back-to-back).
In the case of $\rho$ decay, full sensitivity to the tau lepton polarisation requires reconstruction of the neutrino momenta.  
Attempts were made to use and develop such methods, which so far achieved only limited success. We report on them for completeness.

In the case of back-to-back taus of known energy, 
the neutrino momenta can be estimated by constraining the tau lepton energies
(250 GeV), their being back-to-back, and imposing the known tau mass. Zero or two (possibly identical) solutions occur,
which correspond to the momentum lying along intersection of 2 cones around the visible tau momenta.
It is not clear to me how to choose between these 2 solutions, although the sign of impact parameters may be of use~\cite{kuhn}.
In events with only $\pi$ and $\rho$ decays, a good solution was found in only around one third of events.

An alternative method is based on the impact parameter of the charged particles, as described in \cite{jeans-iptau} and used in \cite{jeans-higgscp}.
We know that the tau must decay somewhere on the charged particle's trajectory. 
If we know the IP position, we can therefore constrain the tau momentum to lie in the plane defined by the trajectory and the IP.
In the events discussed here, the IP cannot be directly measured, since no prompt tracks are produced in the reaction.
However, the small ILC interaction region provides a strong constraint in the transverse plane. To estimate the position in $z$,
we simply take the average of the tau jet seeds' $z_0$ track parameters. 
Better results may come from scanning along $z$, and finding the ``best'' solution, or by requiring a multi-prong decay of one of the taus in an event.
By assuming a single neutrino per tau decay and imposing the tau mass, we find neutrino momenta which 
result in positive tau decay lengths and minimise the transverse momentum of the tau-tau system.
Using this method, again only about a third of events could be reconstructed.

In the future, it would be interesting to combine elements of these two methods, which will hopefully result in a more robust technique
for fully reconstructing the tau momentum.

\end{document}